\documentstyle[12pt,epsf]{article}
\newcommand {\m}{\mu}
\newcommand {\n}{\nu}
\newcommand {\pl}{\partial}
\newcommand {\p} {\phi}

\newcommand {\al}{\alpha}
\newcommand {\be}{\beta}
\newcommand {\ga}{\gamma}
\newcommand {\Ga}{\Gamma}

\newcommand {\la}{\lambda}

\newcommand {\si}{\sigma}
\newcommand {\th}{\theta}

\newcommand {\om}{\omega}

\newcommand {\ep}{\epsilon}

\newcommand {\e} {\mbox{\rm e}}
\newcommand {\na}{\nabla}
\newcommand {\del}  {\delta}

\newcommand {\mn}{{\mu\nu}}
\newcommand {\ls}   {{\lambda\sigma}}
\newcommand {\ab}   {{\alpha\beta}}
\newcommand {\Tr}{{\rm Tr\,}}
\newcommand {\half}{ {\frac{1}{2}} }
\newcommand {\fourth} {\frac{1}{4} }
\newcommand {\sqg} {\sqrt{g}}
\newcommand {\fg}  {\sqrt[4]{g}}
\newcommand {\invfg}  {\frac{1}{\sqrt[4]{g}}}

\newcommand {\Lcal}{{\cal L}}

\newcommand {\Dcal}{{\cal D}}

\newcommand {\Dvec}{{\vec D}}

\newcommand {\Vvec}{{\vec V}}

\newcommand {\ptil}{{\tilde \phi}}

\newcommand {\wbar}  {{\bar w}}
\newcommand {\change} {\leftrightarrow}
\newcommand {\ra} {\rightarrow}
\newcommand {\pr}   {{\quad .}}
\newcommand {\com}  {{\quad ,}}
\newcommand {\q}    {\quad}

\newcommand {\nl}    {\newline}
\newcommand {\nn}    {\nonumber}
\newcommand {\ul}    {\underline}
\newcommand {\vs}[1]  { \vspace*{#1 cm} }
\newcounter{eq}
\newcounter{sc}

\newcommand {\MPL}  {Mod.Phys.Lett.}
\newcommand {\NP}   {Nucl.Phys.}
\newcommand {\PL}   {Phys.Lett.}
\newcommand {\PR}   {Phys.Rev.}
\newcommand {\PRL}   {Phys.Rev.Lett.}

\newcommand {\CQG}  {Class.Quantum.Grav.}
\newcommand {\IJMP}  {Int.J.Mod.Phys.}
\newcommand {\JMP}  {J.Math.Phys.}
\def\overleftrightarrow#1{\vbox{\ialign{##\crcr
 $\leftrightarrow$\crcr\noalign{\kern-1pt\nointerlineskip}
 $\hfil\displaystyle{#1}\hfil$\crcr}}}

\newlength{\minitwocolumn}
\begin{document}

%%%%%%%%%%%%%%%%%%%%%%%%%%%%%%%%%%%%%%%%%%%%%%%%%%%%%%%%%%%%%%%%%%
%%%%%%%%%%%%%%%%%%%%%%%% Title %%%%%%%%%%%%%%%%%%%%%%%%%%%%%%%%%%%
%%%%%%%%%%%%%%%%%%%%%%%%%%%%%%%%%%%%%%%%%%%%%%%%%%%%%%%%%%%%%%%%%%

\begin{flushright}
US-98-07\\
October,1998\\
hep-th/9810256
%hep-th/9707025
\end{flushright}
\vspace{24pt}

%\magnification=\magstep1
\pagestyle{empty}
\baselineskip15pt
%\font\cmssB=cmss17
%\font\cmssS=cmss10

\begin{center}
{\large\bf Weyl Anomaly in Higher Dimensions and \\
Feynman Rules in Coordinate Space
\vskip 1mm
}

\vspace{10mm}

Shoichi ICHINOSE
          \footnote{
On leave of absence from 
Department of Physics, University of Shizuoka,
Yada 52-1, Shizuoka 422, Japan (Address after Nov.6, 1998).\nl          
E-mail address:\ ichinose@u-shizuoka-ken.ac.jp
                  }
and Noriaki IKEDA${}^{\dag}$ 
\footnote
{ E-mail address:\ nori@kurims.kyoto-u.ac.jp }
\\
\vspace{5mm}
Physics Department, Brookhaven National Laboratory,\\
Upton, NY 11973, USA
%Department of Physics, University of Shizuoka,\\
%Yada 52-1, Shizuoka 422, Japan          
\\
and \\
${}^{\dag}$Research Institute for Mathematical Sciences \\
Kyoto University, Kyoto 606-01, Japan

\end{center}

\vspace{15mm}
\begin{abstract}
An algorithm to obtain
the Weyl anomaly in higher dimensions is presented.
It is based on the heat-kernel method. Feynman rules,
such as the vertex rule and the propagator rule, 
are given in (regularized) coordinate space. 
Graphical calculation is introduced.
The 6 dimensional scalar-gravity theory
is taken as an example, and its explicit result is obtained.

\vspace{15mm}

PACS NO:\ 02.70.-c,\ 04.60.-m,\ 11.15.Bt,\ 11.25.Db,\ 11.30.-j\nl
Keywords:
\ Weyl anomaly,\ Higher dimensional field theories,
\ Heat kernel,\ Feynman rule,\ Graphical representation\nl
\end{abstract}

%\pagestyle{empty}
%\newpage
\pagestyle{plain}
\pagenumbering{arabic}
%\setcounter{page}{1}

%%%%%%%%%%%%%%%%%%%%%%%%%%%%%%%%%%%%%%%%%%%%%%%%%%%%%%%%%%%%%%%%%%
%%%%%%%%%%%%%%%%%%%%%%%% Article %%%%%%%%%%%%%%%%%%%%%%%%%%%%%%%%%
%%%%%%%%%%%%%%%%%%%%%%%%%%%%%%%%%%%%%%%%%%%%%%%%%%%%%%%%%%%%%%%%%%

\rm
%%%%%%%%%%%%%%%%%%%%%%%%%%%%%%%%%%%%%%%%%%%%%%%%%%%%%%%%%%%%%%%%%%%%%
%%%%%%%%%%%%%%%%%%%%%%%%%%%%%%   SEC  1    %%%%%%%%%%%%%%%%%%%%%%%%%%
%%%%%%%%%%%%%%%%%%%%%%%%%%%%%%%%%%%%%%%%%%%%%%%%%%%%%%%%%%%%%%%%%%%%%
\section{Introduction}
The anomaly phenomenon is, as well as the renormalization,  
one of important aspects of the quantum field theories. 
It is a local quantum effect originating from  regularization
of the continuous space-time. 
We may say it is the problem of how to define the space-time
as the continuum.
Generally the symmetry, imposed at the classical
level, does not hold at the quantum level due to the anomaly.
As for the chiral anomaly, which is the anomaly concerned with
the $\ga_5$-symmetry(chiral symmetry), 
the general form is well-established in arbitrary even dimensions
\cite{AW}.
It is fixed, except for an overall coefficient, by its cohomology structure 
and is given in a beautiful form
in terms of the differential form:\ 
$\mbox{tr}(R\wedge R\wedge ...\wedge R), 
\mbox{tr}(F\wedge F\wedge ...\wedge F)$.
The topological nature due to the $\ga_5$-projection 
is the origin of its simplicity. On the other hand, 
as for the Weyl anomaly,  
the general form in higher dimensions is not so simple. 
This is the anomaly concerned
with local scale transformations.
Cohomology analysis restricts the Weyl anomaly to some
extent, but there generally remains numerous candidate
terms in  higher dimensions\cite{BPB,AKMM95,KMM96,IO98}. 
The undetermined coefficients of those terms must be
fixed in some way.
%The general form cannot be uniquely fixed
% by the cohomology
The situation originates from the "rich" structure of the
functional space on which the scale transformation acts.
In the development of the string theory or M-theory, 
it becomes more and more necessary to investigate the
Weyl anomaly in higher dimensional field theories\cite{SD97}.
At this circumstance we present a new formalism for it.

The Weyl anomaly itself is not harmful to the construction
of physical theories because we know the realistic theories
have some dimensional parameters, such as the mass or the
cosmological constant, and they break the Weyl symmetry. 
However when we try to understand the origin of those massive
parameters, the Weyl anomaly is so important because it triggers
the symmetry breaking of the conformal invariant vacuum.
In this sense 
the Weyl anomaly can be regarded as a "softer" anomaly than the chiral one. 
The latter threatens the consistency of the theory
whereas the former does not. The nonrenormalization theorem
is valid for the latter, whereas generally not for the former.

In supersymmetric theories,
however, it is also known that both anomalies make a super multiplet
together with the super current\cite{FZ75}. 
Under the supersymmetric treatment
both anomalies are intimately related.
In this connection there has been a lot of work.
Because our interest here is to establish a general algorithm
to treat the Weyl anomaly of the higher dimensional theories,  
we do not consider supersymmetry here. We suppose, using
the present result, we  can investigate how the Weyl anomaly
is constrained by the supersymmetry.

We analyze the Weyl anomaly in the heat-kernel regularization.
The content is essentially the same as, or overlaps with, 
topics with other names such as
the effective action analysis, the background field method, 
the Schwinger-De Witt technique, Seeley series, etc.
They all treat some background functional which represents
(1-loop) quantum effects. 
It has a long history since Schwinger\cite{S}, 
DeWitt\cite{DeW} and Seeley\cite{Seel}.
Especially the first author, in 1951, treated
quantum matter theories in external gauge fields 
keeping the gauge invariance of the effective action. 
He introduced an additional "coordinate"
into the ordinary space(-time) coordinates, 
called "proper time"  or "temperature" in a natural way. 
It was exploited by Alvarez-Gaum\'{e}
and Witten \cite{AW} in the modern problem of the chiral anomaly of the
gravitational theories. 
Later it was generalized to the Weyl anomaly by a SUNY group
\cite{fB91,BN93}. 

The effective action has been analyzed in various
ways. Its divergent part was systematically and algorithmically
analyzed by t'Hooft\cite{tH} in 1973 and with Veltman\cite{tHV} in 1974
using dimensional regularization. 
As will be shown later, the Weyl anomaly
is obtained from the "heat-kernel", which is the 1-loop
effective action with the
proper time as the ultraviolet regularization. 
The evaluation of the heat-kernel
has been well-developed by many people
\cite{Seel,Gil,IGA86,IGA91,BV85,BPSN95,BPSN96,SY97}.
Each of them has their own characteristic formalisms.
Generally they start with expressing a differential operator
by some general background quantities. 
Most of them put emphasis on the covariance in the formulae
in the intermediate stages. As originally done by tHooft\cite{tH},
the "intermediate covariance" is powerful as far as 
the lower-order calculations, 
such as the heat-kernel on 2 and 4 dimensions, are concerned. 
However, at higher-order (this means higher dimensions too), 
it hinders one from performing explicit
calculations because many more "intermediate invariants" appear
than those appearing in the final explicit answer.
The important thing is not the formulae at intermediate
stages, but the final explicit result.
Aiming at the treatment of the Weyl anomaly in higher dimensions,
we present a new algorithm which is based on a simple
weak-field expansion. 
Of course, we assume the invariance of the general coordinate
and gauge symmetries at the final result, 
because it is guaranteed by the background (effective action) formalism
\cite{DeW,tH,BOS92}.

The following are new points of the present approach:\ 
\begin{enumerate}
\item \ We take a formalism in the regularized coordinate 
space, that is the space with the ordinary space(-time) coordinates and
the proper time\cite{foot1}. 
The Feynman rules in the coordinate space are given
for the anomaly (effective action) calculation\cite{foot1bb}.
The ultra-violet regularization is done with the Schwinger's
parameter ( proper time ). The regularization lets us  
properly evaluate
both chiral and Weyl anomalies\cite{II96}.\cite{foot1b}  
The Feynman rules help us to investigate the Weyl
anomaly in higher dimensional theories.
\item \ The graphical representation is taken at all stages. 
The familiar procedures in the ordinary perturbative treatment, such as
the propagator rule and
the vertex rule, are all graphically represented in coordinate space.
The expanded terms themselves are also graphically represented.
Especially we introduce "graphical calculation".
This makes the higher-order calculation transparent and
the Weyl anomaly calculation in the higher dimensions tractable.
\item \ 
We have not introduced any covariant quantities in
intermediate stages.
It is based on the weak-field perturbation. Although loosing 
manifest general covariance in intermediate stages, 
the expanded elements have the
simplest symmetry (w.r.t. the suffix permutation ) and are preferable
for the computer algorithm.
\item \ 
We focus only on a special type of term:\ $(\pl\pl h)^n$, 
in the $n$ dimensional
space. These are the lowest order of the product of $n$ Riemann tensors. 
We demonstrate they can determine the Weyl anomaly up to ``trivial terms''.
\end{enumerate}

In Sec.2 we briefly give the present formalism of the Weyl anomaly.
The anomaly is given by the trace of the heat-kernel. In Sec.3 we
present the Feynman rules to compute the heat-kernel {\it to any
higher order}. They are naturally expressed in the {\it regularized}
coordinate space, that is, the space of n-dim ordinary space-coordinates
plus the the proper time. As always in perturbative approaches, 
a graphical representation helps to 
systematically compute the heat-kernel at the
higher order. The Taylor expansion, with respect
to the ``regularization'' parameter $t$, is explained in Sec.4. 
The more the number of the space dimensions, the more expansion
is necessary. We apply the general algorithm to the 6 dim scalar-gravity
theory in Sec.5. We focus on four special graphs and introduce a 
new graphical calculation. The final result of the Weyl anomaly for
the model is obtained in Sec.6. We conclude in Sec.7.
Four appendices are provided to supplement the text. The propagator
rules are given in App.A. Some supplementary calculation, relegated
by Sec.5 of the text, is done in App.B. Weak field expansion of
the 6 dim scalar-gravity is graphically given in App.C. The table
of coefficients, when some general invariants are weak-field expanded,
is given in App.D.

%%%%%%%%%%%%%%%%%%%%%%%%%%%%%%%%%%%%%%%%%%%%%%%%%%%%%%%%%%%%%%%%%%%%%
%%%%%%%%%%%%%%%%%%%%%%%%%%%%%%   SEC  2    %%%%%%%%%%%%%%%%%%%%%%%%%%
%%%%%%%%%%%%%%%%%%%%%%%%%%%%%%%%%%%%%%%%%%%%%%%%%%%%%%%%%%%%%%%%%%%%%
\section{ Formalism }
Let us explain the present formulation of anomalies taking a simple
example :\ n-dim Euclidean gravity-scalar coupled system.
(see Ref.\cite{II96})
%****(f.1)%%%%%%%%%%%%%%%%%%%%
\begin{eqnarray}
\Lcal[g_\mn,\p]=\sqg \p(-\half \na^2+\half q  R)\p
\equiv\half\ptil\Dvec\ptil\com
\nn\\
q=-\frac{n-2}{4(n-1)}\com\q
{\Dvec}_x\equiv \fg(-\na_x^2-\frac{n-2}{4(n-1)}R(x))\invfg\com 
\label{f.1}
\end{eqnarray}
%%%%%%%%%%%%%%%%%%%%%%%%%%%%%
where $g_\mn$\ and $\p$\ are the metric field and 
the scalar field respectively.
We have introduced $\ptil\equiv \fg~\p$ for the
measure $\Dcal \ptil$ to be general coordinate (BRS) invariant
\cite{F83}.
This Lagrangian is invariant under the local Weyl transformation:
%****(f.2)%%%%%%%%%%%%%%%%%%%%
\begin{eqnarray}
g^\mn(x)'=e^{2\al(x)}g^\mn(x)\com\q 
\ptil(x)'=e^{-\al(x)}\ptil(x)\q
(\ \p(x)'=e^{\frac{n-2}{2}\al(x)}\p(x)\ )
                                                  \com\label{f.2}
\end{eqnarray}
%%%%%%%%%%%%%%%%%%%%%%%%%%%%%
where $\al(x)$ is the parameter of the local Weyl transformation.

Even if the Lagrangian is Weyl invariant, the quantum theory
is generally not (Weyl anomaly). The effective action $\Ga[g_\mn]$
defined by
%****(f.3)%%%%%%%%%%%%%%%%%%%%
\begin{eqnarray}
e^{-\Ga[g_\mn]}=\int\Dcal\ptil~
exp\{~-\int d^nx\Lcal[g_\mn,\p]~\}\com  \label{f.3}
\end{eqnarray}
%%%%%%%%%%%%%%%%%%%%%%%%%%%%%
changes as 
%****(f.4)%%%%%%%%%%%%%%%%%%%%
\begin{eqnarray}
e^{-\Ga[g_\mn']}=\int\Dcal\ptil'exp\{~-\int d^nx~\Lcal[g_\mn',\p']~\}\nn\\
=\int \Dcal\ptil(x)~det\frac{\pl\ptil'(y)}{\pl\ptil(x)}
exp\{~-\int d^nx\Lcal[g_\mn,\p]~\}\pr  \label{f.4}
\end{eqnarray}
%%%%%%%%%%%%%%%%%%%%%%%%%%%%%
The Weyl anomaly is given by the Jacobian\cite{F79,F85}.
%****(f.4b)%%%%%%%%%%%%%%%%%%%%
\begin{eqnarray}
J\equiv
\det\left[ \frac{\pl\ptil'(y)}{\pl\ptil(x)} \right]
=\exp \left( -\Tr\al(x)\del^n(x-y)+O(\al^2) \right)
\pr        \label{f.4b}
\end{eqnarray}
%%%%%%%%%%%%%%%%%%%%%%%%%%%%%
Taking the heat-kernel regularization, we obtain the expression
for the Weyl anomaly as
%****(f.5)%%%%%%%%%%%%%%%%%%%%
\begin{eqnarray}
J
=\exp (-\lim_{t\ra +0}\Tr~[\al(x)G(x,y;t)]+O(\al^2))\com\nn\\
\mbox{Weyl Anomaly}
=\left.\frac{\del\Ga}{\del \al(x)}\right|_{\al=0}
=-\left.\frac{\del\ln J}{\del \al(x)}\right|_{\al=0}
=-2g^\mn <T_\mn>=\lim_{t\ra +0}\Tr G(x,y;t)\com        \label{f.5}
\end{eqnarray}
%%%%%%%%%%%%%%%%%%%%%%%%%%%%%
where $G(x,y;t)$ is the heat-kernel defined by
%****(f.7e)%%%%%%%%%%%%%%%%%%%%
\begin{eqnarray}
(\frac{\pl}{\pl t}+\Dvec_x)G(x,y;t)=0\com\q
G(x,y;t)({\overleftarrow {\frac{\pl}{\pl t}}}
        +{\overleftarrow D}^{\dag}_y)=0\com\nn\\            
\lim_{t\ra +0}~G(x,y;t)=\del^n(x-y)\pr		
		                           \label{f.7e}
\end{eqnarray}
%%%%%%%%%%%%%%%%%%%%%%%%%%%%%
where ${\overleftarrow D}_x$\ means it operates on the left .
The parameter $t$\ \ is regarded here as a regularization parameter
and is called Schwinger's {\it proper time} \cite{S}.
The last equation expresses the regularization of the delta function
$\del^n(x-y)$ \cite{AAR,DH}.
$G(x,y;t)$ can be symbolically written as
%****(f.6)%%%%%%%%%%%%%%%%%%%%
\begin{eqnarray}
G(x,y;t)\equiv <x|e^{-t\Dvec}|y>\com\q t>0\pr\label{f.6}
\end{eqnarray}
%%%%%%%%%%%%%%%%%%%%%%%%%%%%%
We note here the physical dimension of the space coordinate $x^\m$\ 
and the proper-time $t$ are 
%****(f.7)%%%%%%%%%%%%%%%%%%%%
\begin{eqnarray}
[x^\m]=L\com\q [t]=L^2\com
\label{f.7}
\end{eqnarray}
%%%%%%%%%%%%%%%%%%%%%%%%%%%%%
where $L$ is some length. For other (conformally invariant) theories,
the Weyl anomaly is obtained by replacing the operator $\Dvec$ above.
See ref.\cite{II96} for detail.

%%%%%%%%%%%%%%%%%%%%%%%%%%%%%%%%%%%%%%%%%%%%%%%%%%%%%%%%%%%%%%%%%%%%%
%%%%%%%%%%%%%%%%%%%%%%%%%%%%%%   SEC  3    %%%%%%%%%%%%%%%%%%%%%%%%%%
%%%%%%%%%%%%%%%%%%%%%%%%%%%%%%%%%%%%%%%%%%%%%%%%%%%%%%%%%%%%%%%%%%%%%
\section{ Feynman rules in coordinate space  }
\subsection{Heat-kernel and Feynman graph}   %%%%%%%%%%%%%%%%%%%%%%%%%%%
Let us solve Eq.(\ref{f.7e}) in weak-field perturbation theory. 
For a general theory with the derivative couplings up to the second order,
the operator $\Dvec_x^{ij}$\ can be always 
divided into the field-independent (free) part
and the field-dependent (perturbation) one. 
%****(f.10)%%%%%%%%%%%%%%%%%%%%
\begin{eqnarray}
\Dvec_x^{ij} &=& -\del_\mn\del^{ij}\pl_\m\pl_\n-\Vvec^{ij}(x)\com\nn\\
\Vvec^{ij}(x) &
\equiv & W_\mn^{ij}(x)\pl_\m\pl_\n+N_\m^{ij}(x)\pl_\m+M^{ij}(x)\com\nn\\
i,j=1,2,\cdots,N\com
\label{f.10}
\end{eqnarray}
%%%%%%%%%%%%%%%%%%%%%%%%%%%%%%
where $W_\mn^{ij},\ N_\m^{ij}\ \ \mbox{and}\ \ M^{ij}$\ are external fields
( background coefficient fields ) and the field suffixes (such as
a fermion suffix and a vector suffix) $i,j$ are introduced for the general use.
In the present example (\ref{f.1}), 
the above quantities are explicitly written as ($N=1$)
%****(f.10b)%%%%%%%%%%%%%%%%%%%%
\begin{eqnarray}
\Vvec(x) &=& \fg(\na^\mu\na_\m-qR)\invfg-\del_\mn\pl_\m\pl_\n\com\nn\\
W_\mn &=& g^\mn-\del_\mn=-h_\mn+h_{\m\la}h_{\la\n}+O(h^3)\com\label{f.10b}\\
N_\la &=& -g^\mn\Ga^\la_\mn-g^{\la\m}\Ga_\mn^\n
=-\pl_\m h_{\la\m}+O(h^2)\com\nn\\
M &=& -qR+\fourth g^\mn\{\Ga_{\m\la}^\la \Ga_{\n\si}^\si
+2\Ga_{\mn}^\la \Ga_{\la\si}^\si-2\pl_\n \Ga_{\m\la}^\la\}\nn\\
 &=& -q(\pl^2h-\pl_\al\pl_\be h_\ab)-\fourth \pl^2h+O(h^2)\com\nn
\end{eqnarray}
%%%%%%%%%%%%%%%%%%%%%%%%%%%%%
where 
\ $g_\mn=\del_\mn+h_\mn\ ,\ h\equiv h_{\m\m}$. 
The graphical representation of $W_\mn,\ N_\la$ and $M$ above
is given, up to $h^3$-order, for the n=6 dim case in App.C:
\ Fig.16 for $W_\mn$ and $N_\la$, Fig.17-21 for $M$. 
The usage of general coefficients 
$W_\mn^{ij},N_\m^{ij}$\ and
$M^{ij}$, instead of their concrete contents, 
makes it possible to do the general treatment valid for many theories
\cite{II96,DAMTP9687,foot2}
. 
All equations below are valid for all theories so far as
they have no higher-derivative interactions.
Let us  solve the differential equation (\ref{f.7e}) 
perturbatively for the case of weak external fields 
($W_\mn^{ij},N_\m^{ij},M^{ij}$). (In the present example, this corresponds
to perturbation around {\it flat} space.) 
The differential equation (\ref{f.7e}) becomes
%****(f.11)%%%%%%%%%%%%%%%%%%%%
\begin{eqnarray}
(\frac{\pl}{\pl t}-\pl^2)G^{ij}(x,y;t) = \Vvec^{ik}(x)G^{kj}(x,y;t)\com\nn\\
\pl^2\equiv \del_\mn\pl_\m\pl_\n=\sum^n_{\mu=1}(\frac{\pl}{\pl x^\m})^2\pr
                                                          \label{f.11}
\end{eqnarray}
%%%%%%%%%%%%%%%%%%%%%%%%%%%%%
In the following we suppress
the field suffixes $i,j,\cdots$ and take the matrix notation.
This equation (\ref{f.11}) 
is the n-dim heat equation with the small perturbation $\Vvec$. 
We give two important quantities in order to
obtain the solution.
( This approach is popular in the perturbative quantum field theory
under the name {\it propagator approach}\cite{BD}. In \cite{BD} the momentum
representation is taken, which is to be compared with the coordinate one of 
the present approach.
The Weyl anomaly in the string theory was analyzed in this approach
by Alvarez\cite{Al}. )
%%%%%%%%%%%%%%%%%%%%%%%%  i)  %%%%%%%%%%%%%%%%%%%%%%%%%%%%%%%%%%%%%%
\begin{flushleft}
i)\ Heat Equation
\end{flushleft}
The  heat equation:
%****(f.12)%%%%%%%%%%%%%%%%%%%%
\begin{eqnarray}
(\frac{\pl}{\pl t}-\pl^2)G_0(x,y;t)=0\com\q t>0\label{f.12}
\end{eqnarray}
%%%%%%%%%%%%%%%%%%%%%%%%%%%%%
has the solution
%****(f.13)%%%%%%%%%%%%%%%%%%%%
\begin{eqnarray}
G_0(x,y;t)=G_0(x-y;t)=\int\frac{d^nk}{(2\pi)^n}exp\{-k^2t+ik^\m(x-y)^\m\}
                                              ~I_N\nn\\
=\frac{1}{(4\pi t)^{n/2}}e^{-\frac{(x-y)^2}{4t}}~I_N\com\q 
k^2\equiv\sum^n_{\m=1}(k^\m)^2\com \label{f.13}
\end{eqnarray}
%%%%%%%%%%%%%%%%%%%%%%%%%%%%%
where $I_N$ is the identity matrix of the size $N\times N$.
$G_0$ satisfies the initial condition:\ $
\lim_{t\ra +0}~G_0(x-y;t)=\del^n(x-y)~I_N
$\ .
We define
%****(f.13b)%%%%%%%%%%%%%%%%%%%%
\begin{eqnarray}
G_0(x,y;t)=0 \mbox{ for } t\leq 0\pr\label{f.13b}
\end{eqnarray}
%%%%%%%%%%%%%%%%%%%%%%%%%%%%%
%%%%%%%%%%%%%%%%%%%%%%%%  ii)  %%%%%%%%%%%%%%%%%%%%%%%%%%%%%%%%%%%%%%
\begin{flushleft}
ii)\ Heat Propagator
\end{flushleft}
The heat equation with the delta-function source defines the heat propagator.
%****(f.14)%%%%%%%%%%%%%%%%%%%%
\begin{eqnarray}
(\frac{\pl}{\pl t}-\pl^2)S(x,y;t-s)=\del(t-s)\del^n(x-y)~I_N\com\nn\\
S(x,y;t)=S(x-y;t)=\int\frac{d^nk}{(2\pi)^n}\frac{dk^0}{2\pi}
\frac{exp\{-ik^0t+ik\cdot(x-y)\} }{-ik^0+k^2}~I_N  \nn\\
=\th(t)G_0(x-y;t)\com
                                                           \label{f.14}\\
k^2\equiv \sum^n_{\m=1}k^\m k^\m\com\q
k\cdot x\equiv \sum^n_{\m=1}k^\m x^\m\pr\nn
\end{eqnarray}
%%%%%%%%%%%%%%%%%%%%%%%%%%%%%
$\th(t)$\ is the {\it step function} defined by :
$\th(t)=1\ \mbox{for}\ t>0\ ,\ \th(t)=0\ \mbox{for}\ t<0\ $.
$S(x-y;t)$\ satisfies the initial condition:\ $
\lim_{t\ra+0}S(x-y;t)=\del^n(x-y)~I_N$\ 
and $
S(x,y;t)=0 \mbox{ for } t\leq 0$\ .

Now the formal solution of (\ref{f.11}) with the initial condition 
(\ref{f.7e}) is given by
%****(f.15)%%%%%%%%%%%%%%%%%%%%
\begin{eqnarray}
G(x,y;t)=G_0(x-y;t)+\int d^nz\int^\infty_{-\infty}ds~
S(x-z;t-s)\Vvec(z)G(z,y;s)               \pr\label{f.15}
\end{eqnarray}
%%%%%%%%%%%%%%%%%%%%%%%%%%%%%
$G(x,y;t)$\ appears in both sides above.
We can iteratively solve (\ref{f.15}) as\cite{S}
%****(f.16)%%%%%%%%%%%%%%%%%%%%
\begin{eqnarray}
G(x,y;t)&=&
G_0(x-y;t)+\int S\Vvec G_0+\int S\Vvec\int S\Vvec G_0+\cdots\com\nn\\
G_1(x,y;t)&\equiv&\int S\Vvec G_0=\int d^nzds S(x-z;t-s)\Vvec(z)G_0(z-y;s)\nn\\
&=&\int d^nz\int^t_0ds G_0(x-z;t-s)\Vvec(z)G_0(z-y;s)\com\label{f.16}\\
G_2(x,y;t)&\equiv&\int S\Vvec\int S\Vvec G_0
=\int d^nz'ds'S(x-z';t-s')\Vvec(z')\nn\\
&&\times\int d^nz ds S(z'-z;s'-s)\Vvec(z)G_0(z-y;s)\nn\\
&=&\int d^nz'\int^t_0ds' G_0(x-z';t-s')\Vvec(z')\nn\\
&&\times\int d^nz\int^{s'}_0ds G_0(z'-z;s'-s)\Vvec(z)G_0(z-y;s)\com\nn\\
G_3(x,y;t)&=&\int d^nz''\int^t_0ds'' G_0(x-z'';t-s'')\Vvec(z'')\nn\\
&&\times\int d^nz'\int^{s''}_0ds' G_0(z''-z';s''-s')\Vvec(z')\nn\\
&&\times\int d^nz\int^{s'}_0ds G_0(z'-z;s'-s)\Vvec(z)G_0(z-y;s)\pr\nn
\end{eqnarray}
%%%%%%%%%%%%%%%%%%%%%%%%%%%%%
Expressions for higher-order terms are similarly obtained.

\q As in the ordinary field theory, it is illuminating to represent
this perturbative solution in graphs. We do it in the (n+1) dimensional
space, which is composed of the space of the ordinary 
n-dim (Euclidean) space plus 1-dim positive proper time.  
In the above solution, $G_0(x-y;t)$ plays the role of "propagator".
It is represented by a directed\cite{foot3}
 straight line as in Fig.1. 
%%%%%%%%%%%%%%%%%%%%%%%  Fig.1  %%%%%%%%%%%%%%%%%%%%%%%%%%%%%%%%%%%%
\begin{figure}
\centerline{\epsfysize=6cm\epsfbox{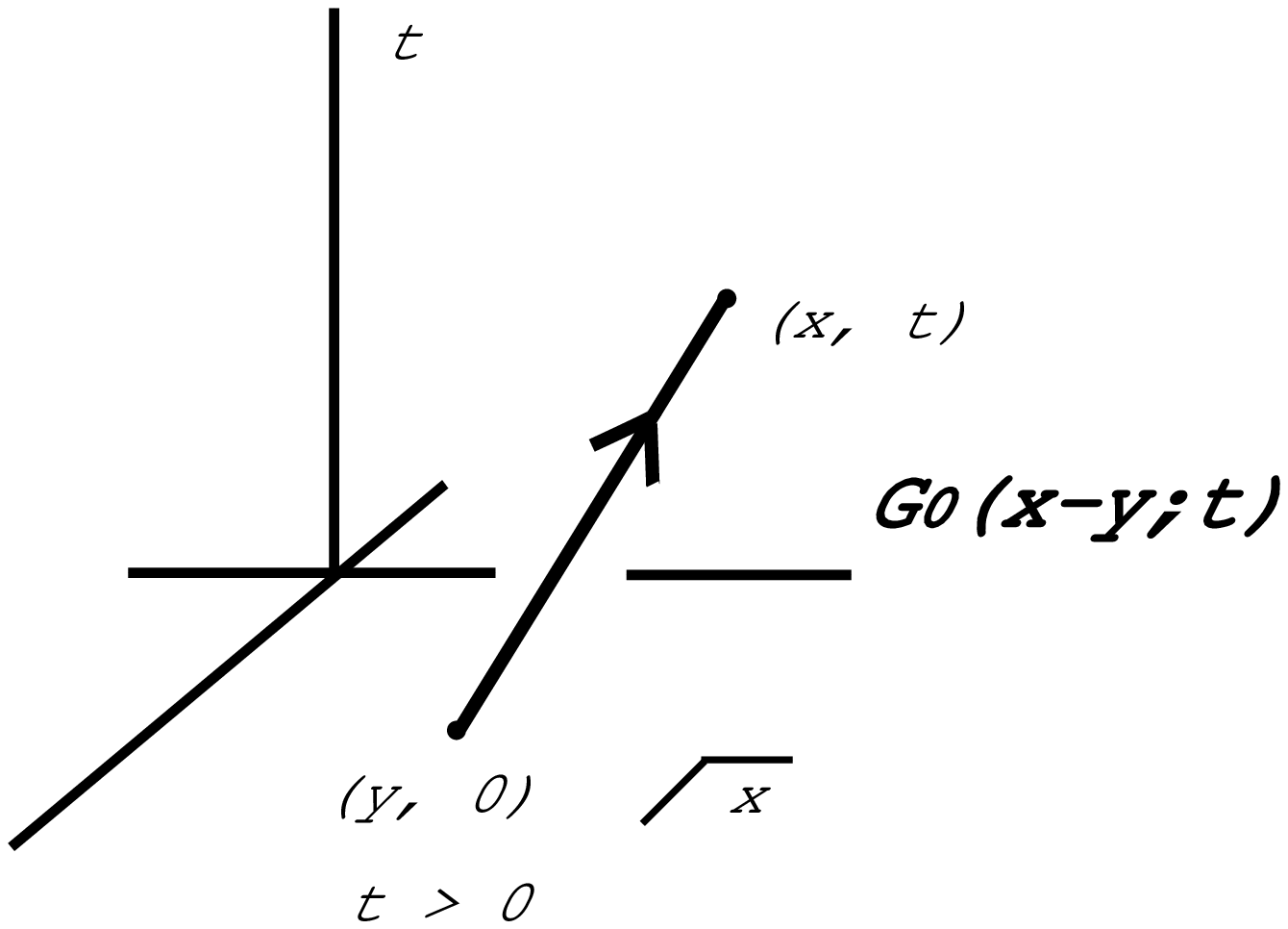}}
%\begin{center}\psbox[height=6cm]{g0xy}\end{center}
   \begin{center}
Fig.1\ 
Graph of $G_0(x-y;t)$
%***g0xy.eps
   \end{center}
\end{figure}
%%%%%%%%%%%%%%%%%%%%%%%%%%%%%%%%%%%%%%%%%%%%%%%%%%%%%%%%%%%%%%%%%%%%%
$\Vvec(z)$
plays the role of "vertex", which acts on the system,  
during the process of "propagation of particles", some times
according to the perturbation order. The situation
up to the 3-rd order is represented in Fig.2-4. We can easily
write down the expression of any higher order, $G_k(x,y;t)$, with
the help of this graphical representation.
%%%%%%%%%%%%%%%%%%%%%%%  Fig.2  %%%%%%%%%%%%%%%%%%%%%%%%%%%%%%%%%%%%
\begin{figure}
\centerline{\epsfysize=6cm\epsfbox{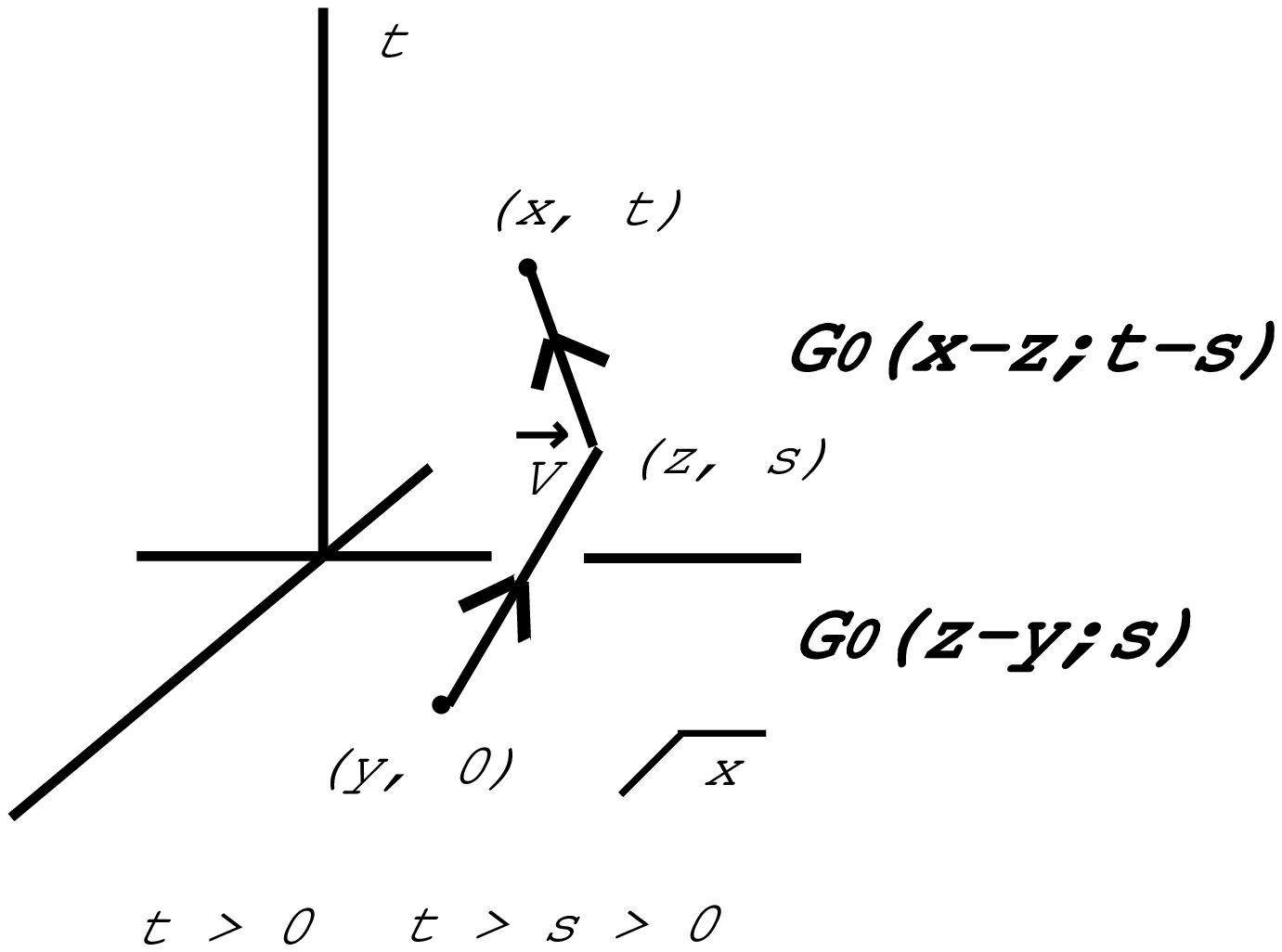}}
%\begin{center}\psbox[height=6cm]{g1xy}\end{center}
   \begin{center}
Fig.2\ 
Graph of $G_1(x,y;t)$
%***g1xy.eps
   \end{center}
\end{figure}
%%%%%%%%%%%%%%%%%%%%%%%%%%%%%%%%%%%%%%%%%%%%%%%%%%%%%%%%%%%%%%%%%%%%%
%%%%%%%%%%%%%%%%%%%%%%%  Fig.3  %%%%%%%%%%%%%%%%%%%%%%%%%%%%%%%%%%%%
\begin{figure}
\centerline{\epsfysize=6cm\epsfbox{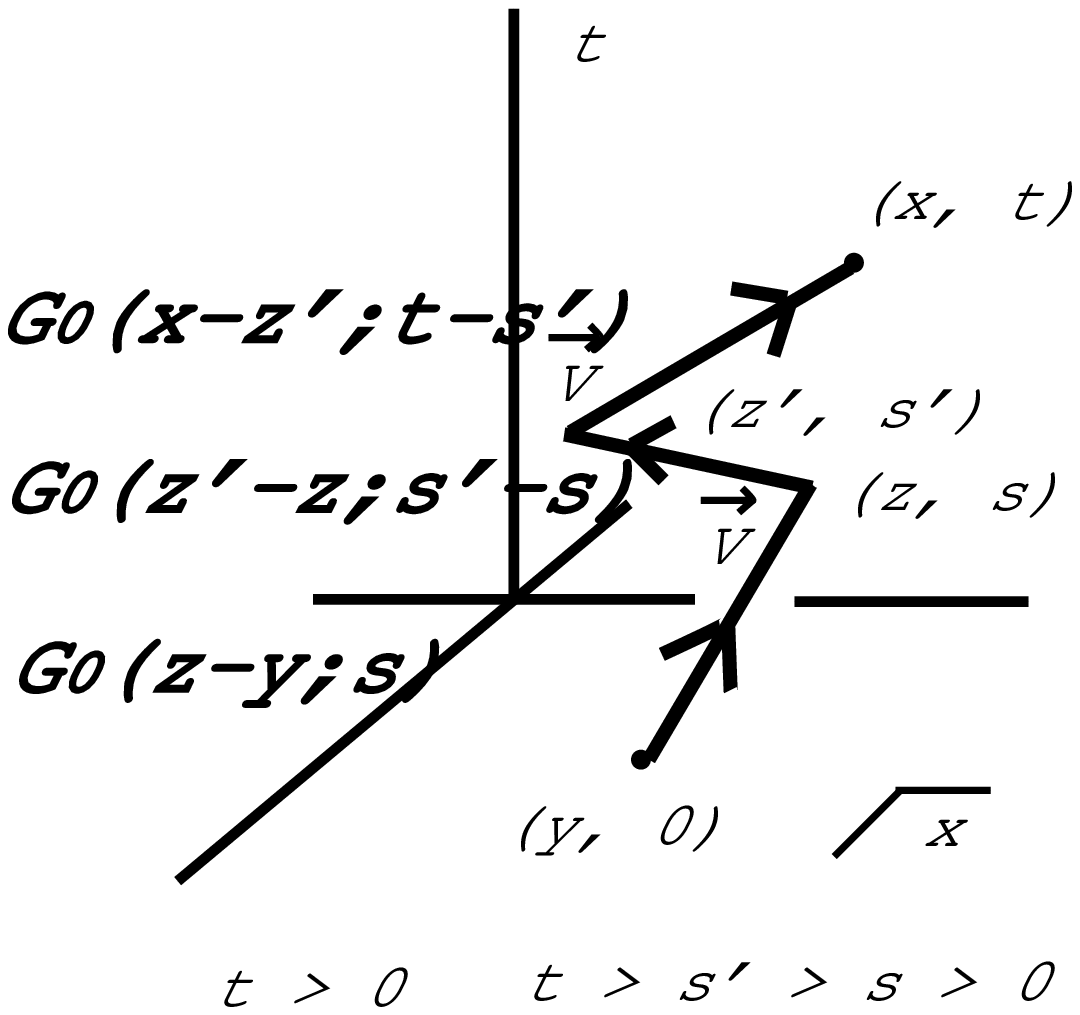}}
%\begin{center}\psbox[height=6cm]{g2xy}\end{center}
   \begin{center}
Fig.3\ 
Graph of $G_2(x,y;t)$
%***g2xy.eps
   \end{center}
\end{figure}
%%%%%%%%%%%%%%%%%%%%%%%%%%%%%%%%%%%%%%%%%%%%%%%%%%%%%%%%%%%%%%%%%%%%%
%%%%%%%%%%%%%%%%%%%%%%%  Fig.4  %%%%%%%%%%%%%%%%%%%%%%%%%%%%%%%%%%%%
\begin{figure}
\centerline{\epsfysize=6cm\epsfbox{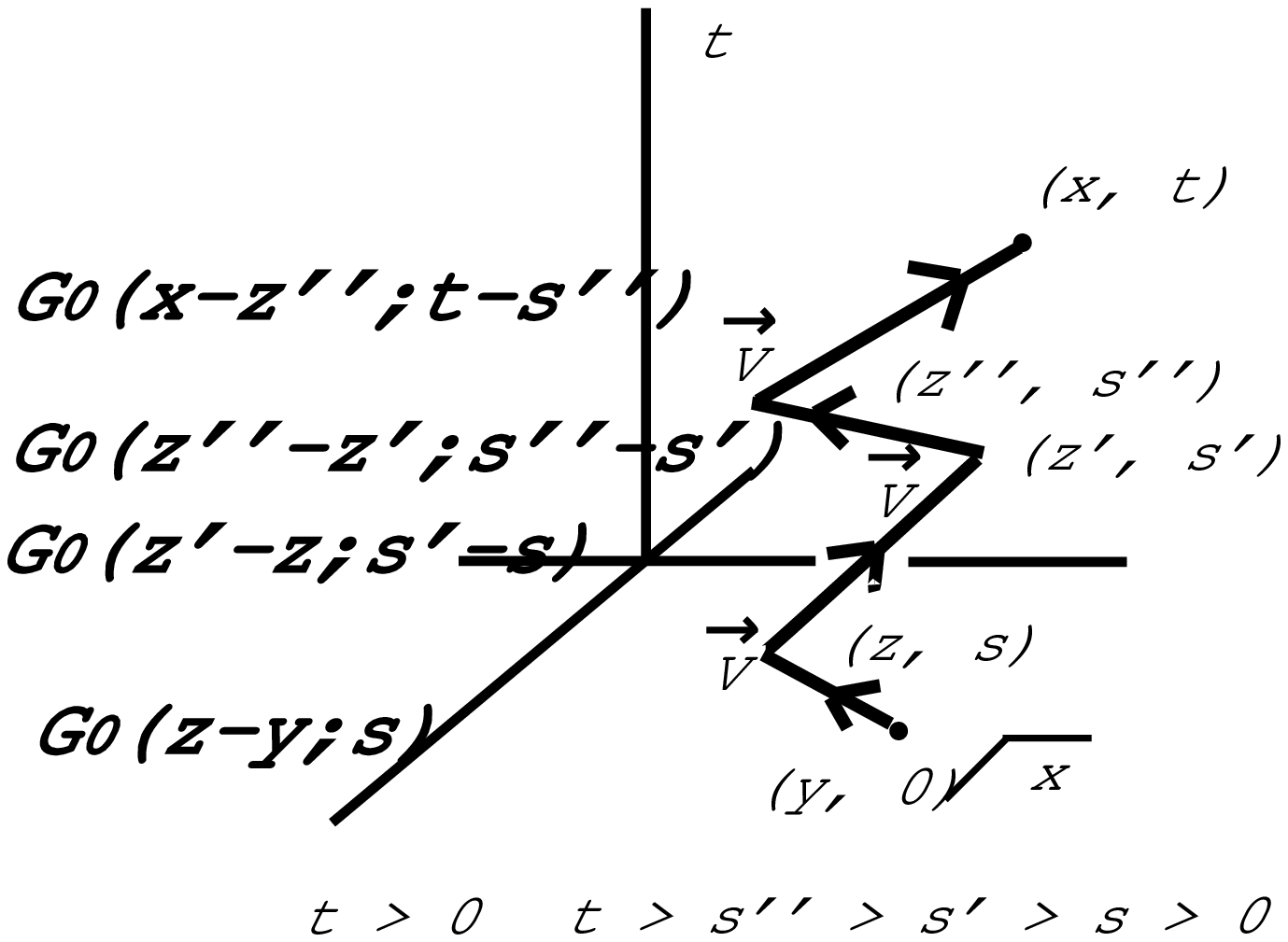}}
%\begin{center}\psbox[height=6cm]{g3xy}\end{center}
   \begin{center}
Fig.4\ 
Graph of $G_3(x,y;t)$
%***g3xy.eps
   \end{center}
\end{figure}
%%%%%%%%%%%%%%%%%%%%%%%%%%%%%%%%%%%%%%%%%%%%%%%%%%%%%%%%%%%%%%%%%%%%%

Generally, in n-dim, the terms up to $G_{n/2}$\ are 
sufficient for the anomaly calculation\cite{II96}. In this sense, 
the present expansion looks like that with respect to the space(-time)
dimension.

\subsection{Factoring out the scale parameter $t$}   %%%%%%%%%%%%%%%%%%%%%%%%%

Because of the presence of the positive regularization parameter
$t$, we can safely (without singularity) take
the trace-part 
%of (\ref{f.5}) is given by putting 
, $x=y$, in the above equations.
%****(f.17)%%%%%%%%%%%%%%%%%%%%
\begin{eqnarray}
G(x,x;t) &=& G_0(0;t)+G_1(x,x;t)+G_2(x,x;t)+G_3(x,x;t)
+\cdots\com\q \nn \\
G_0(0;t) &=&\frac{1}{(4\pi t)^{n/2}}~I_N\pr \label{f.17}
\end{eqnarray}
%%%%%%%%%%%%%%%%%%%%%%%%%%%%%
As shown in (\ref{f.5}), the $t^0$-part of $G(x,x;t)$
is the Weyl anomaly. In order to see their $t$-dependence, 
it is best to replace the 
dimensional coordinate variables such as (z,s) in (\ref{f.16}) by the dimensionless
ones (w,r).
%****(f.17b1)%%%%%%%%%%%%%%%%%%%%  
\begin{eqnarray}
G_1(x,x;t)&\equiv&\int \left.S\Vvec G_0\right|_{x=y}      \nn\\
          &=& \frac{1}{t^{(n/2)-1}}\int d^nw\int^1_0dr
G_0(w;1-r)~\Vvec(x+\sqrt{t}w)G_0(w;r),                  
%\frac{1}{(4\pi)^nt^{(n/2)-1}}\int d^nw\int^1_0dr\frac{1}{\{(1-r)r\}^{n/2}}
%e^{-\frac{w^2}{4(1-r)}}~\Vvec(x+\sqrt{t}w)e^{-\frac{w^2}{4r}},
                                             \label{f.17b1}\\
     && 1> r = \frac{s}{t}\ >0,\ w=(z-x)/\sqrt{t}      \com\nn\\
\end{eqnarray}
%%%%%%%%%%%%%%%%%%%%%%%%%%%%%
where the relation (\ref{a.2}) is used. In the same way, we obtain
%****(f.17b2,3)%%%%%%%%%%%%%%%%%%%%  
\begin{eqnarray}
G_2(x,x;t) &\equiv& \left.\int S\Vvec\int S\Vvec G_0\right|_{x=y}\nn\\
&=&\frac{1}{t^{(n/2)-2}}\int d^nw' \int^1_0dr' G_0(w';1-r')\Vvec(x+\sqrt{t}w')\nn\\
&& \times\int d^nw\int^{r'}_0dr 
             G_0(w'-w;r'-r)\Vvec(x+\sqrt{t}w)G_0(w;r)\com
%&=&\frac{1}{(4\pi)^{3n/2}}\frac{1}{t^{(n/2)-2}}\int d^nv \int d^nu
%\int^1_0dk\int^k_0dl\frac{1}{\{(1-k)(k-l)l\}^{n/2}}
%e^{-\frac{v^2}{4(1-k)}}    \label{f.18}\\
%&&\times \Vvec(x+\sqrt{t}v)e^{-\frac{(v-u)^2}{4(k-l)}}
%\Vvec(x+\sqrt{t}u)e^{-\frac{u^2}{4l}}\com\nn
                                                    \label{f.17b2}\\
\mbox{where}&&\nn\\
&& 1>r'=s'/t\ >\ r=s/t\ >0,\ w'=(z'-x)/\sqrt{t}\ ,
                                      \ w=(z-x)/\sqrt{t}\com	\nn\\												
G_3(x,x;t) &\equiv& \left.\int S\Vvec\int S\Vvec \int S\Vvec G_0\right|_{x=y}\nn\\
&=&\frac{1}{t^{(n/2)-3}}\int d^nw'' \int^1_0dr'' G_0(w'';1-r'')\Vvec(x+\sqrt{t}w'')\nn\\
&& \times\int d^nw'\int^{r''}_0dr' 
                      G_0(w''-w';r''-r')\Vvec(x+\sqrt{t}w')   \nn\\
&& \times\int d^nw\int^{r'}_0dr G_0(w'-w;r'-r)\Vvec(x+\sqrt{t}w)G_0(w;r)\com
													\label{f.17b3}\\
\mbox{where}&&\nn\\
&& 1>r''=s''/t\ >\ r'=s'/t\ >\ r=s/t\ >0,\ \nn\\
&& w''=(z''-x)/\sqrt{t}\ ,\ w'=(z'-x)/\sqrt{t}\ ,\ w=(z-x)/\sqrt{t}\pr	\nn												
\end{eqnarray}
%%%%%%%%%%%%%%%%%%%%%%%%%%%%%
The above quantities of $G_k(x,x;t), (k=0,\cdots,3)$ are depicted in Fig.5-8 with
the dimensionless quantities. We can easily write down higher-order
terms with the help of graphs.
%%%%%%%%%%%%%%%%%%%%%%%  Fig.5  %%%%%%%%%%%%%%%%%%%%%%%%%%%%%%%%%%%%
\begin{figure}
\centerline{\epsfysize=6cm\epsfbox{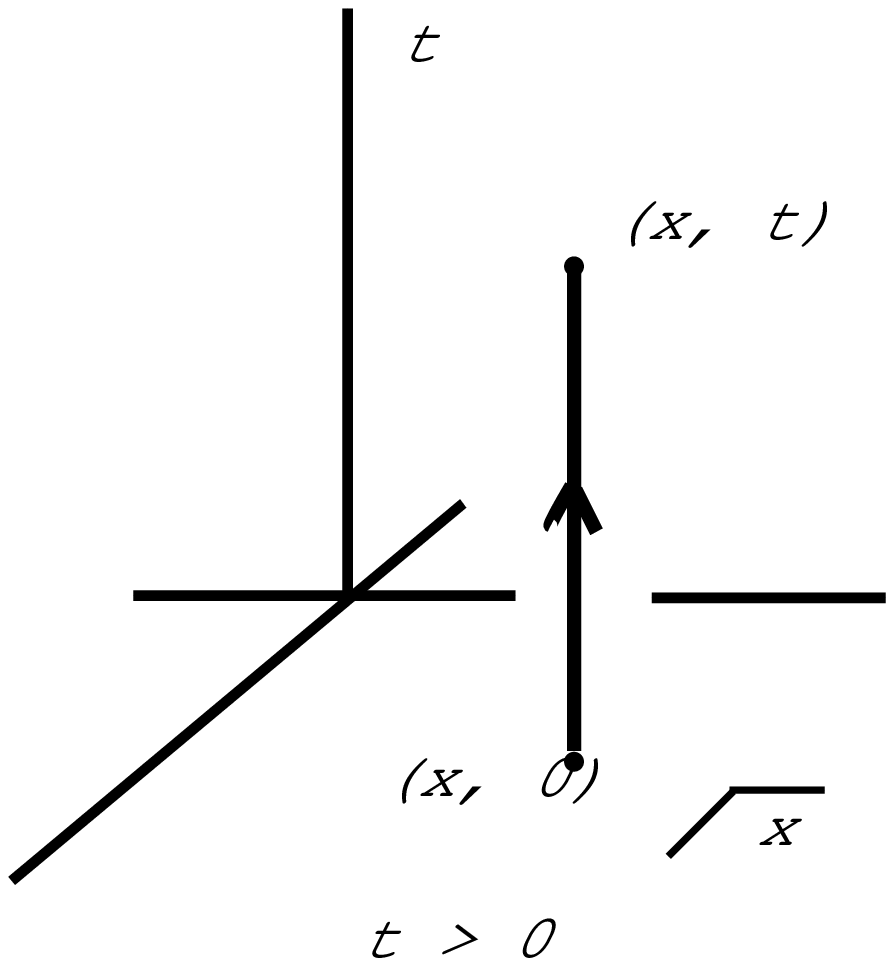}}
%\begin{center}\psbox[height=6cm]{g0xx}\end{center}
   \begin{center}
Fig.5\ 
Graph of $G_0(0;t)$
%***g0xx.eps
   \end{center}
\end{figure}
%%%%%%%%%%%%%%%%%%%%%%%%%%%%%%%%%%%%%%%%%%%%%%%%%%%%%%%%%%%%%%%%%%%%%
%%%%%%%%%%%%%%%%%%%%%%%  Fig.6  %%%%%%%%%%%%%%%%%%%%%%%%%%%%%%%%%%%%
\begin{figure}
\centerline{\epsfysize=6cm\epsfbox{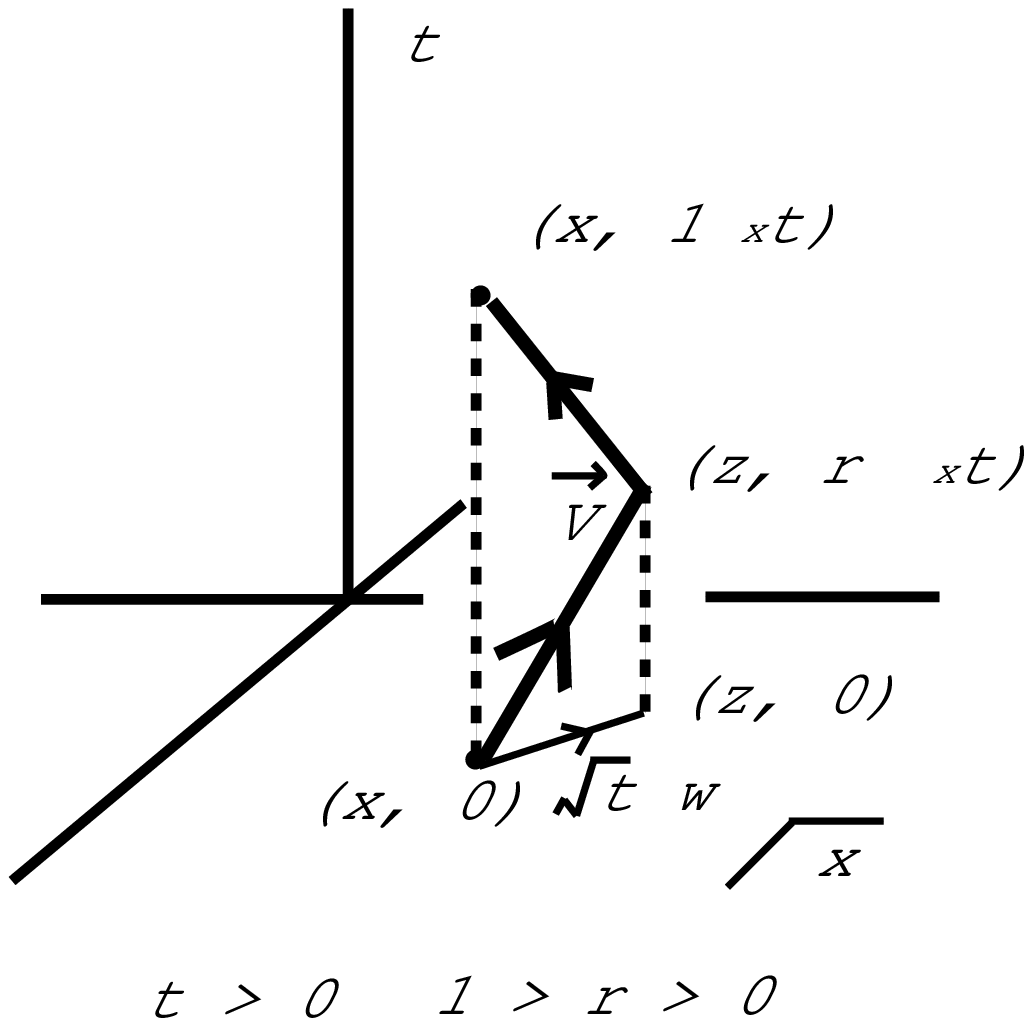}}
%\begin{center}\psbox[height=6cm]{g1xx}\end{center}
   \begin{center}
Fig.6\ 
Graph of $G_1(x,x;t)$
%***g1xx.eps
   \end{center}
\end{figure}
%%%%%%%%%%%%%%%%%%%%%%%%%%%%%%%%%%%%%%%%%%%%%%%%%%%%%%%%%%%%%%%%%%%%%
%%%%%%%%%%%%%%%%%%%%%%%  Fig.7  %%%%%%%%%%%%%%%%%%%%%%%%%%%%%%%%%%%%
\begin{figure}
\centerline{\epsfysize=6cm\epsfbox{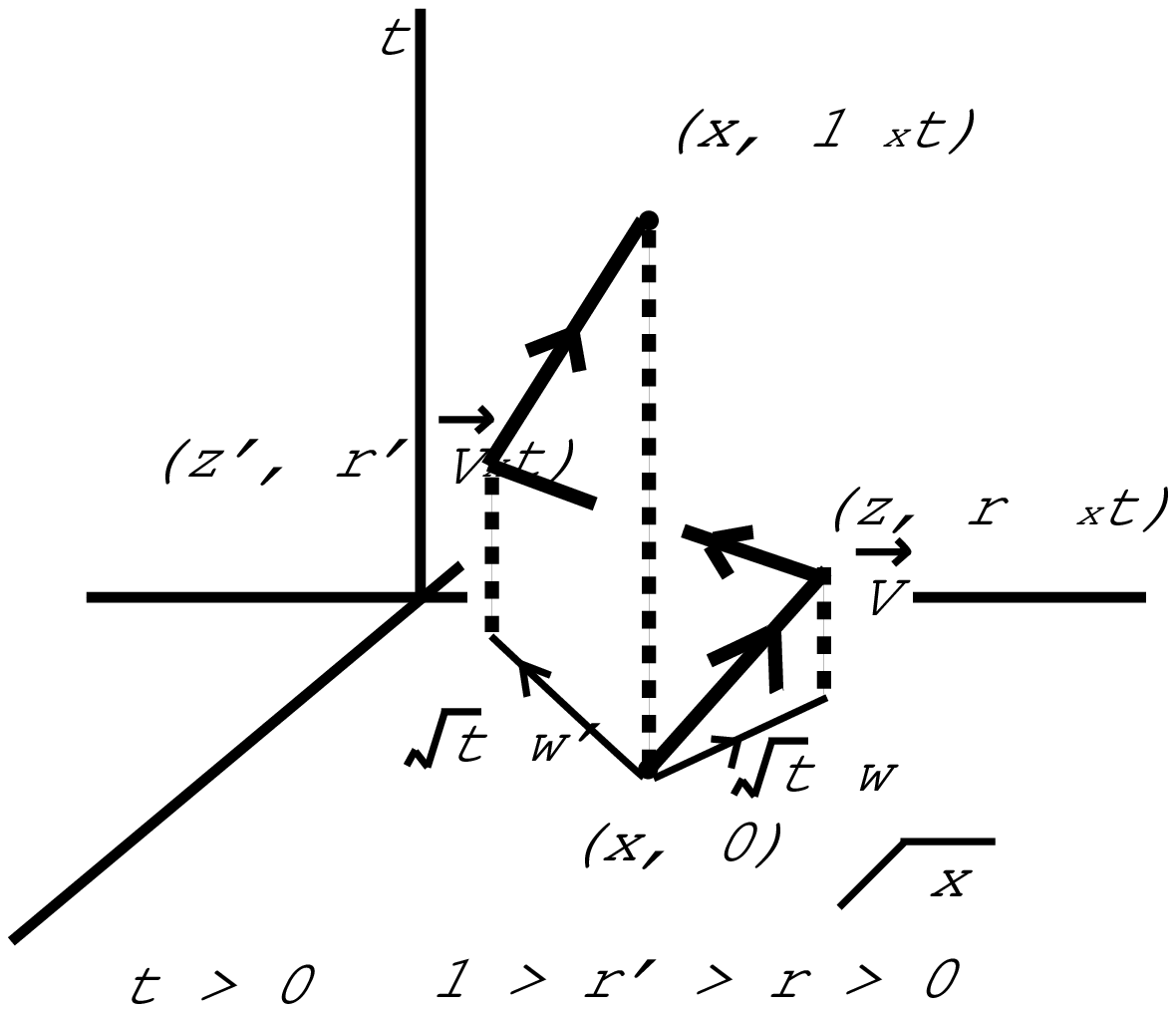}}
%\begin{center}\psbox[height=6cm]{g2xx}\end{center}
   \begin{center}
Fig.7\ 
Graph of $G_2(x,x;t)$
%***g2xx.eps
   \end{center}
\end{figure}
%%%%%%%%%%%%%%%%%%%%%%%%%%%%%%%%%%%%%%%%%%%%%%%%%%%%%%%%%%%%%%%%%%%%%
%%%%%%%%%%%%%%%%%%%%%%%  Fig.8  %%%%%%%%%%%%%%%%%%%%%%%%%%%%%%%%%%%%
\begin{figure}
\centerline{\epsfysize=6cm\epsfbox{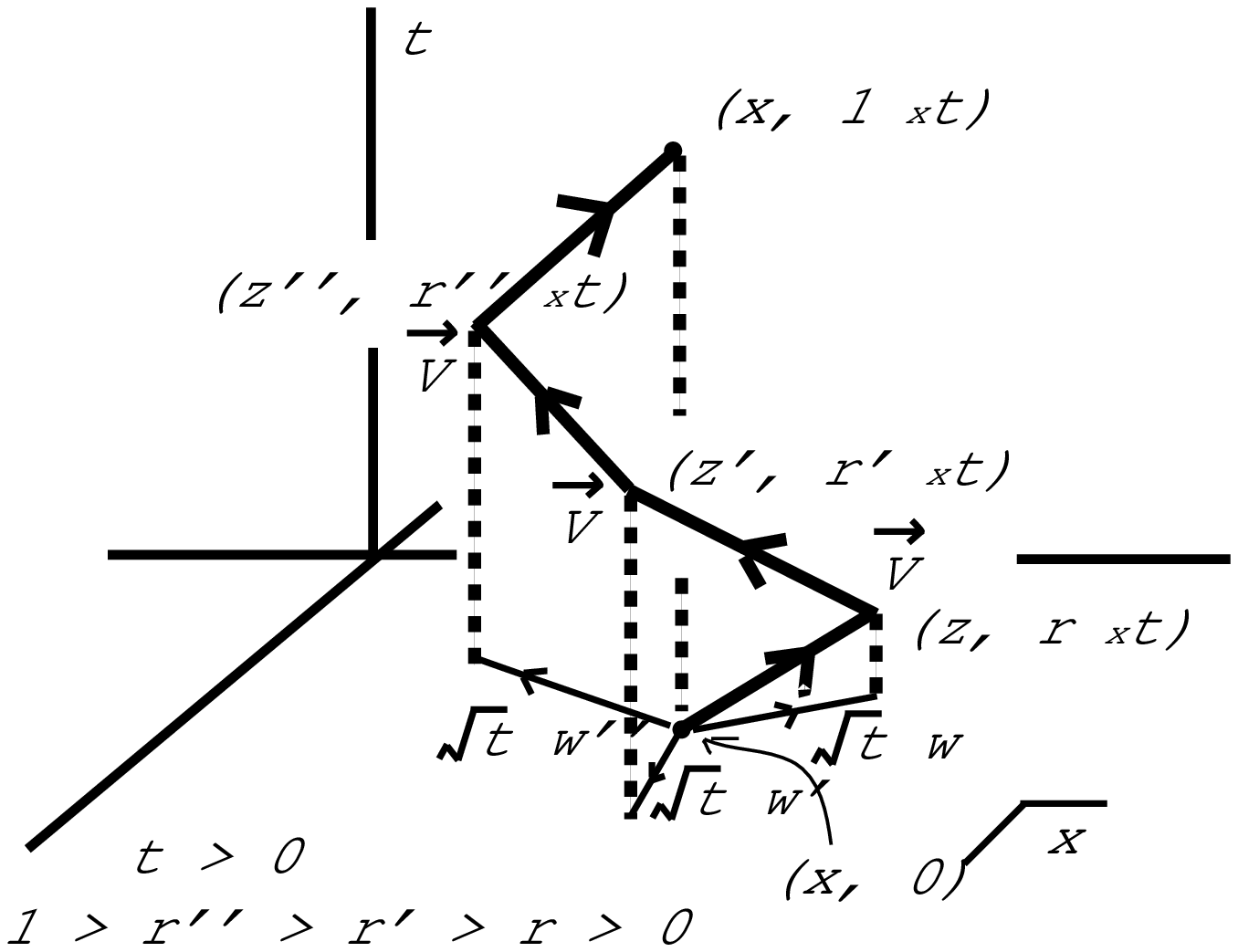}}
%\begin{center}\psbox[height=6cm]{g3xx}\end{center}
   \begin{center}
Fig.8\ 
Graph of $G_3(x,x;t)$
%***g3xx.eps
   \end{center}
\end{figure}
%%%%%%%%%%%%%%%%%%%%%%%%%%%%%%%%%%%%%%%%%%%%%%%%%%%%%%%%%%%%%%%%%%%%%

%
\subsection{Vertex and propagator rules}  %%%%%%%%%%%%%%%%%%%
The vertex operator $\Vvec(x+\sqrt{t}w)$ in the expression
$G_k(x,x;t)$ has differentials
%****(ver.1)%%%%%%%%%%%%%%%%%%%%  
\begin{eqnarray}
\Vvec(x+\sqrt{t}w)&=&
\frac{1}{t}W_\mn(x+\sqrt{t}w)\frac{\pl}{\pl w^\m}\frac{\pl}{\pl w^\n}
+\frac{1}{\sqrt{t}}N_\m(x+\sqrt{t}w)\frac{\pl}{\pl w^\m}+M(x+\sqrt{t}w)\ , 
\end{eqnarray}
%%%%%%%%%%%%%%%%%%%%%%%%%%%%%%%%
and acts only on the "adjacently right" $G_0$ in 
Eqs.(\ref{f.17b1})-(\ref{f.17b3}).
Here we have the following {\it vertex rule}. 
%****(ver.2)%%%%%%%%%%%%%%%%%%%%  
\begin{eqnarray}
& \mbox{Vertex Rule 1}  & \nn\\
& \Vvec(x+\sqrt{t}w')G_0(w'-w;r'-r)=
\frac{1}{ \{4\pi(r'-r)\}^{n/2} }\Vvec(x+\sqrt{t}w')
\e^{ -\frac{(w'-w)^2}{4(r'-r)} }              &    \nn\\
& =\left\{ 
\frac{1}{t}W_\mn(x+\sqrt{t}w')
          \left( -\frac{\del_\mn}{2(r'-r)}
               +\frac{(w'-w)_\m(w'-w)_\n}{4(r'-r)^2} \right)
\right.                                       &  \nn\\			   
& \left. +\frac{1}{\sqrt{t}}N_\m(x+\sqrt{t}w')
          \left( -\frac{(w'-w)_\m}{2(r'-r)} \right)
+M(x+\sqrt{t}w') 
\right\}
G_0(w'-w;r'-r)                               & \nn\\
& = V(x+\sqrt{t}w';w'-w,r'-r;t)G_0(w'-w;r'-r) \com &\nn\\
& \mbox{where}                        &\nn\\
&V(x;w,r;t)\equiv
\frac{1}{t}W_\mn(x)
          \left( -\frac{\del_\mn}{2r}+\frac{w_\m w_\n}{4r^2} \right)
+\frac{1}{\sqrt{t}}N_\m(x) \left( -\frac{w_\m}{2r} \right)+M(x) 
                                     \com    &  
									 \label{ver.2}
\end{eqnarray}
%%%%%%%%%%%%%%%%%%%%%%%%%%%%%%%%
where $V$ does {\it not} have differentials. 
Especially taking $r=0, w=0$ in the above and dropping primes ($'$) , 
we obtain a relation. 
%****(ver.3)%%%%%%%%%%%%%%%%%%%%  
\begin{eqnarray}
& \mbox{Vertex Rule 2}             & \nn\\
& \Vvec(x+\sqrt{t}w)G_0(w;r)       &       \nn\\
& =\left\{ 
\frac{1}{t}W_\mn(x+\sqrt{t}w)
          \left( -\frac{\del_\mn}{2r}
               +\frac{w_\m w_\n}{4r^2} \right)
%\right.                              &           \nn\\			   
%& \left. 
+\frac{1}{\sqrt{t}}N_\m(x+\sqrt{t}w)
          \left( -\frac{w_\m}{2r} \right)
+M(x+\sqrt{t}w) 
\right\}
G_0(w;r)                             &   \nn\\
& = V(x+\sqrt{t}w;w,r;t)G_0(w;r)    \pr &  \label{ver.3}
\end{eqnarray}
%%%%%%%%%%%%%%%%%%%%%%%%%%%%%%%%

Making use of the above rule, we can evaluate $G_1(x,x;t)$ of
Eq.(\ref{f.17b1}) as
%****(ver.4)%%%%%%%%%%%%%%%%%%%%  
\begin{eqnarray}
G_1(x,x;t)
=\frac{1}{t^{(n/2)-1}}\int d^nw\int^1_0drG_0(w;1-r)G_0(w;r)V(x+\sqrt{t}w;w,r;t) 
                                                      \nn\\                  
=\frac{1}{(4\pi)^{n/2}t^{(n/2)-1}}\int d^nw\int^1_0dr
G_0(w;(1-r)r)V(x+\sqrt{t}w;w,r;t) 
\com \label{ver.4}
\end{eqnarray}
%%%%%%%%%%%%%%%%%%%%%%%%%%%%%%%%
where the relation:\ 
$G_0(w;1-k)G_0(w;r)=\frac{1}{(4\pi)^{n/2}}G_0(w;(1-r)r),\ 1>r>0$
, which is derived from
a {\it propagator rule} 4 (\ref{a.6z}), is used in order to
reduce the number of $G_0$'s by one.
In Appendix A, we list the propagator rules (\ref{a.5}-\ref{a.6z}), and
their graphical ("geometrical") expressions in the coordinate
space:\ Fig.10-13. 
$G_2(x,x;t)$ of Eq.(\ref{f.17b2}) is written, in terms of $V$, as
%****(ver.5)%%%%%%%%%%%%%%%%%%%%  
\begin{eqnarray}
G_2(x,x;t)
=\frac{1}{t^{(n/2)-2}}\int d^nw' \int^1_0dr' G_0(w';1-r')  \nn\\
\times\int d^nw\int^{r'}_0dr V(x+\sqrt{t}w';w'-w,r'-r;t)G_0(w'-w;r'-r)
V(x+\sqrt{t}w;w,r;t)G_0(w;r)\pr
                                                    \label{ver.5}
\end{eqnarray}
%%%%%%%%%%%%%%%%%%%%%%%%%%%%%
By changing the integration variable $w'$ to a new one $\wbar$ defined below, 
we can make all $G_0$'s have space-variables independent of each other. 
Using the following relations,
%****(ver.6)%%%%%%%%%%%%%%%%%%%%
\begin{eqnarray}
G_0(w';1-r')G_0(w'-w;r'-r)=G_0({\bar w};\frac{(1-r')(r'-r)}{1-r})
G_0(w;1-r)\com\nn\\
1>r'>r\com   {\bar w}=w'-\frac{1-r'}{1-r}w\com  \nn\\
G_0(w;1-r)G_0(w;r)=\frac{1}{(4\pi)^{n/2}}G_0(w;(1-r)r)
\com\q 1>r>0\com
                                                      \label{ver.6}
\end{eqnarray}
%%%%%%%%%%%%%%%%%%%%%%%%%%%%%%
which are obtained from Rule 2 ,(\ref{a.6x}), and Rule 3 ,(\ref{a.6z}), 
respectively, we can finally evaluate eq.(\ref{ver.5}) as
%****(ver.7)%%%%%%%%%%%%%%%%%%%%  
\begin{eqnarray}
G_2(x,x;t)
=\frac{1}{(4\pi)^{n/2}t^{(n/2)-2}}\int d^n\wbar \int^1_0dr' 
\int d^nw\int^{r'}_0dr                \nn\\
\times G_0(\wbar;\frac{(1-r')(r'-r)}{1-r})G_0(w;(1-r)r)  
 V(x+\sqrt{t}w';w'-w,r'-r;t)V(x+\sqrt{t}w;w,r;t)\com  \nn\\
 w'=\wbar+\frac{1-r'}{1-r}w\com
                                                    \label{ver.7}
\end{eqnarray}
%%%%%%%%%%%%%%%%%%%%%%%%%%%%%
where the integration variable $w'$ is changed to $\wbar$
($\det \frac{\pl(w',w)}{\pl(\wbar,w)}=1$). 
We note the following things in the above evaluation. 
\begin{enumerate}
\item The number of propagators
,$G_0$'s, decreases by one and becomes the same as that of the space
integration variables. 
\item All "Gaussian" coordinates in $G_0$'s ( $\wbar$ and
$w$ in the present case ) can be taken independent 
\item In the above reduction, we have applied the propagator
rules on $G_0$'s from "left" to "right". There are some other choices
( for example, from "right" to "left") which give expressions different 
from (\ref{ver.7}) in appearance. 
%It is checked****** that those expressions give the same result. 
\end{enumerate}
These properties are valid for any order of $G_k(x,x;t)$.

Similarly we can evaluate $G_3$ as 
%****(ver.8)%%%%%%%%%%%%%%%%%%%%  
\begin{eqnarray}
G_3(x,x;t)
=\frac{1}{(4\pi)^{n/2}t^{(n/2)-3}}
\int d^n{\bar \wbar}\int d^n\wbar \int d^nw
\int^1_0dr'' \int^{r''}_0dr' \int^{r'}_0dr                \nn\\
\times G_0({\bar \wbar};\frac{(1-r'')(r''-r')}{1-r'})
G_0(\wbar;\frac{(1-r')(r'-r)}{1-r})
G_0(w;(1-r)r)                                            \nn\\
 \times V(x+\sqrt{t}w'';w''-w',r''-r';t)V(x+\sqrt{t}w';w'-w,r'-r;t)
                                         V(x+\sqrt{t}w;w,r;t)\com  \nn\\
 w''={\bar \wbar}+\frac{1-r''}{1-r'}\wbar+\frac{1-r''}{1-r}w\com
 w'=\wbar+\frac{1-r'}{1-r}w\com
                                                    \label{ver.8}
\end{eqnarray}
%%%%%%%%%%%%%%%%%%%%%%%%%%%%%
where the integration variables $w''$ and $w'$ are changed to 
${\bar \wbar}$ and $\wbar$
($\det \frac{\pl(w'',w',w)}{\pl({\bar \wbar},\wbar,w)}=1$). As in
$G_2$, there are some different expressions depending on how
we apply the propagator rules in $G_0$'s.

Because of the integration formulae (\ref{a.3}) of App.A, 
the space
integrations in (\ref{ver.4}),(\ref{ver.7}) and (\ref{ver.8}) 
give a product of Kronecker's deltas 
times some rational function of dimensionless parameters. 
%\footnote{
%The "intricacy" of the key quantity
%$G(x,x;t)$ reduces to the parameter integrals.
%         }
From above results we can write down $G_k(x,x;t)$ 
for any higher $k$. 
$G_k(x,x;t)$ has $k$-fold parameter integrals.

%%%%%%%%%%%%%%%%%%%%%%%%%%%%%%%%%%%%%%%%%%%%%%%%%%%%%%%%%%%%%%%%%%%%%
%%%%%%%%%%%%%%%%%%%%%%%%%%%%%%   SEC  4    %%%%%%%%%%%%%%%%%%%%%%%%%%
%%%%%%%%%%%%%%%%%%%%%%%%%%%%%%%%%%%%%%%%%%%%%%%%%%%%%%%%%%%%%%%%%%%%%
\section{ Evaluation of $t^0$-Part of $G(x,x;t)$\\
-- Taylor expansion for taking the limit $t\ra +0$ ---}
Let us evaluate $G(x,x;t)|_{t^0}$. 
We consider $n=6$ dim space. We focus on the
$(\pl\pl h)^3$-type terms, because that part has the sufficient
information to determine the  Weyl anomaly in 6 dim.
Taking the Taylor-expansion of 
$W_\mn(x+\sqrt{t}v),N_\m(x+\sqrt{t}v)$ and $M(x+\sqrt{t}v)$
around $\sqrt{t}v=0$, $V(x+\sqrt{t}v;w,r;t)$ can be expanded as
%****(exp.1)%%%%%%%%%%%%%%%%%%%%  
\begin{eqnarray}
& V(x+\sqrt{t}v;w,r;t)= 
\frac{1}{t}W_\mn(x+\sqrt{t}v)
          \left( -\frac{\del_\mn}{2r}
               +\frac{w_\m w_\n}{4r^2} \right)
+\frac{1}{\sqrt{t}}N_\m(x+\sqrt{t}v)
          \left( -\frac{w_\m}{2r} \right)
+M(x+\sqrt{t}v)                              &   \nn\\
& 
=\frac{1}{t}V_{-1}(x,v;w,r)
+V_{0}('')+tV_1('')+\cdots                      
+\frac{1}{\sqrt{t}}V_{-1/2}('')+\sqrt{t}V_{1/2}('')+\cdots
                                              \com &  \label{exp.1}
\end{eqnarray}
%%%%%%%%%%%%%%%%%%%%%%%%%%%%%%%%
where $V_{-1/2},V_{1/2},\cdots$ are irrelevant parts because
they all have odd-time derivatives of $h$'s:\ 
$\pl h, \pl\pl\pl h, \cdots$. 
$V_{-1}('')=W_\mn (x)(-\frac{\del_\mn}{2r}+\frac{w_\m w_\n}{4r^2})$
is also irrelevant because it has no derivatives of $h'$s.\cite{foot4} 
$V_0,V_1,V_2$ are given by
%****(exp.2)%%%%%%%%%%%%%%%%%%%%  
\begin{eqnarray}
V_0(x,v;w,r) &=\frac{1}{2!}\pl_{\al_1}\pl_{\al_2}W_\mn
\cdot v^{\al_1}v^{\al_2}
(-\frac{\del_\mn}{2r}+\frac{w_\m w_\n}{4r^2})
+\pl_{\al_1}N_\m\cdot v^{\al_1}(-\frac{w_\m}{2r})+M   \com &\nn\\
V_1(x,v;w,r) &=\frac{1}{4!}\pl_{\al_1}\pl_{\al_2}\cdots\pl_{\al_4}W_\mn
\cdot v^{\al_1}v^{\al_2}\cdots v^{\al_4}
(-\frac{\del_\mn}{2r}+\frac{w_\m w_\n}{4r^2})              &\nn\\
 & +\frac{1}{3!}\pl_{\al_1}\pl_{\al_2}\pl_{\al_3}N_\m
\cdot v^{\al_1}v^{\al_2}v^{\al_3}(-\frac{w_\m}{2r})+
\frac{1}{2!}\pl_{\al_1}\pl_{\al_2}M
\cdot v^{\al_1}v^{\al_2}\com                               &\nn\\
V_2(x,v;w,r) &=\frac{1}{6!}\pl_{\al_1}\pl_{\al_2}\cdots\pl_{\al_6}W_\mn
\cdot v^{\al_1}v^{\al_2}\cdots v^{\al_6}
(-\frac{\del_\mn}{2r}+\frac{w_\m w_\n}{4r^2})               &\nn\\
& +\frac{1}{5!}\pl_{\al_1}\pl_{\al_2}\cdots\pl_{\al_5}N_\m
\cdot v^{\al_1}v^{\al_2}\cdots v^{\al_5}
(-\frac{w_\m}{2r})                                     &\nn\\
& +\frac{1}{4!}\pl_{\al_1}\pl_{\al_2}\cdots\pl_{\al_4}M
\cdot v^{\al_1}v^{\al_2}\cdots v^{\al_4}\pr        &
                                                    \label{exp.2}
\end{eqnarray}
%%%%%%%%%%%%%%%%%%%%%%%%%%%%%
The Taylor-expansion of $W_\mn,N_\m,M$ 
, for the case of the 6 dim scalar-gravity theory, 
is graphically shown in App.C:\ 
(\ref{GraphExp.1}) for $V_0$, (\ref{GraphExp.2}) for $V_1$ and
(\ref{GraphExp.3}) for $V_2$.

Now we pick up the $t^0$-part\cite{foot5} 
and focus on the $(\pl\pl h)^3$-terms in the final form. 
For $G_1(x,x;t)$
%****(g1)%%%%%%%%%%%%%%%%%%%%  
\begin{eqnarray}
G_1(x,x;t)
=\frac{1}{(4\pi)^{3}}\int^1_0dr
\int d^6w~ G_0(w;(1-r)r)~\left[\frac{1}{t^{2}}V(x+\sqrt{t}w;w,r;t)\right]
                                                  \com\nn\\
%\frac{1}{t^{2}}V(x+\sqrt{t}w;w,r;t)|_{t^0} 
\left[\frac{1}{t^{2}}V('')\right]|_{t^0}=V_2(x,w;w,r)\pr
                                                    \label{g1}
\end{eqnarray}
%%%%%%%%%%%%%%%%%%%%%%%%%%%%%
We notice the $w^{\al_1}w^{\al_2}\cdots$ parts give 
, after the $w$-integration, different products
of Kronecker's deltas times powers of $(1-r)r$.
See (\ref{a.3}).

As for $G_2(x,x;t)$,from (\ref{ver.7}),
%
%****(g2)%%%%%%%%%%%%%%%%%%%%  
\begin{eqnarray}
G_2(x,x;t)
=\frac{1}{(4\pi)^{3}} \int^1_0dr' \int^{r'}_0dr
\int d^6\wbar \int d^6w                \nn\\
\times G_0(\wbar;\frac{(1-r')(r'-r)}{1-r})G_0(w;(1-r)r) \left[ 
 \frac{1}{t}V(x+\sqrt{t}w';w'-w,r'-r;t)V(x+\sqrt{t}w;w,r;t)
                                                   \right]\ ,  \nn\\
%\frac{1}{t}V(x+\sqrt{t}w';w'-w,r'-r;t)V(x+\sqrt{t}w;w,r;t)|_{t^0}\nn\\
\left[ \frac{1}{t}V('')V(''') \right]|_{t^0}
 =V_1(x,w';w'-w,r'-r)V_0(x,w;w,r)+V_0(x,w';w'-w,r'-r)V_1(x,w;w,r)
                                                             \nn\\
+\mbox{irrel. terms}\com                 \nn\\
w'= \wbar+R_1w\com\q 
w'-w= \wbar-S_1w\com \nn\\
R_1\equiv\frac{1-r'}{1-r}\equiv R(r,r')\com\q
S_1\equiv\frac{r'-r}{1-r}\equiv S(r,r')\com\q
R_1>0\com\q S_1>0\com\q R_1+S_1=1\pr
                                                    \label{g2}
\end{eqnarray}
%%%%%%%%%%%%%%%%%%%%%%%%%%%%%
The functions
$R(r,r')$ and $S(r,r')$ will be used in (\ref{g3.1}).
The irrelevant terms ("irrel. terms") are such ones as 
$V_2V_{-1},\ V_{1/2}V_{1/2}$, etc.

As for $G_3(x,x;t)$, from (\ref{ver.8}), 
%
%****(g3)%%%%%%%%%%%%%%%%%%%%  
\begin{eqnarray}
G_3(x,x;t)
=\frac{1}{(4\pi)^{3}}
\int^1_0dr'' \int^{r''}_0dr' \int^{r'}_0dr
\int d^6{\bar \wbar}\int d^6\wbar \int d^6w            \nn\\
\times G_0({\bar \wbar};\frac{(1-r'')(r''-r')}{1-r'})
G_0(\wbar;\frac{(1-r')(r'-r)}{1-r})G_0(w;(1-r)r)        \nn\\
\times \left[
V(x+\sqrt{t}w'';w''-w',r''-r';t)V(x+\sqrt{t}w';w'-w,r'-r;t)
                              V(x+\sqrt{t}w;w,r;t)\right]\com  \nn\\
%V(x+\sqrt{t}w'';w''-w',r''-r';t)V(x+\sqrt{t}w';w'-w,r'-r;t)
%                              V(x+\sqrt{t}w;w,r;t)|_{t^0}\nn\\
\left[V('')V(''')V('''')\right]|_{t^0}
 =V_0(x,w'';w''-w',r''-r')V_0(x,w';w'-w,r'-r)V_0(x,w;w,r)							  
                                                        \nn\\
+\mbox{irrel. terms}\com
                                                    \label{g3}
\end{eqnarray}
%%%%%%%%%%%%%%%%%%%%%%%%%%%%%
where "irrel. terms" above are such ones as $V_1V_{-1}V_0,V_{1/2}V_{1/2}V_0$,
etc. 
In the above, the coordinate variables in the integrand are
written by the integration variables ${\bar \wbar},\wbar$ and $w$ as
%****(g3.1)%%%%%%%%%%%%%%%%%%%%  
\begin{eqnarray}
w''= {\bar \wbar}+R_2\wbar+R_3w\com\q
w''-w'= {\bar \wbar}-S_2\wbar-T_1w\com     \nn\\
w'= \wbar +R_1w\com\q
w'-w= \wbar-S_1w\com                       \nn\\
\mbox{where}                               \nn\\
R_1=R(r,r')=\frac{1-r'}{1-r}\com\q
R_2=R(r',r'')=\frac{1-r''}{1-r'}\com\q
R_3=R(r,r'')=\frac{1-r''}{1-r}\com                \nn\\
S_1=S(r,r')=\frac{r'-r}{1-r}\com\q
S_2=S(r',r'')=\frac{r''-r'}{1-r'}\com\q
T_1=\frac{r''-r'}{1-r}\equiv T(r,r',r'')\com
                                                    \label{g3.1}
\end{eqnarray}
%%%%%%%%%%%%%%%%%%%%%%%%%%%%%
where $R(r,r')$ and $S(r,r')$ was introduced in the previous order
(\ref{g2}), and
$T(r,r',r'')$ is a new function. $R_i,S_i$ and $T_1$ have the
following relations.
%****(g3.2)%%%%%%%%%%%%%%%%%%%%  
\begin{eqnarray}
R_1+S_1=1\com\q R_2+S_2=1\com\q R_3+T_1=R_1\com
\q\frac{R_3}{R_2}=R_1\com\q \frac{T_1}{S_2}=R_1  \com\nn\\
R_1,R_2,R_3,S_1,S_2,T_1>0                   \pr
                                                    \label{g3.2}
\end{eqnarray}
%%%%%%%%%%%%%%%%%%%%%%%%%%%%%

Here we finish treating the Weyl anomaly $G(x,x;t)|_{t^0}$
in terms of the general ones:\ $(W_\mn, N_\la, M)$ and their
derivatives. Their explicit forms depend on each model.
The content of App.C comes from 
the 6 dim scalar-gravity theory (\ref{f.10b}). 
In the following two sections, we explain
how we obtain the explicit form of the Weyl anomaly
using the obtained formulae .

%%%%%%%%%%%%%%%%%%%%%%%%%%%%%%%%%%%%%%%%%%%%%%%%%%%%%%%%%%%%%%%%%%%%%
%%%%%%%%%%%%%%%%%%%%%%%%%%%%%%   SEC  5    %%%%%%%%%%%%%%%%%%%%%%%%%%
%%%%%%%%%%%%%%%%%%%%%%%%%%%%%%%%%%%%%%%%%%%%%%%%%%%%%%%%%%%%%%%%%%%%%
\section{ Four Special Graphs }
In the 6 dim space, there are four ``important general invariants'' as
the Weyl anomaly terms, which will be explained in the
next section. In order to obtain the four coefficients
of those terms, 
let us determine the coefficients of the following
four $(\pl\pl h)^3$-terms which appear in $G(x,x;t)|_{t^0}$.
%****(coef.0)%%%%%%%%%%%%%%%%%%%%  
\begin{eqnarray}
\mbox{Graph3}=\pl_\si \pl_\tau h_\mn\cdot \pl_\n \pl_\la h_{\tau\om}\cdot
                                                      \pl_\om \pl_\m h_\ls  \com   \nn\\
\mbox{Graph67}=\pl_\tau \pl_\om h_\mn\cdot \pl_\m \pl_\n h_{\ls}\cdot
                                                      \pl_\la \pl_\si h_{\tau\om}  \com   \nn\\
\mbox{Graph1}=\pl_\m \pl_\n h_{\n\la}\cdot \pl_\la \pl_\si h_{\si\tau}\cdot
                                     \pl_\tau \pl_\om h_{\om\m}   \com\nn\\
\mbox{Graph2}=\pl_\m \pl_\n h_{\tau\si}\cdot \pl_\si \pl_\la h_{\la\n}\cdot
                                                      \pl_\tau \pl_\om h_{\om\m}  \pr
                                                                                     \label{coef.0}
\end{eqnarray}
%%%%%%%%%%%%%%%%%%%%%%%%%%%%%
Here we introduce a useful graphical representation to express
SO(n) invariants like given above.
(See ref.\cite{II97} for detail.)
We graphically express the basic ingredient\ $\pl_\m\pl_\n h_\ls$ as
%****(graph.1)%%%%%%%%%%%%%%%%%%%%  
\begin{eqnarray}
     \begin{array}{c} 
\mbox{
%%%%%%%%%%%%%%%%%%%%  FIG  %%%%%%%%%%%%
{\epsfysize=30mm\epsfbox{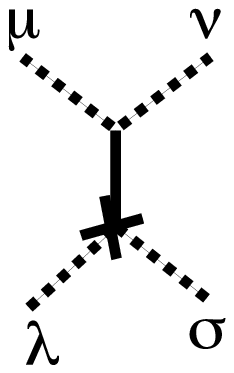}}
%\psbox[height=15mm]{S5g}
      }
      \end{array}	  
\equiv\pl_\m\pl_\n h_\ls\ \pr  
                                                    \label{graph.1}
\end{eqnarray}
%%%%%%%%%%%%%%%%%%%%%%%%%%%%%
The contraction of suffixes is expressed by gluing the corresponding
suffix-lines. Then the above four terms (\ref{coef.0})
are graphically shown in Fig.9. The advantage of this representation
is that the way suffixes are contracted can be read by the topology
of the graph. We need not bother about the dummy contracted suffixes.
For later use, we introduce the following usage too.
%****(graph.2)%%%%%%%%%%%%%%%%%%%%  
\begin{eqnarray}
     \begin{array}{c} 
\mbox{
%%%%%%%%%%%%%%%%%%%% FIG %%%%%%%%%%%%%%%
{\epsfysize=30mm\epsfbox{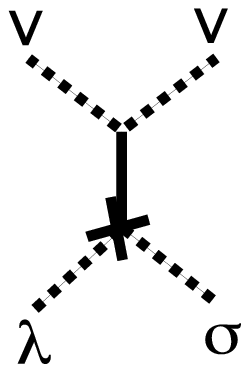}}
%\psbox[height=30mm]{S5h}
      }
      \end{array}	  
\equiv\pl_\m\pl_\n h_\ls\cdot v^\m v^\n\ ,\ 
     \begin{array}{c} 
\mbox{
%%%%%%%%%%%%%%%%%%%%% FIG %%%%%%%%%%%%%%
{\epsfysize=30mm\epsfbox{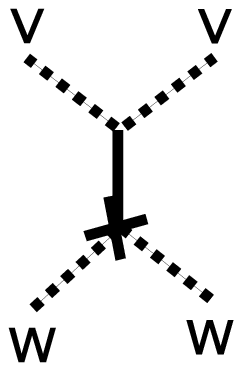}}
%\psbox[height=30mm]{S5i}
      }
      \end{array}	  
\equiv\pl_\m\pl_\n h_\ls\cdot v^\m v^\n w^\la w^\si\pr
                                                    \label{graph.2}
\end{eqnarray}
%%%%%%%%%%%%%%%%%%%%%%%%%%%%%
 
The choice of those graphs (\ref{coef.0}) 
is an important step of this algorithm. As will be explained in Sec.6, 
it is done by looking the table of App.D 
and the content of the ``important invariants''.
%
%in such a way that
%the chosen graphs ($(\pl\pl h)^3$-terms) satisfy the following properties:
%\begin{enumerate}
%\item
%They must appear in the weak-field expansion of the important invariants
%:\ the Euler term and the conformal invariants.
%\item
%The important invariants only contribute them.
%\item
%They are so "characteristic" that some special parts of
%$G=G_0+G_1+\cdots$ only contribute to them.
%\end{enumerate}
%The item 1 is indispensable to fix the coefficients. 
%Others are for calculational simplicity but
%become practically more important as we treat the higher order.
%
In the evaluation, we exploit the topology of graphs in order
to efficiently select relevant terms.
%%%%%  FIGURE of 4 GRAPHS  %%%%%%%%%%%
%%%%%%%%%%%%%%%%%%%%%%%    Fig.9   %%%%%%%%%%%%%%%%%%%%%%%%%%%%%%%%%%%%
\begin{figure}
\centerline{\epsfysize=4cm\epsfbox{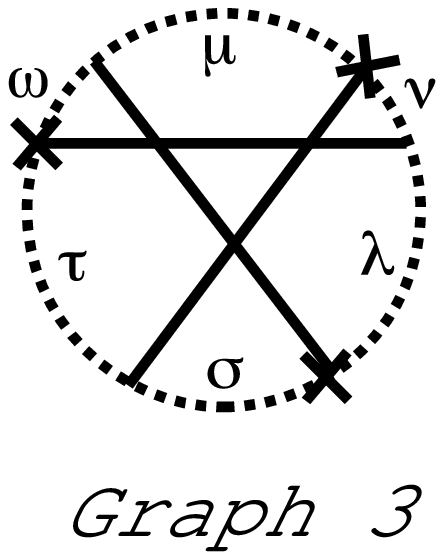},\ 
\epsfysize=4cm\epsfbox{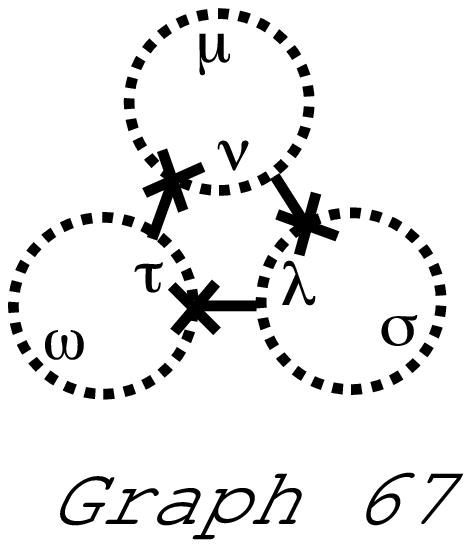},\ 
\epsfysize=4cm\epsfbox{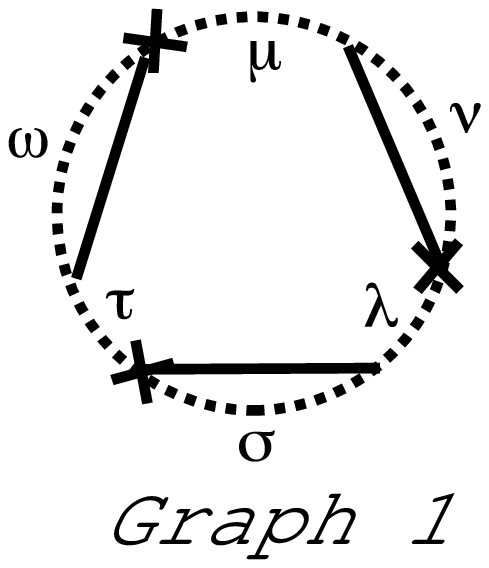},\ 
\epsfysize=4cm\epsfbox{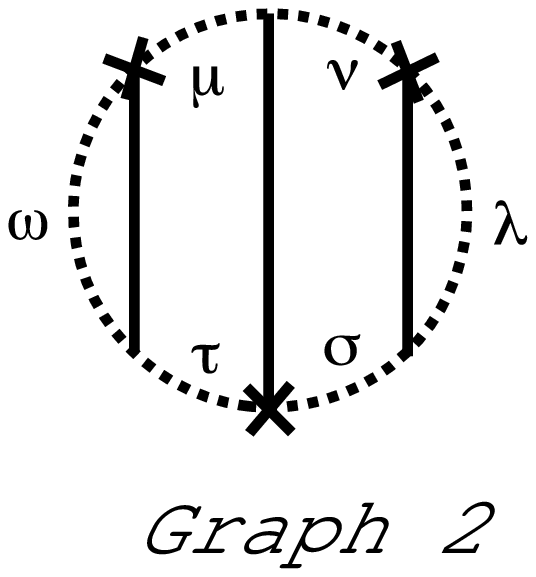}
             }
%\begin{center}\psbox[height=4cm]{Graph3},\ 
%\psbox[height=4cm]{Graph67},\ 
%\psbox[height=4cm]{Graph1},\ 
%\psbox[height=4cm]{Graph2}
%\end{center}
   \begin{center}
Fig.9\ 
Four Graphs of 3,67,1 and 2. See the (\ref{graph.1})
 for the definition of the graph.
%***Graph3.eps, Graph67.eps, Graph1.eps, Graph2.eps
   \end{center}
\end{figure}
%%%%%%%%%%%%%%%%%%%%%%%%%%%%%%%%%%%%%%%%%%%%%%%%%%%%%%%%%%%%%%%%%%%%%
All four graphs above come only from $G_3(x,x;t)|_{t^0}$. It is
graphically seen by the following common features:
%%%%%%%%%%%%%%%%%%%%
\begin{itemize}
\item
There is no 
%%%%%%%%%%%%%%  FIG  %%%%%%%%%%%%%%%%
{\epsfysize=5mm\epsfbox{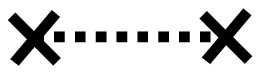}}\ .
%\psbox[height=5mm]{S5a}\ .
\item
No tadpoles (
%%%%%%%%%%%%%  FIG %%%%%%%%%%%%%
{\epsfysize=5mm\epsfbox{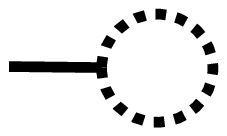}}\ ,\ {\epsfysize=5mm\epsfbox{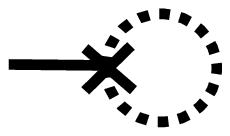}} 
%\psbox[height=5mm]{S5b1}\ ,\ \psbox[height=5mm]{S5b2}
)\ .
\end{itemize}
%%%%%%%%%%%%%%%%%%%%
and the structure of $G_1$ and $G_2$:\ 
%****(coef.0b)%%%%%%%%%%%%%%%%%%%%  
\begin{eqnarray}
G_1(x,x;t)|_{t^0}
\sim\int dr
\int d^6w~ G_0(w;(1-r)r)\,V_2(x,w;w,r)\com\nn\\
G_2(x,x;t)|_{t^0}
\sim\int dr' \int dr
\int d^6\wbar \int d^6w \,
G_0(\wbar;\frac{(1-r')(r'-r)}{1-r})G_0(w;(1-r)r) \nn\\
\left\{ 
V_1(x,w';w'-w,r'-r)V_0(x,w;w,r)+V_0(x,w';w'-w,r'-r)V_1(x,w;w,r)
                                                   \right\}\ ,  \label{coef.0b}
\end{eqnarray}
%%%%%%%%%%%%%%%%%%%%%%%%%%%%%
where $V_2,\ V_1$ and $V_0$ are graphically shown in App.C.2.
Both $G_1|_{t=0}$ and $G_2|_{t=0}$ contain, at least, one
of two graph-ingredients itemized above. 
For simplicity we focus, in this section,
only on Graphs3 and 67.
Graph1 and 2 
are evaluated in App.B.
Because of the following common features of Graph3 and 67,
%%%%%%%%%%%%%%%%%%%%%
\begin{itemize}
\item
No tadpoles.
\item
There is no \ \ 
%%%%%%%%%%%%%%%%  FIG  %%%%%%%%%%%%
{\epsfysize=8mm\epsfbox{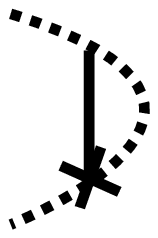}}
%\psbox[height=8mm]{S5c}
\ .
\end{itemize}
%%%%%%%%%%%%%%%%%%%%%
we can reduce the expression of $G_3(x,x;t)|_{t^0}$, 
(\ref{g3}) with $V_0$ given by (\ref{C2.2}) and (\ref{GraphExp.1}), to the
following form as the relevant part.
%****(coef.1)%%%%%%%%%%%%%%%%%%%%  
\begin{eqnarray}
G_3(x,x;t)|_{t^0}
=\frac{1}{(4\pi)^{3}}
\int^1_0dr'' \int^{r''}_0dr' \int^{r'}_0dr
\int d^6{\bar \wbar}\int d^6\wbar \int d^6w            \nn\\
\times G_0({\bar \wbar};\frac{(1-r'')(r''-r')}{1-r'})
G_0(\wbar;\frac{(1-r')(r'-r)}{1-r})G_0(w;(1-r)r)        \nn\\
\times (-\half)^3\times\frac{1}{4(r''-r')^2}
\times\frac{1}{4(r'-r)^2}\times\frac{1}{4r^2}
\times (R_2)^2(S_2)^2\times
                                                        \nn\\
     \begin{array}[t]{c} 
\mbox{
%%%%%%%%%%%%%%%  FIG   %%%%%%%%%%%%%%
{\epsfysize=30mm\epsfbox{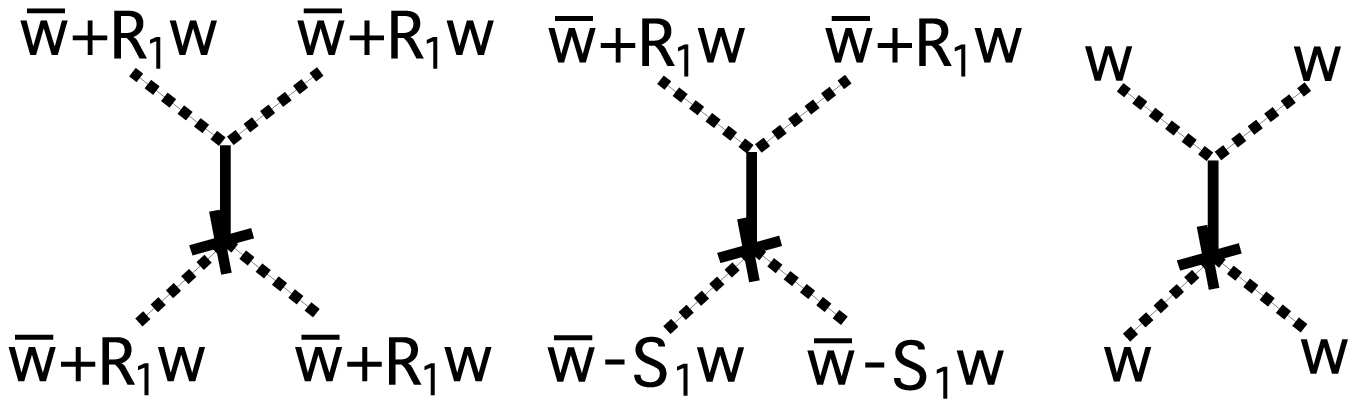}}
%\psbox[height=30mm]{S5d}
      }
     \end{array}	  
	               \nn\\
+\mbox{irrelevant terms}\com
                                                    \label{coef.1}
\end{eqnarray}
%%%%%%%%%%%%%%%%%%%%%%%%%%%%%
where the relations $\frac{R_3}{R_2}=R_1,\ \frac{T_1}{S_2}=R_1 $
are used (see (\ref{g3.2})). 
Furthermore, again 
by the common features of the graphs3 and 67, we see 
no contribution comes from $\wbar^8$- and $\wbar^6$-terms.\cite{foot5b}
%****(coef.2)%%%%%%%%%%%%%%%%%%%%  
\begin{eqnarray}
     \begin{array}[t]{c} 
\mbox{
%%%%%%%  FIG   %%%%%%%%%%%%%%%%%
{\epsfysize=30mm\epsfbox{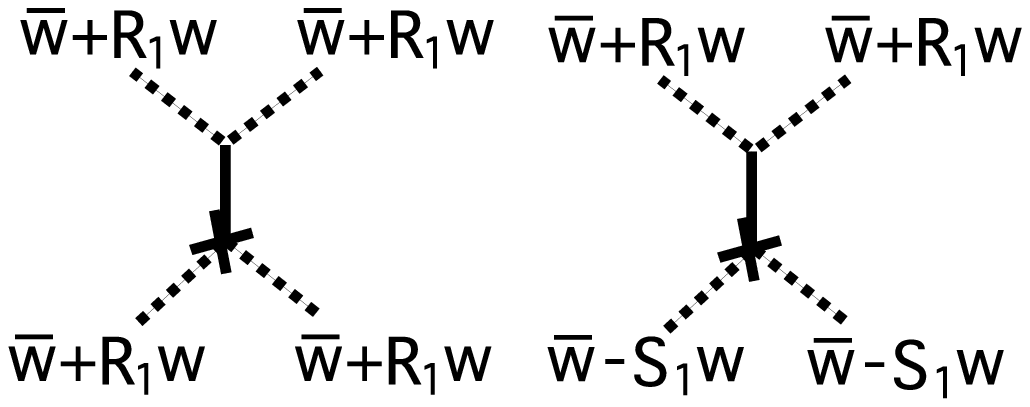}}
%\psbox[height=30mm]{S5e}
      }
      \end{array}	  
\nn\\
=
\begin{array}{c} 
\mbox{
%%%%%%%%%%%%  FIG  %%%%%%%%%%%%%
{\epsfysize=30mm\epsfbox{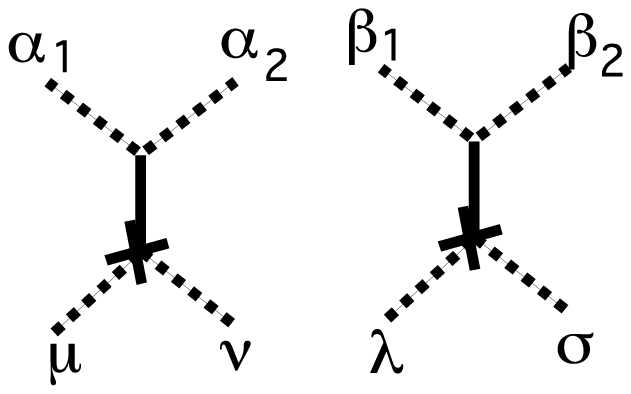}}
%\psbox[height=30mm]{S5f}
      }
      \end{array}	  
\nn\\
\times \{ (R_1)^6(S_1)^2
w^{\al_1}w^{\al_2}w^{\m}w^{\n}w^{\be_1}w^{\be_2}w^{\la}w^{\si}\nn\\
+(R_1)^6
w^{\al_1}w^{\al_2}w^{\m}w^{\n}w^{\be_1}w^{\be_2}
                                        \wbar^{\la}\wbar^{\si}\nn\\
-(R_1)^5 S_1
(w^{\al_1}w^{\al_2}w^{\m}w^{\n}w^{\be_1}\wbar^{\be_2}+\mbox{}_6C_1)
                                        \wbar^{\la}w^{\si}\times 2\nn\\
+(R_1)^4 (S_1)^2
(w^{\al_1}w^{\al_2}w^{\m}w^{\n}\wbar^{\be_1}\wbar^{\be_2}+\mbox{}_6C_2)
                                        w^{\la}w^{\si}\nn\\
+(R_1)^2 (S_1)^2
(\wbar^{\al_1}\wbar^{\al_2}\wbar^{\m}\wbar^{\n}w^{\be_1}w^{\be_2}
                                    +\mbox{}_6C_4) w^{\la}w^{\si}\nn\\
-(R_1)^3 (S_1)
(\wbar^{\al_1}\wbar^{\al_2}\wbar^{\m}w^{\n}w^{\be_1}w^{\be_2}
                          +\mbox{}_6C_3) \wbar^{\la}w^{\si}\times 2\nn\\
+(R_1)^4
(\wbar^{\al_1}\wbar^{\al_2}w^{\m}w^{\n}w^{\be_1}w^{\be_2}
                          +\mbox{}_6C_2) \wbar^{\la}\wbar^{\si} \}\nn\\
											+O(\wbar^6)+O(\wbar^8)\com
                                                    \label{coef.2}
\end{eqnarray}
%%%%%%%%%%%%%%%%%%%%%%%%%%%%%
where $\mbox{}_mC_r$ means "other choices of $r$ $\wbar$'s among
$m$ suffixes appearing in the first term within the brackets".
The ${\bar \wbar},\wbar,w$-integrations 
are three independent Gaussian integrations, and they
give the product of two parts for each term.
One part is
a rational function
of the parameters and the other is 
a symmetrized product of the Kronecker's deltas. In terms
of a "components" notation:   
%****(coef.3)%%%%%%%%%%%%%%%%%%%%  
\begin{eqnarray}
<c_1,c_2>\equiv 
c_1\mbox{(Graph3)}+c_2\mbox{(Graph67)}
               \com
                                                    \label{coef.3}
\end{eqnarray}
%%%%%%%%%%%%%%%%%%%%%%%%%%%%%
the final result is evaluated as
%****(coef.4)%%%%%%%%%%%%%%%%%%%%  
\begin{eqnarray}
G_3(x,x;t)|_{t^0}
=\frac{1}{(4\pi)^{3}}
\int^1_0dr'' \int^{r''}_0dr' \int^{r'}_0dr
\times (-\half)^3\times\frac{1}{4(r''-r')^2}
\times\frac{1}{4(r'-r)^2}\times\frac{1}{4r^2}           \nn\\
\times (R_2)^2(S_2)^2
\times \left[\ \ 
        (R_1)^6(S_1)^2\{2(1-r)r\}^6\times <64,16> \right. \nn\\
+2\frac{(1-r')(r'-r)}{1-r} \{2(1-r)r\}^5 \times 
\{ (R_1)^6 <0,0>                            
-2(R_1)^5S_1 <32,8>      \nn\\
+(R_1)^4(S_1)^2 <64,16>  \}\nn\\
+\{2\frac{(1-r')(r'-r)}{1-r}\}^2 \{2(1-r)r\}^4 \times 
\{ (R_1)^2(S_1)^2 <0,8>                            
-2(R_1)^3S_1 <32,0>           \nn\\
\left. +(R_1)^4 <0,8>  \} \ \ 
       \right]                \nn\\
+\mbox{other graph-terms}\nn\\
=\frac{1}{(4\pi)^{3}}<\frac{4}{27\times 105},-\frac{1}{36\times 63}>
+\mbox{other graph-terms}\pr
                                                    \label{coef.4}
\end{eqnarray}
%%%%%%%%%%%%%%%%%%%%%%%%%%%%%
The above numbers $<c_1,c_2>$ are obtained by a computer
calculation\cite{SI98}.
The numbers $c_1$ and $c_2$\ 
show the "weights" when a symmetrized product
of Kronecker's deltas are multiplied to a $(\pl\pl h)^3$-tensor.
For example the first one of (\ref{coef.4}), $<64,16>$, says 
%****(coef.4b)%%%%%%%%%%%%%%%%%%%%  
\begin{eqnarray}
     \begin{array}{c} 
\mbox{
%%%%%%%%%%%%  FIG  %%%%%%%%%%
{\epsfysize=30mm\epsfbox{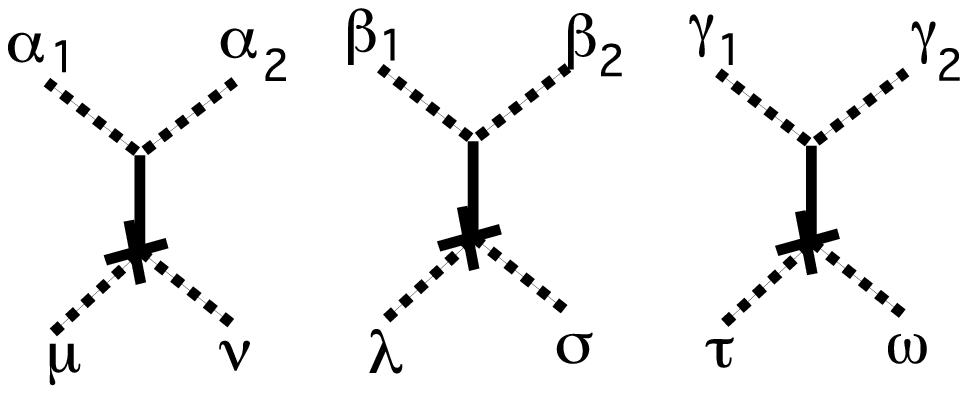}}
%\psbox[height=30mm]{S5j}
      }
      \end{array}
\times [\al_1\al_2\mn\be_1\be_2\ls\ga_1\ga_2\tau\om]\nn\\
=64\mbox{(Graph3)}+16\mbox{(Graph67)}+\mbox{other graph-terms}
\com                                           \label{coef.4b}
\end{eqnarray}
%%%%%%%%%%%%%%%%%%%%%%%%%%%%%
where the notation 
$[\al_1\al_2\cdots]$ is the symmetrized product of Kronecker's deltas
and is defined in (\ref{a.4}) of App.A.
Note that in the expression just before the parameter integral,
poles at $r=1,\ r'=1$ cancel out and there remain no divergences
of the parameter integrals. This occurs also in the calculation
in App.B.
Adding the result for
Graph1 and 2 explained in App.B, we obtain
finally the total contribution of $G(x,x;t)|_{t^0}$ to the four graphs as 
%****(coef.5)%%%%%%%%%%%%%%%%%%%%  
\begin{eqnarray}
G(x,x;t)|_{t^0}
=\frac{1}{(4\pi)^{3}}
\{ \frac{4}{27\times 105}\mbox{(Graph3)}
-\frac{1}{36\times 63}\mbox{(Graph67)}
+\frac{1}{36\times 630}\mbox{(Graph1)}\nn\\
+\frac{1}{36\times 35}\mbox{(Graph2)} \}
+\mbox{other terms}\pr
                                                    \label{coef.5}
\end{eqnarray}
%%%%%%%%%%%%%%%%%%%%%%%%%%%%%

%%%%%%%%%%%%%%%%%%%%%%%%%%%%%%%%%%%%%%%%%%%%%%%%%%%%%%%%%%%%%%%%%%%%%
%%%%%%%%%%%%%%%%%%%%%%%%%%%%%%   SEC  6    %%%%%%%%%%%%%%%%%%%%%%%%%%
%%%%%%%%%%%%%%%%%%%%%%%%%%%%%%%%%%%%%%%%%%%%%%%%%%%%%%%%%%%%%%%%%%%%%
\section{ Final Invariant Form of 6 Dim Weyl Anomaly}
The general structure of the Weyl anomaly has been analyzed by
refs.\cite{MD77,DS93,SI98b}. 
They claim that
the Weyl anomaly is composed of three types of term:\ 
A)\ the Euler term;\ B)\ Conformal invariants;\ C)\ Trivial terms.
Trivial terms are those which can be absorbed by local counter-terms
made of the metric.
This statement is checked by the cohomology analysis up to
6 dimensions\cite{BPB}. We called, in Sec.5, the A and B terms
``important invariants''. Therefore we may write, 
for the present 6 dim case, 
%****(inv.1)%%%%%%%%%%%%%%%%%%%%  
\begin{eqnarray}
G(x,x;t)|_{t^0}
=\frac{1}{(4\pi)^{3}}\sqg\{
xC_1+yC_2+zC_3+wE+\mbox{trivial terms} \}\nn\\
C_1=C_{\mn\ls}C^{\si\la\ab}C_{\be\al}^{~~\n\m}\com\q
C_2=C_{\mn\al\si}C^{\n\la\be\al}C_{\la~~\be}^{~\m\si}\com\nn\\
C_3\sim C_{\mn\ab}\na^2C^{\mn\ab}+\cdots\com\q 
E\sim R_{\mn\ab}R_{\ls\ga\del}R_{\tau\om\ep\th}
\ep^{\mn\ls\tau\om}\ep^{\ab\ga\del\ep\th}\com\nn\\
C^\la_{~\mn\si}=R^\la_{~\mn\si}+\frac{1}{4}(\del^\la_\n R_{\m\si}
+g_{\m\si}R^\la_{~\n}-\n\change\si)
+\frac{1}{20}(\del^\la_\si g_\mn-\n\change\si)R\com
                                                    \label{inv.1}
\end{eqnarray}
%%%%%%%%%%%%%%%%%%%%%%%%%%%%%
where $C^\la_{~\mn\si}$ is the Weyl tensor, $C_1,C_2$ and $C_3$ are
three independent conformal invariants, and $E$ is the Euler term.
$x,y,z$, and $w$ are some constants to be determined.
It is well-established that all possible independent invariants
(parity even) with mass-dimension 6 are, in the 6 and higher 
space-dimensions,
given by the following 17 terms\cite{BPB,FKWC,SI}.
%****(inv.2)%%%%%%%%%%%%%%%%%%%%
\begin{eqnarray}
P_1=RRR\com\ \ P_2=RR_\mn R^\mn\com\ \ P_3=RR_{\mn\ls}R^{\mn\ls}\com \nn\\
P_4=R_\mn R^{\n\la}R_\la^{~\mu}\com\ \ P_5=-R_{\mn\ls}R^{\mu\la}R^{\nu\si}
\com\ \ P_6=R_{\mn\ls}R_\tau^{~\nu\ls}R^{\mu\tau}\com\nn\\
A_1=R_{\mn\ls}R^{\si\la}_{~~~\tau\om}R^{\om\tau\n\m}\com\ \ 
B_1=R_{\mn\tau\si}R^{\n~~~\tau}_{~\la\om}R^{\la\mu\si\om}\com\nn\\
O_1=\na^\mu R\cdot \na_\mu R\com\ \ 
O_2=\na^\mu R_\ls\cdot \na_\mu R^\ls\com\nn\\ 
O_3=\na^\mu R^{\la\rho\si\tau}\cdot \na_\mu R_{\la\rho\si\tau}\com\ \ 
O_4=\na^\mu R_{\la\n}\cdot \na^\n R^\la_{~\m}\com\nn\\ 
T_1=\na^2R\cdot R\com\ \ T_2=\na^2R_\ls\cdot R^\ls\com\ \ 
T_3=\na^2R_{\la\rho\si\tau}\cdot R^{\la\rho\si\tau}\com\ \ \nn\\
T_4=\na^\m\na^\n R\cdot R_\mn\com\ \nn\\ 
%T_5=\na^\m\na^\n R^{\la\rho}\cdot R_{\mu\la\n\rho}\com\nn\\
S=\na^2\na^2R\pr
                               \label{inv.2}
\end{eqnarray}
%%%%%%%%%%%%%%%%%%%%%%%%%%%%%
In terms of above 17 "basis", we can rewrite (\ref{inv.1}) as follows.
%****(inv.3)%%%%%%%%%%%%%%%%%%%%
\begin{eqnarray}
C_1=\frac{9}{200}P_1-\frac{27}{40}P_2+\frac{3}{10}P_3
+\frac{5}{4}P_4+\frac{3}{2}P_5-3P_6+A_1\com\nn\\
C_2=-\frac{19}{800}P_1+\frac{57}{160}P_2-\frac{3}{40}P_3
-\frac{7}{16}P_4-\frac{9}{8}P_5+\frac{3}{4}P_6+B_1\com\nn\\
C_3=P_1-8P_2-2P_3
+10P_4+10P_5-\frac{1}{2}T_1+5T_2-5T_3\com\nn\\
E=P_1-12P_2+3P_3+16P_4+24P_5-24P_6+4A_1-8B_1\com\nn\\
G(x,x;t)|_{t^0}
=\frac{1}{(4\pi)^{3}}\sqg\{
               (x+4w)A_1+(y-8w)B_1               \nn\\
+(\frac{5}{4}x-\frac{7}{16}y+10z+16w)P_4
+(\frac{3}{2}x-\frac{9}{8}y+10z+24w)P_5
+\mbox{other invariants} 
                          \}\com
                               \label{inv.3}
\end{eqnarray}
%%%%%%%%%%%%%%%%%%%%%%%%%%%%%
where "other invariants" means other than $A_1, B_1, P_4$ and $P_5$.
Now we see how nicely we have chosen the four graphs in Sec.5.
\flushleft{(i)}\nl
We note that the trivial terms are written in the total
derivative form, therefore they do not contain any one of 
$A_1, B_1, P_4$ and $P_5$.\nl
\flushleft{(ii)}\nl
Now we know, from App.D,
%****(inv.4)%%%%%%%%%%%%%%%%%%%%
\begin{eqnarray}
P_4|_{(\pl\pl h)^3}& =-\fourth \mbox{(Graph1)}
+\mbox{other\ }(\pl\pl h)^3\mbox{-terms}\com\q  & \nn\\
P_5|_{(\pl\pl h)^3}& =\fourth \mbox{(Graph2)}
+\mbox{other\ }(\pl\pl h)^3\mbox{-terms}\com\q  & \nn\\
A_1|_{(\pl\pl h)^3}& =- \mbox{(Graph3)}
+\mbox{other\ }(\pl\pl h)^3\mbox{-terms}\com\q  & \nn\\
B_1|_{(\pl\pl h)^3}&=-\fourth \mbox{(Graph3)}
+\fourth \mbox{(Graph67)}
+\mbox{other\ }(\pl\pl h)^3\mbox{-terms}\com   & 
                               \label{inv.4}
\end{eqnarray}
%%%%%%%%%%%%%%%%%%%%%%%%%%%%%
where "other $(\pl\pl h)^3$-terms" means ``other terms than 
Graph1,2,3 and 67''.\cite{foot7}\nl
\flushleft{(iii)}\nl 
We also note, again from App.D, 
Graphs1,2,3 and 67 do not come from other invariants than
($A_1,B_1,P_4,P_5$). \nl
These properties are due to the chosen four graphs.
Then we can rewrite (\ref{inv.3}) as
%****(inv.5)%%%%%%%%%%%%%%%%%%%%
\begin{eqnarray}
G(x,x;t)|_{t^0,(\pl\pl h)^3}
=\frac{1}{(4\pi)^{3}}\{
(-x-\fourth y-2w)\mbox{(Graph3)}+\fourth (y-8w)\mbox{(Graph67)}\nn\\
+\fourth(\frac{3}{2}x-\frac{9}{8}y+10z+24w)\mbox{(Graph2)}                
-\fourth(\frac{5}{4}x-\frac{7}{16}y+10z+16w)\mbox{(Graph1)}\nn\\
+\mbox{other\ }(\pl\pl h)^3\mbox{-terms} 
                          \}\pr
                               \label{inv.5}
\end{eqnarray}
%%%%%%%%%%%%%%%%%%%%%%%%%%%%%
Comparing the result of Sec.5, we finally obtain
%****(inv.6)%%%%%%%%%%%%%%%%%%%%  
\begin{eqnarray}
x=-\frac{83}{189\times 60}\com\q 
y=\frac{31}{189\times 15}\com\q 
z=-\frac{11}{189\times 50}\com\q 
w=\frac{3}{189\times 10}\com \nn\\
G(x,x;t)|_{t^0}=-2g^\mn <T_\mn>
=\frac{1}{(4\pi)^{3}}\frac{1}{189}\sqg\{
-\frac{83}{60}C_1+\frac{31}{15}C_2-
\frac{11}{50}C_3+\frac{3}{10}E+\mbox{trivial terms} \}\pr
                                                    \label{inv.6}
\end{eqnarray}
%%%%%%%%%%%%%%%%%%%%%%%%%%%%%
The trivial terms can be similarly obtained, but they do change by
introducing the local counterterms in the action and seem
to be unimportant. This result (\ref{inv.6}) is similar to
that of Ref.\cite{IGA91}.

%%%%%%%%%%%%%%%%%%%%%%%%%%%%%%%%%%%%%%%%%%%%%%%%%%%%%%%%%%%%%%%%%%%%%
%%%%%%%%%%%%%%%%%%%%%%%%%%%%%%   SEC  7    %%%%%%%%%%%%%%%%%%%%%%%%%%
%%%%%%%%%%%%%%%%%%%%%%%%%%%%%%%%%%%%%%%%%%%%%%%%%%%%%%%%%%%%%%%%%%%%%
\section{ Discussion and Conclusion  }

We have presented a new algorithm to obtain the Weyl anomaly
in the higher-dimensions. The Feynman rules in coordinate
space are presented. The graphical representation is exploited
at all stages. Especially the graphical calculation is
taken to efficiently compute the coefficients of the four graphs
given in Sec.5. Explicit result of the Weyl anomaly
for the 6 dim scalar-gravity is obtained. 
%The final result 
%partially agrees with that of Avramidi\cite{IGA91}. 

As the space-time dimension increases, the number of suffixes
to deal with increases because we must treat higher
products of Riemann tensors. The aid of the computer
is indispensable in this algebraic calculation. 
As mentioned at some places in the text and the appendices, 
the weight calculation in the
contraction of multi-suffixed quantities relies on a computer
(a C-program\cite{SI98}). 
Although we do not touch
on the computer algorithm in the present paper,
it helps to obtain the concrete results.  

In the recent rapid development of string theory, the low-energy
effective actions play an important role. They are field theories
(supergravities) on various (higher) dimensions.
The chiral anomalies of gravitational and gauge symmetries
were well analyzed for those low-energy effective actions.
The famous one is the pioneering work by Green and Schwarz\cite{GS84}
where they found the SO(32) and $E_8\times E_8$ gauge group
from the analysis of 
the chiral anomaly structure of 10 dim N=1
supergravity. 
%In the string field theory, the "quantization" of the string field 
%is still obscure.
%The analysis at present
%heavily depends on the supersymmetry and some duality relations
%in the effctive field theories. 
%As the purpose of finding a "solvable model" or a "critical theory", 
%it might be a good approach. 
%The approach, however, 
%looks difficult to understand the meaning of the string-field
%"quantization".
%As in the past experience in 2 dim quantum gravity, the symmetry
%argument helps to understand only the critical model but
%not its dynamics.
%In order to see the dynamical properties, we needed 
%the lattice calculation, the semiclassical analysis, etc.
Clearly the analysis from the Weyl anomaly side is lacking.
From the present standpoint, we stress the importance
of examining the string from 
the structure of the Weyl anomaly in 
the low-energy effective theories.\cite{MJD94}
One reason is we know some important models often have 
vanishing Weyl anomaly due to conformal symmetry.
Another is that the supersymmetry surely
relates the Weyl anomaly to the chiral one. 
We can examine the role of supersymmetry from this point.
Ref.\cite{HS98} recently analyzed the Weyl anomaly of conformal field theories
using the adS/CFT correspondence conjectured by \cite{Malda98}.
Weyl anomalies are explicitly obtained for some theories
in 2, 4 and 6 dimensions.
We hope the present result
will become useful when we go over the quantum field theory through
the string theory.

We have most experience with quantization only in 0 to 4 dim
field theories. 
Higher dimensional quantum field theories have not been so thoroughly 
examined (at least systematically) so far, 
because they are unrenormalizable except for free theories. 
In the text we consider free theories on curved space. 
It is meaningful if we ignore the trivial terms and the
the divergent massive terms, which should be explained by the
quantum gravitational mode. It might be possible to find
a clue to make meaningful the higher dimensional field theories
through the present analysis.

Although the non-perturbative aspect is recently stressed 
in string theory, 
the perturbative analysis is still important because it is 
one of few
reliable approaches to analyze dynamical aspects systematically.
We suppose the analysis will become important to understand
how the quantum field theory is generalized in the coming
new era of Planck physics. 
Finally we point out that the graphs shown in Fig.10 and Fig.11
are expected to become some of the "important graphs" in the Weyl
anomaly calculation in 8 dim and 10 dim respectively.

%%%%%  Two (ddh)**4 Graphs in 8 dim  %%%%%%%%%%%
%%%%%%%%%%%%%%%%%%%%%%%    Fig.10   %%%%%%%%%%%%%%%%%%%%%%%%%%%%%%%%%%%%
  \begin{figure}
\centerline{\epsfysize=4cm\epsfbox{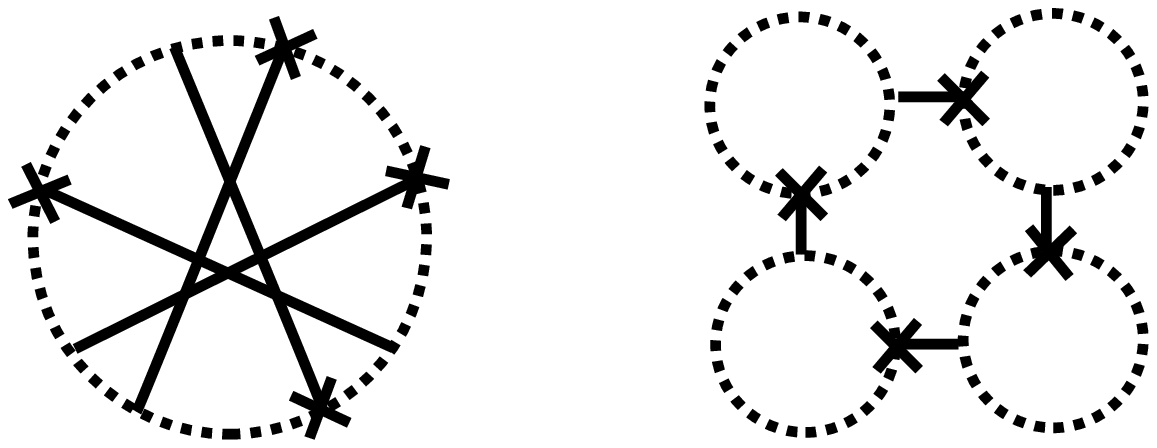} }
%\begin{center}\psbox[height=4cm]{8Dgraph}\end{center}
   \begin{center}
Fig.10\
Two important $(\pl\pl h)^4$-graphs in 8 dim. 
%***8Dgraph.eps
   \end{center}
\end{figure}
%%%%%%%%%%%%%%%%%%%%%%%%%%%%%%%%%%%%%%%%%%%%%%%%%%%%%%%%%%%%%%%%%%%%%

%%%%%  Two (ddh)**5 Graphs in 10 dim  %%%%%%%%%%%
%%%%%%%%%%%%%%%%%%%%%%%    Fig.11   %%%%%%%%%%%%%%%%%%%%%%%%%%%%%%%%%%%%
  \begin{figure}
\centerline{\epsfysize=4cm\epsfbox{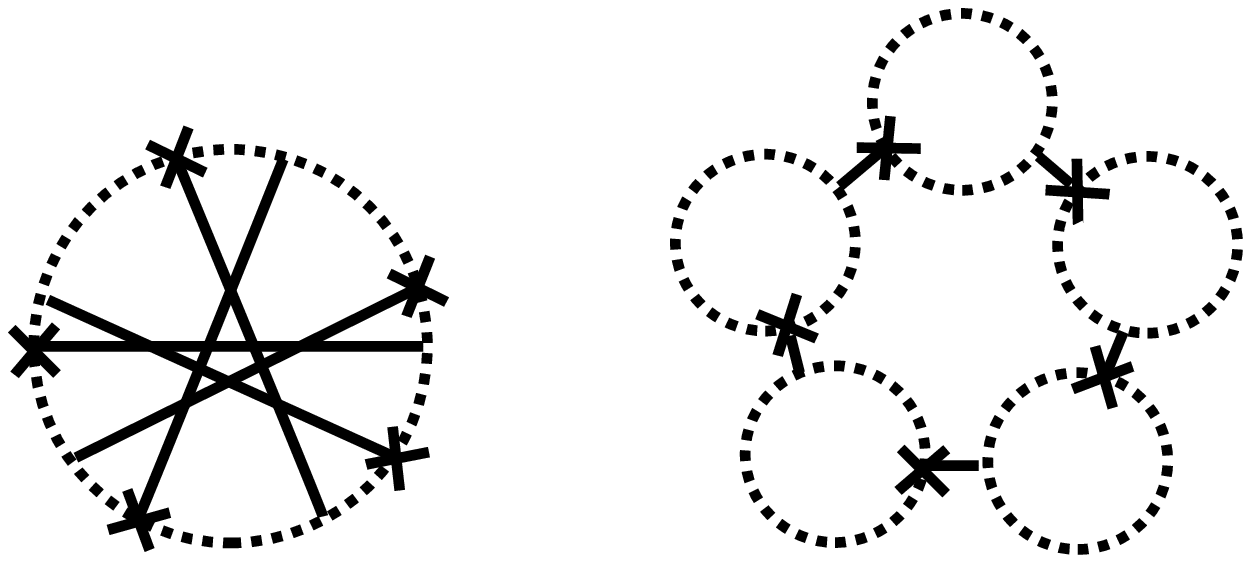} }
%\begin{center}\psbox[height=4cm]{10Dgraph}\end{center}
   \begin{center}
Fig.11\
Two important $(\pl\pl h)^5$-graphs in 10 dim. 
%***10Dgraph.eps
   \end{center}
\end{figure}
%%%%%%%%%%%%%%%%%%%%%%%%%%%%%%%%%%%%%%%%%%%%%%%%%%%%%%%%%%%%%%%%%%%%%

\vs 1
%%%%%%%%%%%%%%%%%%%%%%%%%%%%%%%%%%%%%%%%%%%%%%%%%%%%%%%%%%%%%%%%%%
%%%%%%%%%%%%%%%%%%%%%%%% Acknowledgment %%%%%%%%%%%%%%%%%%%%%%%%%%%%%
%%%%%%%%%%%%%%%%%%%%%%%%%%%%%%%%%%%%%%%%%%%%%%%%%%%%%%%%%%%%%%%%%%
\begin{flushleft}
{\bf Acknowledgment}
\end{flushleft}
The authors thank Prof. M. Creutz for reading the manuscript
carefully. One of the authors (S.I.) thanks the hospitality
at Physics Department, Brookhaven National Laboratory
where this work is finished.
%informing their
%stimulating results and discussions about this subject. 
%He also thanks Prof. 
%for an encouraging comment and some corrections in the original 
%manuscript.

\vs 2

\newpage
%%%%%%%%%%%%%%%%%%%%%%%%%%  App. A  %%%%%%%%%%%%%%%%%%%%%%%%%%%%%%%
%%%%%%%%%%%%%%%%%%%%%%%%%%%%%%%%%%%%%%%%%%%%%%%%%%%%%%%%%%%%%%%%%%%
\begin{flushleft}
{\Large\bf Appendix A.\ Propagator Rules}
\end{flushleft}

$G_0(x;t)$ is the solution of the heat equation Eq.(\ref{f.12})
and is given  by
%****(a.1)%%%%%%%%%%%%%%%%%%%%
\begin{eqnarray}
G_0(x;t) &=& \frac{1}{(4\pi t)^{n/2}}e^{-\frac{x^2}{4t}}~I_N\com
\q t>0\com\q x^2=(x_1)^2+(x_2)^2+\cdots +(x_n)^2\com
                                                      \label{a.1}
\end{eqnarray}
%%%%%%%%%%%%%%%%%%%%%%%%%%%%%%
where $I_N$ is the $N\times N$ unit matrix and
$x=(x_1,x_2,\cdots,x_n)$ is the n-dim Euclidean space coordinates.
It has the basic properties:
%****(a.2)%%%%%%%%%%%%%%%%%%%%
\begin{eqnarray}
G_0(x;t) &=& G_0(-x;t)\ :\mbox{even function of }x\com\nn\\
G_0(\sqrt{a}x;at) &=& a^{-\frac{n}{2}}G_0(x;t)\com\q a>0\pr
                                                      \label{a.2}
\end{eqnarray}
%%%%%%%%%%%%%%%%%%%%%%%%%%%%%%
The integral formula is given by
%****(a.3)%%%%%%%%%%%%%%%%%%%%
\begin{eqnarray}
\int d^nx~x^{\m_1}\cdots x^{\m_{2s}}G_0(x;t) &=&
[\m_1\cdots \m_{2s}](2t)^s~I_N\com
                                                      \label{a.3}
\end{eqnarray}
%%%%%%%%%%%%%%%%%%%%%%%%%%%%%%
where $[\m_1\cdots \m_{2s}]$ is the totally symmetric sum of
products of Kronecker's deltas and is concretely defined by
%****(a.4)%%%%%%%%%%%%%%%%%%%%
%%%98.10.16 the first [ after \\ do bad thing in BNL-latex %%%
\begin{eqnarray}
[\mn]&\equiv&\del^\mn\com\nn\\
{[}\mn\ls]&\equiv&\del^\mn\del^\ls+\del^{\m\si}\del^{\n\la}
+\del^{\m\la}\del^{\n\si}\com\nn\\
{[}\m_1\cdots\m_6]&\equiv&\del^{\mu_1\mu_2}[\mu_3\cdots\mu_6]
+\del^{\mu_1\mu_3}[\cdots]+\cdots+\del^{\mu_1\mu_6}[\cdots]\com\nn\\
&\cdots&\nn\\
&\cdots&\nn\\
{[}\m_1\cdots\m_{2s}]
&\equiv&\del^{\mu_1\mu_2}[\mu_3\cdots\mu_{2s}]
+\del^{\mu_1\mu_3}[\cdots]+\cdots+\del^{\mu_1\mu_{2s}}[\cdots]\pr
\label{a.4}
\end{eqnarray}
%%%%%%%%%%%%%%%%%%%%%%%%%%%%%%
$G_0(x,t)$ has the "convolution" property.
%****(a.5)%%%%%%%%%%%%%%%%%%%%
\begin{eqnarray}
& \mbox{Propagator Rule 1} &\nn\\
& G_0(v-w;r-k)G_0(w-u;k-l)=G_0(\wbar;\frac{(r-k)(k-l)}{r-l})
G_0(v-u;r-l)\com                   & \nn\\ 
& r>k>l\com\q \wbar=w-\frac{(k-l)v+(r-k)u}{r-l}\pr &
                                                      \label{a.5}
\end{eqnarray}
%%%%%%%%%%%%%%%%%%%%%%%%%%%%%%
This relation is geometrically expressed in the coordinate space
as in Fig.12. 
%%%%%%%%%%%%%%%%%%%%%%%  Fig.12  %%%%%%%%%%%%%%%%%%%%%%%%%%%%%%%%%%%%
\begin{figure}
\centerline{\epsfysize=6cm\epsfbox{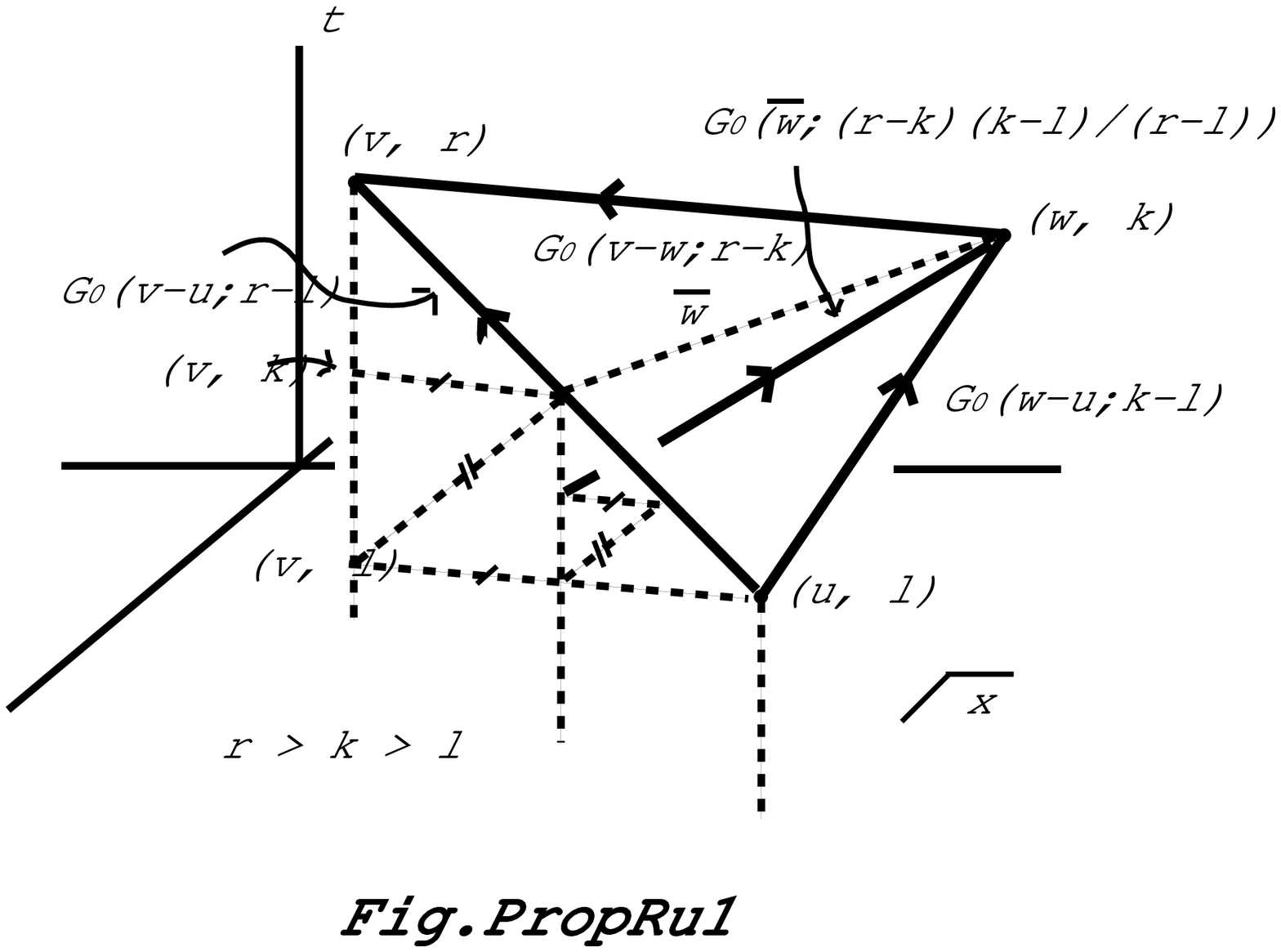}}
%\begin{center}\psbox[height=6cm]{prop1}\end{center}
   \begin{center}
Fig.12\ 
Propagator Rule 1, Eq.(\ref{a.5}).
%***prop1.eps
   \end{center}
\end{figure}
%%%%%%%%%%%%%%%%%%%%%%%%%%%%%%%%%%%%%%%%%%%%%%%%%%%%%%%%%%%%%%%%%%%%%
Some special cases of above are given by
%****(a.6x,a.6y,a.6z)%%%%%%%%%%%%%%%%%%%%
\begin{eqnarray}
& \mbox{Propagator Rule 2} \com\q\q
v=0,\ r=1\ \mbox{in Eq.(\ref{a.5})}     & \nn\\
& G_0(w;1-k)G_0(w-u;k-l)=G_0(\wbar;\frac{(1-k)(k-l)}{1-l})
G_0(u;1-l)\com  &  \nn\\ 
& 1>k>l\com\q \wbar=w-\frac{1-k}{1-l}u\pr
                                                      \label{a.6x}\\
& \mbox{Propagator Rule 3}\com\q\q
u=0,\ l=0\ \mbox{in Eq.(\ref{a.5})}    & \nn\\
& G_0(v-w;r-k)G_0(w;k)=G_0(\wbar;\frac{(r-k)k}{r})
G_0(v;r)\com &\nn\\ 
& r>k>0\com\q\wbar=w-\frac{k}{r}v\pr &
                                                      \label{a.6y}\\
& \mbox{Propagator Rule 4}\com\q\q
u=0,\ v=0,\ l=0,\ \mbox{in Eq.(\ref{a.5})}   & \nn\\
& G_0(w;r-k)G_0(w;k)=\frac{1}{(4\pi r)^{n/2}}G_0(w;\frac{(r-k)k}{r})
\com\q r>k>0\pr  &
                                                      \label{a.6z}
\end{eqnarray}
%%%%%%%%%%%%%%%%%%%%%%%%%%%%%%
The above three relations are geometrically represented as 
in Fig.13-15.
%%%%%%%%%%%%%%%%%%%%%%%  Fig.13  %%%%%%%%%%%%%%%%%%%%%%%%%%%%%%%%%%%%
\begin{figure}
\centerline{\epsfysize=6cm\epsfbox{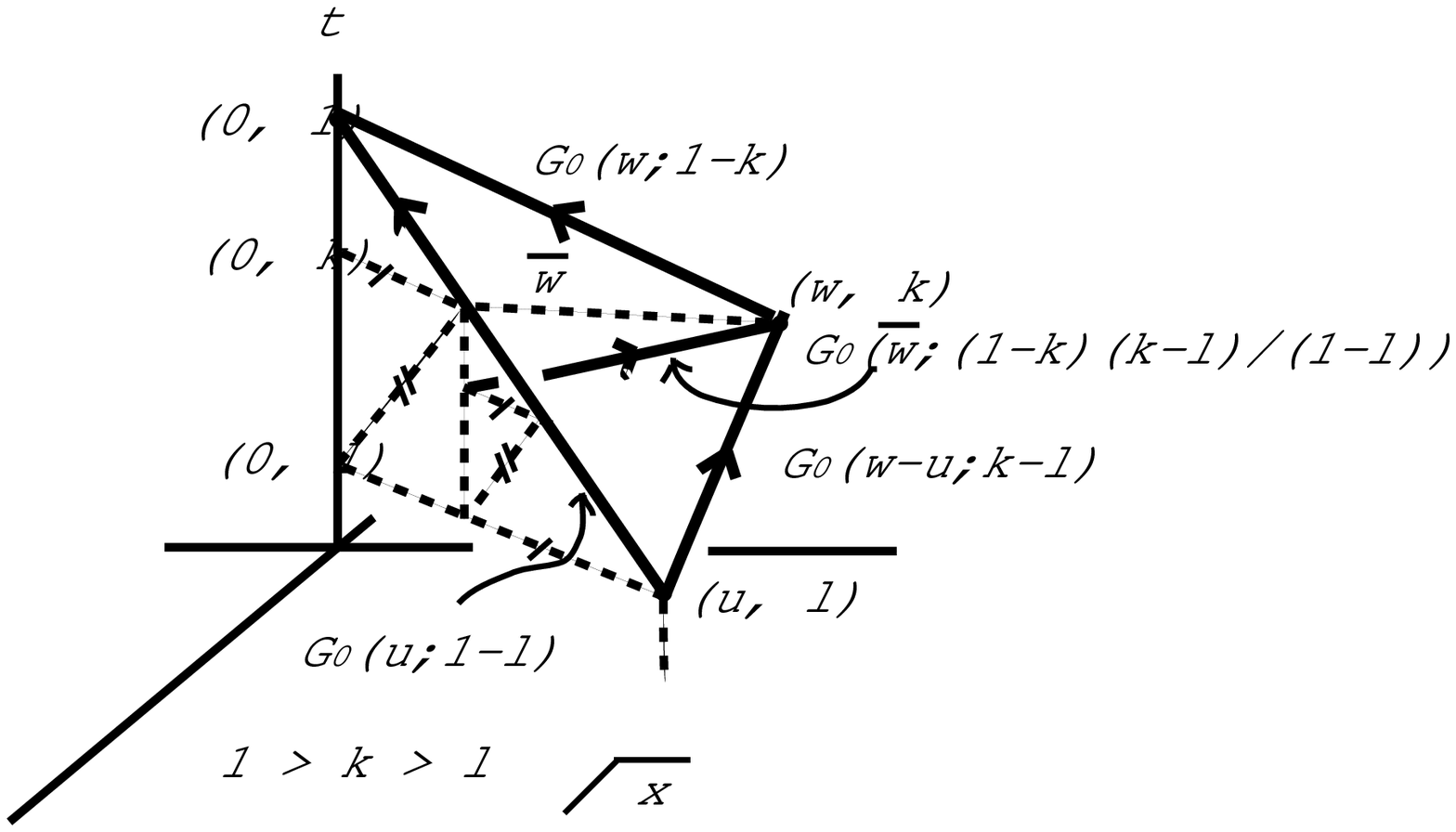}}
%\begin{center}\psbox[height=6cm]{prop2}\end{center}
   \begin{center}
Fig.13\ 
Propagator Rule 2, Eq.(\ref{a.6x}).
%***prop2.eps
   \end{center}
\end{figure}
%%%%%%%%%%%%%%%%%%%%%%%%%%%%%%%%%%%%%%%%%%%%%%%%%%%%%%%%%%%%%%%%%%%%%
%%%%%%%%%%%%%%%%%%%%%%%  Fig.14  %%%%%%%%%%%%%%%%%%%%%%%%%%%%%%%%%%%%
\begin{figure}
\centerline{\epsfysize=6cm\epsfbox{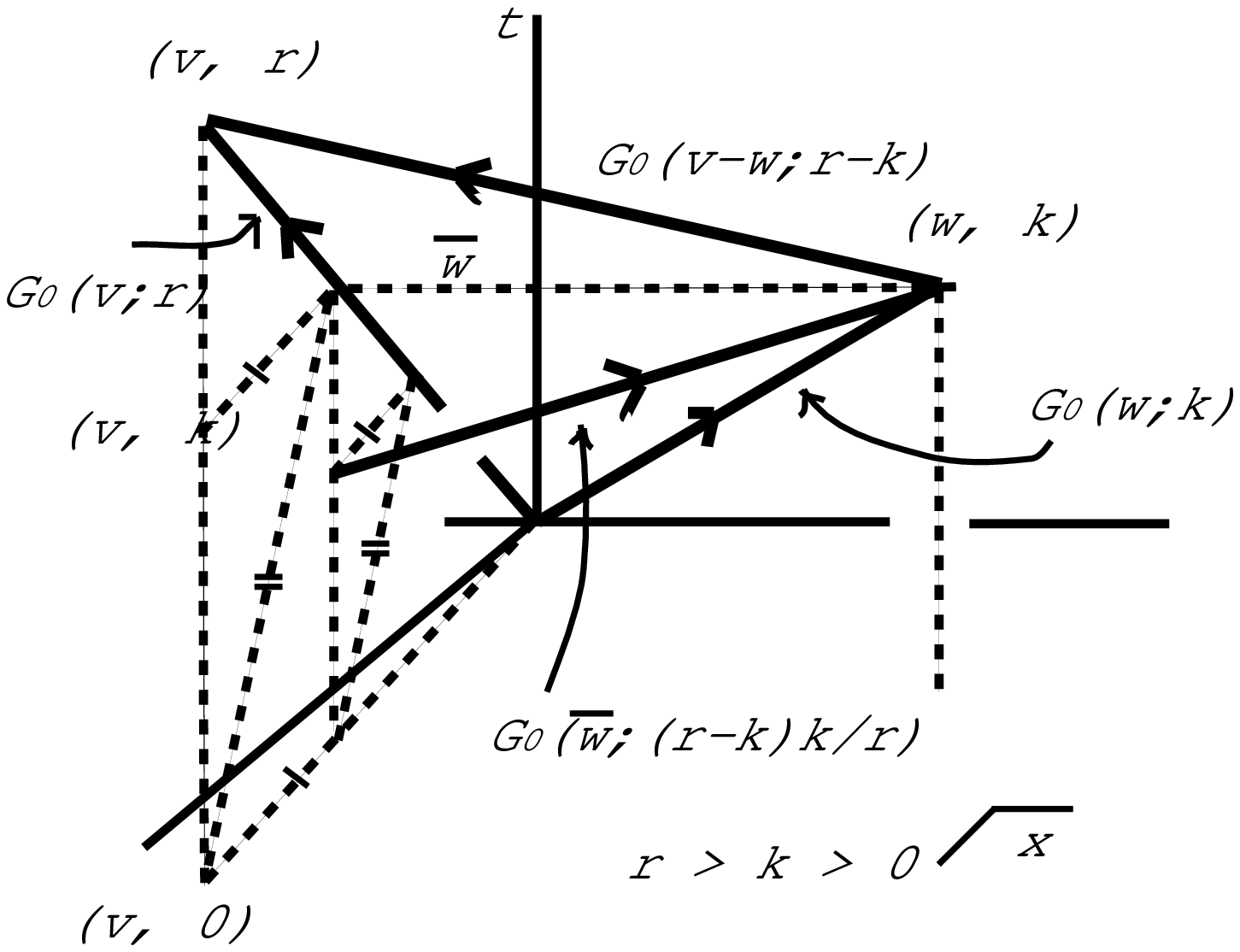}}
%\begin{center}\psbox[height=6cm]{prop3}\end{center}
   \begin{center}
Fig.14\ 
Propagator Rule 3, Eq.(\ref{a.6y}).
%***prop3.eps
   \end{center}
\end{figure}
%%%%%%%%%%%%%%%%%%%%%%%%%%%%%%%%%%%%%%%%%%%%%%%%%%%%%%%%%%%%%%%%%%%%%
%%%%%%%%%%%%%%%%%%%%%%%  Fig.15  %%%%%%%%%%%%%%%%%%%%%%%%%%%%%%%%%%%%
\begin{figure}
\centerline{\epsfysize=6cm\epsfbox{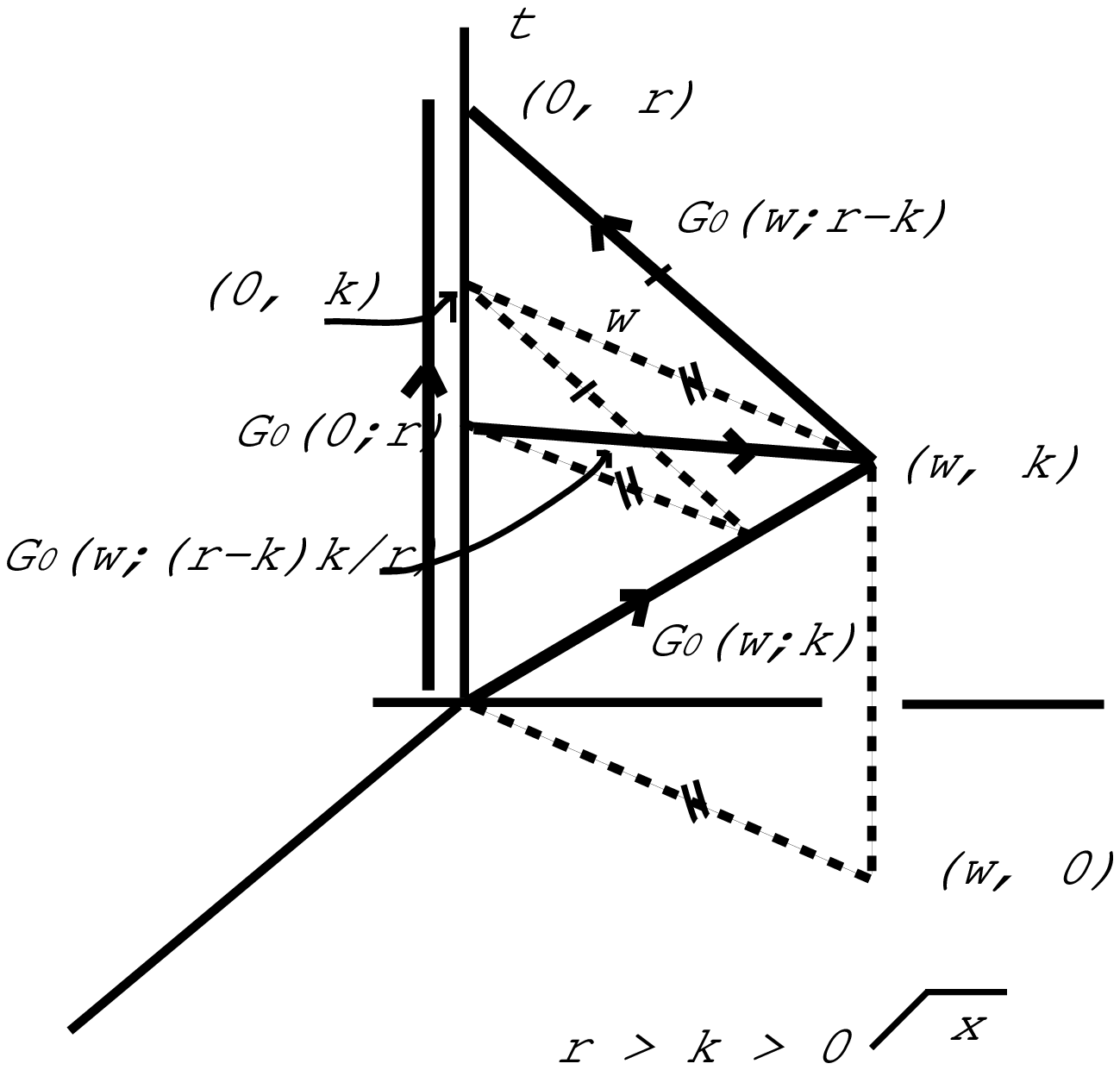}}
%\begin{center}\psbox[height=6cm]{prop4}\end{center}
   \begin{center}
Fig.15\ 
Propagator Rule 4, Eq.(\ref{a.6z}).
%***prop4.eps
   \end{center}
\end{figure}
%%%%%%%%%%%%%%%%%%%%%%%%%%%%%%%%%%%%%%%%%%%%%%%%%%%%%%%%%%%%%%%%%%%%%

%%%%%%%%%%%%%%%%%%%%%%%%%%  App. B  %%%%%%%%%%%%%%%%%%%%%%%%%%%%%%%
%%%%%%%%%%%%%%%%%%%%%%%%%%%%%%%%%%%%%%%%%%%%%%%%%%%%%%%%%%%%%%%%%%%
\begin{flushleft}
{\Large\bf Appendix B.\ Supplementary Calculation of Sec.5}
\end{flushleft}
In Sec.5 of the text, we have evaluated $G(x,x;t)|_{t^0}$ 
focusing on 4 special
$(\pl\pl h)^3$-terms(graphs) in order to fix 4 coefficients
of the main Weyl anomaly terms given in (\ref{inv.1}). 
Among the four graphs, Graph3 and 67 only are explained
for simplicity. We evaluate the remaining graphs, Graph1 and
2 to supplement Sec.5. 

From (\ref{g3}),
%****(B2.1)%%%%%%%%%%%%%%%%%%%%  
\begin{eqnarray}
G_3(x,x;t)|_{t^0}
=\frac{1}{(4\pi)^{3}}
\int d^6{\bar \wbar}\int d^6\wbar \int d^6w
\int^1_0dr'' \int^{r''}_0dr' \int^{r'}_0dr                \nn\\
\times G_0({\bar \wbar};\frac{(1-r'')(r''-r')}{1-r'})
G_0(\wbar;\frac{(1-r')(r'-r)}{1-r})
G_0(w;(1-r)r)                                            \nn\\
 \times V_0(x,w'';w''-w',r''-r')V_0(x,w';w'-w,r'-r)
                                V_0(x,w;w,r)
+\mbox{irrel}\com\nn\\
w''= {\bar \wbar}+R_2\wbar+R_3w\com\q
w''-w'= {\bar \wbar}-S_2\wbar-T_1w\com     \nn\\
w'= \wbar +R_1w\com\q
w'-w= \wbar-S_1w\com
                                                    \label{B2.1}
\end{eqnarray}
%%%%%%%%%%%%%%%%%%%%%%%%%%%%%
where $R_i, S_i$ and $T_1$ are defined in (\ref{g3.1}).
Because Graph1 and 2 do not have tadpoles and\  
%%%%%%%%%%%%%%%%%%%  FIG  %%%%%%%%%%%
\epsfysize=5mm\epsfbox{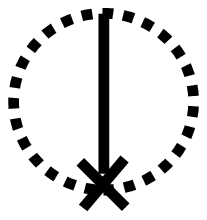}
%\psbox[height=5mm]{B2a}
\ subgraph,
we can reduce $V_0$ of (\ref{C2.2}) with (\ref{GraphExp.1}) 
to the following relevant form.
%****(B2.2)%%%%%%%%%%%%%%%%%%%%  
\begin{eqnarray}
V_0(x,v;w,r)=-\half\times\frac{1}{4r^2}\times
           \begin{array}{c}
\mbox{
%%%%%%%%%%%%%  FIG  %%%%%%%%%% 
\epsfysize=2cm\epsfbox{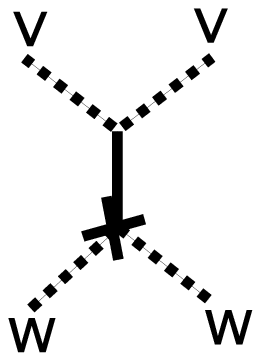}
%\psbox[height=2cm]{B1d}
      }
         \end{array}                        
		 +\frac{1}{2r}
         \begin{array}{c}
\mbox{ 
%%%%%%%%%%%%%  FIG  %%%%%%%%%%
\epsfysize=2cm\epsfbox{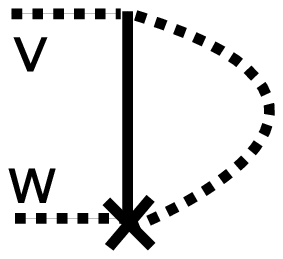}
%\psbox[height=2cm]{B2c}
      }
         \end{array}
		 +\mbox{irrel. terms}\pr                       
                                                   \label{B2.2}
\end{eqnarray}
%%%%%%%%%%%%%%%%%%%%%%%%%%%%%
The first $V_0$ in (\ref{B2.1}) is evaluated as
%****(B2.25, B2.3)%%%%%%%%%%%%%%%%%%%%  
\begin{eqnarray}
V_0(x,w'';w''-w',r''-r')=-\frac{1}{8(r''-r')^2}\times
           \begin{array}{c}
\mbox{
%%%%%%%%%%%%%%  FIG  %%%%%%%% 
\epsfysize=2cm\epsfbox{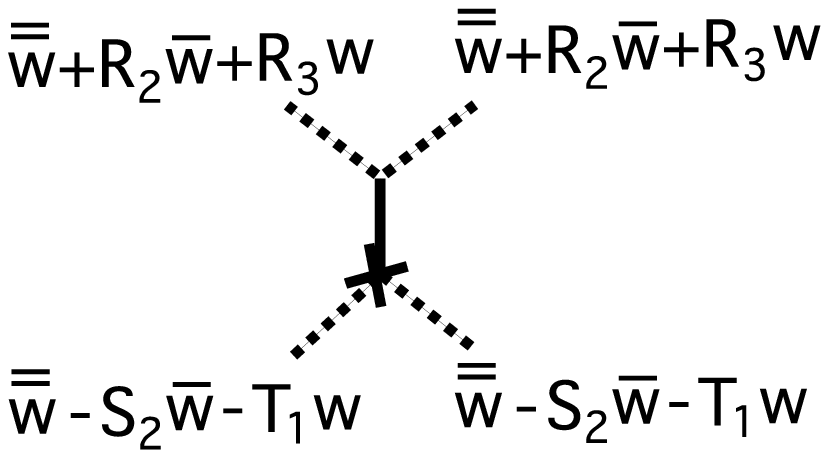}
%\psbox[height=2cm]{B2d}
      }
         \end{array}                     \nn\\                       
		 +
		 \frac{1}{2(r''-r')}\times
         \begin{array}{c}
\mbox{
%%%%%%%%%%%%%%% FIG %%%%%%%%% 
\epsfysize=2cm\epsfbox{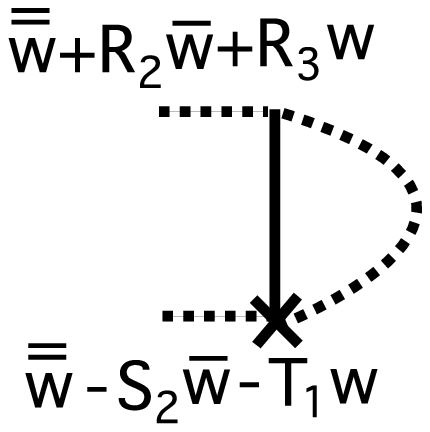}
%\psbox[height=2cm]{B2g}
      }
         \end{array}
		 +\mbox{irrel. terms}\com                    \label{B2.25}
                                                                               \\               
\mbox{second term of (\ref{B2.25})}=
         \frac{1\times (-1)}{2(r''-r')}\times
         \begin{array}{c}
\mbox{
%%%%%%%%%%%%%%  FIG %%%%%%%% 
\epsfysize=2cm\epsfbox{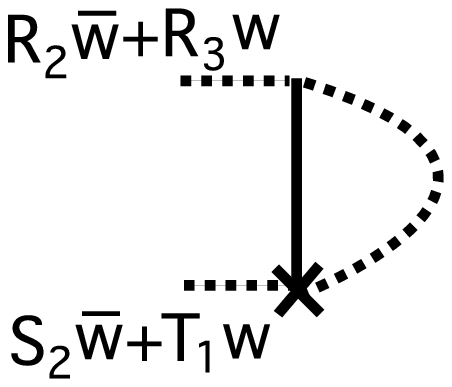}
%\psbox[height=2cm]{B2h}
      }
         \end{array}
		 +\mbox{irrel. terms}\com                               \nn\\
\mbox{first term of (\ref{B2.25})}=
        -\frac{1}{8(r''-r')^2}\left(
		2\times 2\times (-1)\times           
        \begin{array}{c}
\mbox{
%%%%%%%%%%%%%%% FIG %%%%%%%% 
\epsfysize=2cm\epsfbox{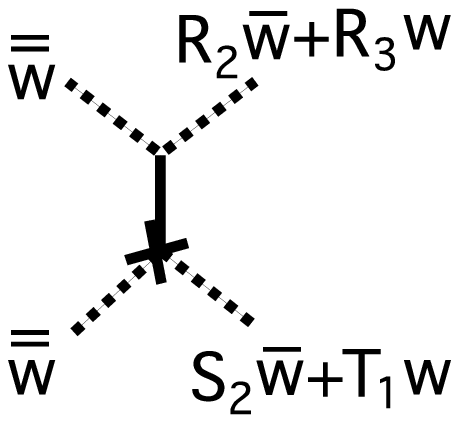}
%\psbox[height=2cm]{B2e}
      }
         \end{array}                        
		 +
         \begin{array}{c}
\mbox{
%%%%%%%%%%%%%% FIG %%%%%%%%% 
\epsfysize=2cm\epsfbox{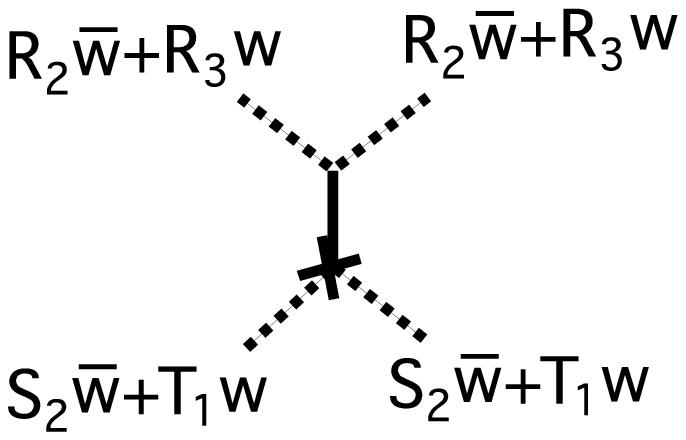}
%\psbox[height=2cm]{B2f}
      }
         \end{array}          \right)            \nn\\
		 +\mbox{irrel. terms}\pr
                                                    \label{B2.3}
\end{eqnarray}
%%%%%%%%%%%%%%%%%%%%%%%%%%%%%
The above $V_0$, with the ${\bar \wbar}$-integration, gives
%****(B2.4)%%%%%%%%%%%%%%%%%%%%  
\begin{eqnarray}
\int d^6{\bar \wbar}\, V_0(x,w'';w''-w',r''-r')=
        -\frac{1}{8(r''-r')^2}\times
        \begin{array}{c}
\mbox{
%%%%%%%%%%  FIG %%%%%%%%%%%% 
\epsfysize=2cm\epsfbox{B2f.eps}
%\psbox[height=2cm]{B2f}
      }
         \end{array}                        \nn\\
		 +
		 \left\{ \frac{-1}{2(r''-r')}+\frac{2\times 2}{8(r''-r')^2}
		 \times 2\frac{(1-r'')(r''-r')}{(1-r')}
		 \right\}
         \begin{array}{c}
\mbox{
%%%%%%%%%%  FIG  %%%%%%%%%% 
\epsfysize=2cm\epsfbox{B2h.eps}
%\psbox[height=2cm]{B2h}
      }
         \end{array}
		 +\mbox{irrel. terms}                    \nn\\
=       -\frac{(R_2)^2(S_2)^2}{8(r''-r')^2}\times
        \begin{array}{c}
\mbox{
%%%%%%%%%  FIG  %%%%%%%%%% 
\epsfysize=2cm\epsfbox{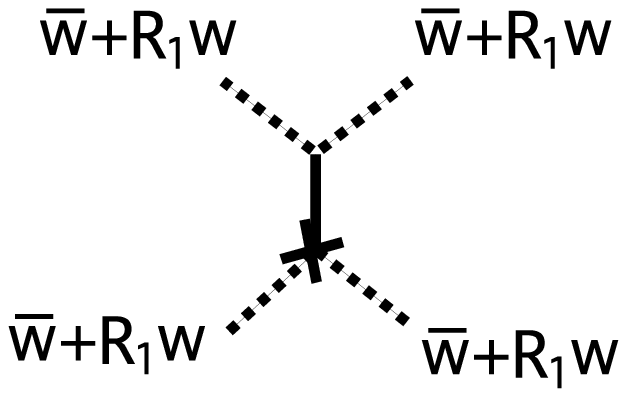}
%\psbox[height=2cm]{B2i}
      }
         \end{array}                        
		 +
		 \frac{1+r'-2r''}{2(r''-r')(1-r')}R_2S_2\times
         \begin{array}{c}
\mbox{ 
%%%%%%%%%  FIG  %%%%%%%%
\epsfysize=2cm\epsfbox{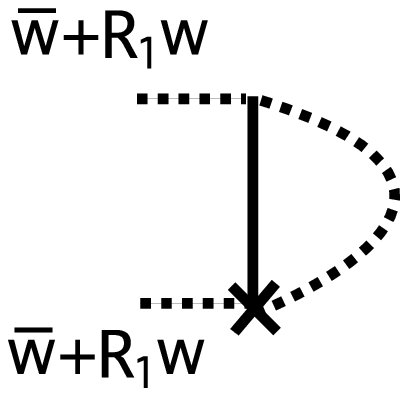}
%\psbox[height=2cm]{B2j}
      }
         \end{array}                     \nn\\
		 +\mbox{irrel. terms} \com
                                                    \label{B2.4}
\end{eqnarray}
%%%%%%%%%%%%%%%%%%%%%%%%%%%%%
where ${\bar \wbar}$ is integrated out and the relations
$R_3/R_2=T_1/S_2=R_1$ are used.
Therefore the $V_0V_0V_0$-part of (\ref{B2.1}), with the
${\bar \wbar}$-integration, can be written as
%****(B2.5)%%%%%%%%%%%%%%%%%%%%  
\begin{eqnarray}
\int d^6{\bar \wbar}\, 
V_0(x,w'';w''-w',r''-r')V_0(x,w';w'-w,r'-r)V_0(x,w;w,r)=\nn\\
\left\{
        -\frac{(1-r'')^2}{8(1-r')^4}\times
        \begin{array}{c}
\mbox{
%%%%%%%%%%%%  FIG %%%%%%%%% 
\epsfysize=2cm\epsfbox{B2i.eps}
%\psbox[height=2cm]{B2i}
      }
         \end{array}                        
		 +\frac{(1+r'-2r'')(1-r'')}{2(1-r')^3}
         \begin{array}{c}
\mbox{
%%%%%%%%%%%%  FIG  %%%%%%%% 
\epsfysize=2cm\epsfbox{B2j.eps}
%\psbox[height=2cm]{B2j}
      }
         \end{array}
\right\}                                                \nn\\
\times \left\{
        -\frac{1}{8(r'-r)^2}
        \begin{array}{c}
\mbox{
%%%%%%%%%%%%  FIG %%%%%%%%% 
\epsfysize=2cm\epsfbox{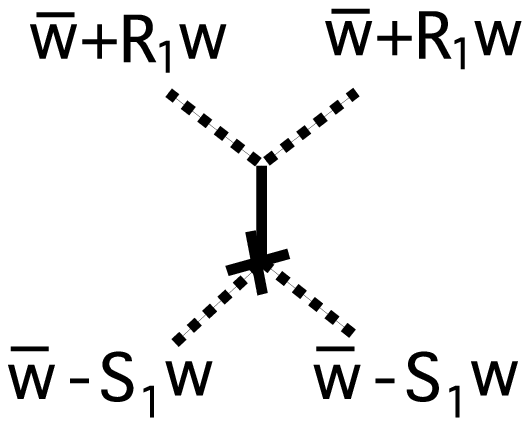}
%\psbox[height=2cm]{B2k}
      }
         \end{array}                        
		 +\frac{1}{2(r'-r)}
         \begin{array}{c}
\mbox{
%%%%%%%%%%%%  FIG  %%%%%%%%% 
\epsfysize=2cm\epsfbox{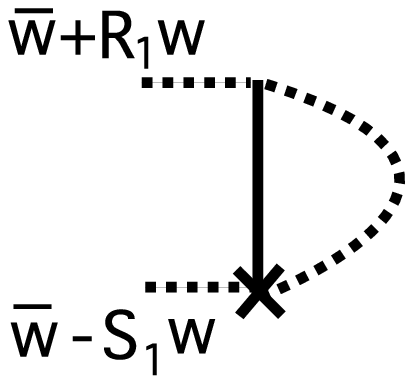}
%\psbox[height=2cm]{B2l}
      }
         \end{array}
\right\}                                                \nn\\
\times \left\{
        -\frac{1}{8r^2}\times
        \begin{array}{c}
\mbox{ 
%%%%%%%%%%%%  FIG  %%%%%%%%%
\epsfysize=2cm\epsfbox{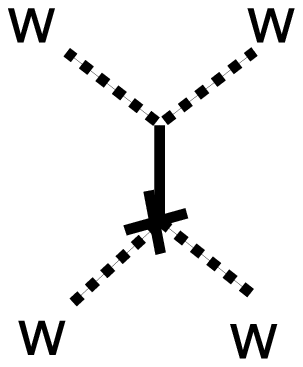}
%\psbox[height=2cm]{B2m}
      }
         \end{array}                        
		 +\frac{1}{2r}\times
         \begin{array}{c}
\mbox{
%%%%%%%%%%%  FIG  %%%%%%%% 
\epsfysize=2cm\epsfbox{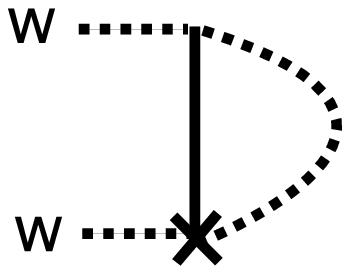}
%\psbox[height=2cm]{B2n}
      }
         \end{array}
\right\}                                                \nn\\
+\mbox{irrel}\pr
                                                    \label{B2.5}
\end{eqnarray}
%%%%%%%%%%%%%%%%%%%%%%%%%%%%%
Now we can further evaluate (\ref{B2.1}) by picking up relevant
powers of $\wbar$ in (\ref{B2.5}) in  the same way
as Sec.5.  After integration of the remaining coordinates $(\wbar,w)$
and the parameters $(r,r',r'')$, the final result is
%****(B2.6)%%%%%%%%%%%%%%%%%%%%  
\begin{eqnarray}
G_3(x,x;t)|_{t^0}
=\frac{1}{(4\pi)^{3}}
\{ \frac{1}{36\times 630}\mbox{(Graph1)} 
+\frac{1}{36\times 35}\mbox{(Graph2)} 
\}+\mbox{other terms}\pr
                                                    \label{B2.6}
\end{eqnarray}
%%%%%%%%%%%%%%%%%%%%%%%%%%%%%

%%%%%%%%%%%%%%%%%%%%%%%%%%  App. C  %%%%%%%%%%%%%%%%%%%%%%%%%%%%%%%
%%%%%%%%%%%%%%%%%%%%%%%%%%%%%%%%%%%%%%%%%%%%%%%%%%%%%%%%%%%%%%%%%%%
\begin{flushleft}
{\Large\bf Appendix C.\ Weak Field Expansion of 6 Dim Scalar-Gravity Theory}
\end{flushleft}

In the text, we have taken 6 dim scalar-gravity theory (\ref{f.1}) as an higher-dimensional
model.  The expressions for the general theories are expressed in terms of
$W_\mn, N_\la$ and $M$ (\ref{f.10}). In this appendix we graphically express 
those general expressions for the case of the explicit model.

\begin{flushleft}
{\Large C.1\ Graphs of W,N and M}
\end{flushleft}

For the 6 dim scalar-gravity theory, $W_\mn, N_\la$ and $M$ are given by (\ref{f.10b}).
Their weak-field expansions ($g_\mn=\del_\mn+h_\mn$) up to $O(h^3)$
are given in Fig.16 for 
$W_\mn$ and $N_\la$, and in Fig.17-21 for $M$. Especially they are classified
by the number of closed suffix-loops.

%%%%%%%%%%%%%%%%%%%%%%%  Fig.16  %%%%%%%%%%%%%%%%%%%%%%%%%%%%%%%%%%%%
\begin{figure}
\centerline{\epsfysize=10cm\epsfbox{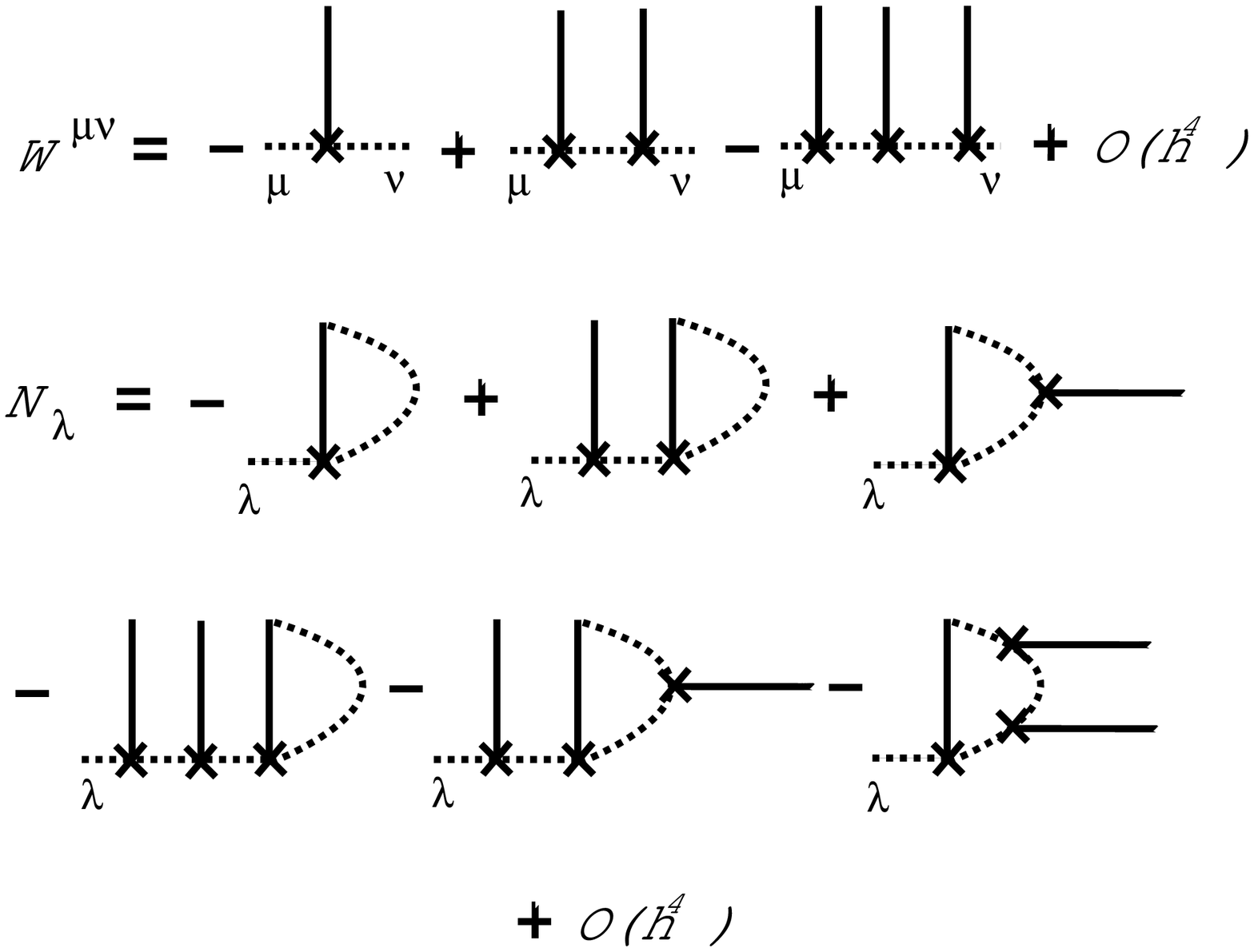}}
%\begin{center}\psbox[height=10cm]{WaN}\end{center}
   \begin{center}
Fig.16\ 
$W^\mn$ and $N_\la$.
%***WaN.eps
   \end{center}
\end{figure}
%%%%%%%%%%%%%%%%%%%%%%%%%%%%%%%%%%%%%%%%%%%%%%%%%%%%%%%%%%%%%%%%%%%%%
%%%%%%%%%%%%%%%%%%%%%%%  Fig.17  %%%%%%%%%%%%%%%%%%%%%%%%%%%%%%%%%%%%
\begin{figure}
\centerline{\epsfysize=5cm\epsfbox{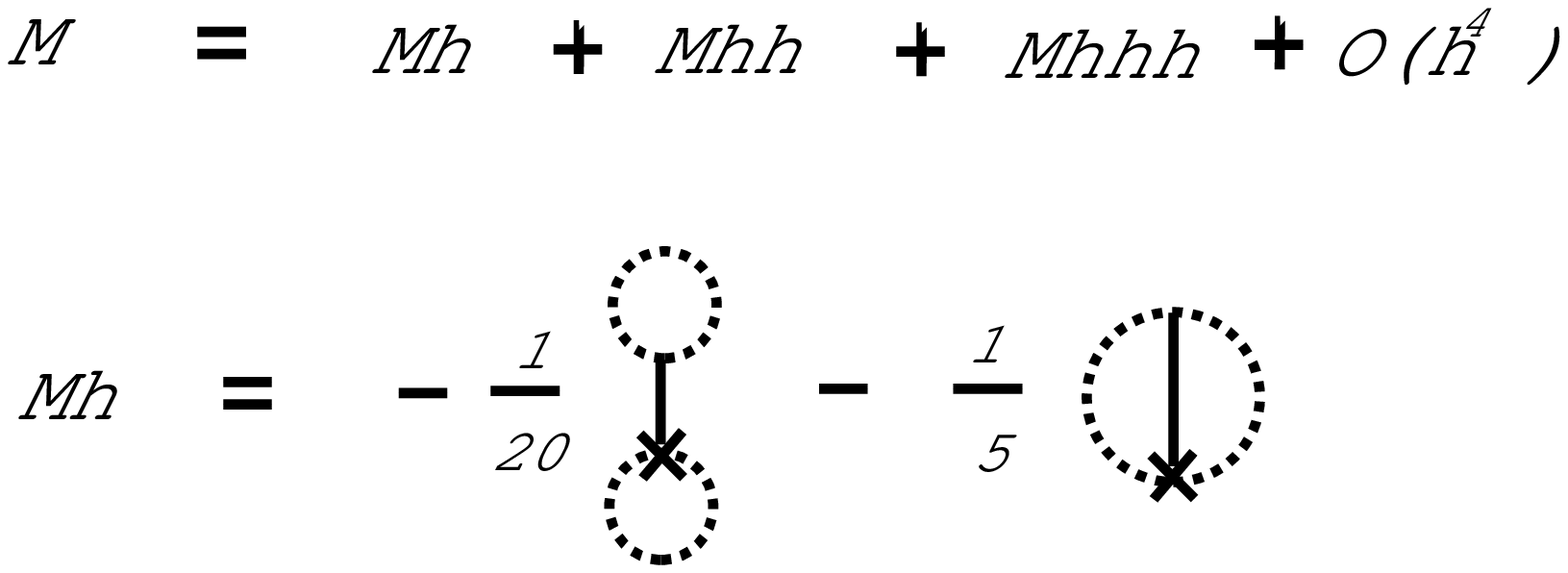}}
%\begin{center}\psbox[height=5cm]{Mh}\end{center}
   \begin{center}
Fig.17\ 
$M$, Order of $h$.
%***Mh.eps
   \end{center}
\end{figure}
%%%%%%%%%%%%%%%%%%%%%%%%%%%%%%%%%%%%%%%%%%%%%%%%%%%%%%%%%%%%%%%%%%%%%
%%%%%%%%%%%%%%%%%%%%%%%  Fig.18  %%%%%%%%%%%%%%%%%%%%%%%%%%%%%%%%%%%%
\begin{figure}
\centerline{\epsfysize=10cm\epsfbox{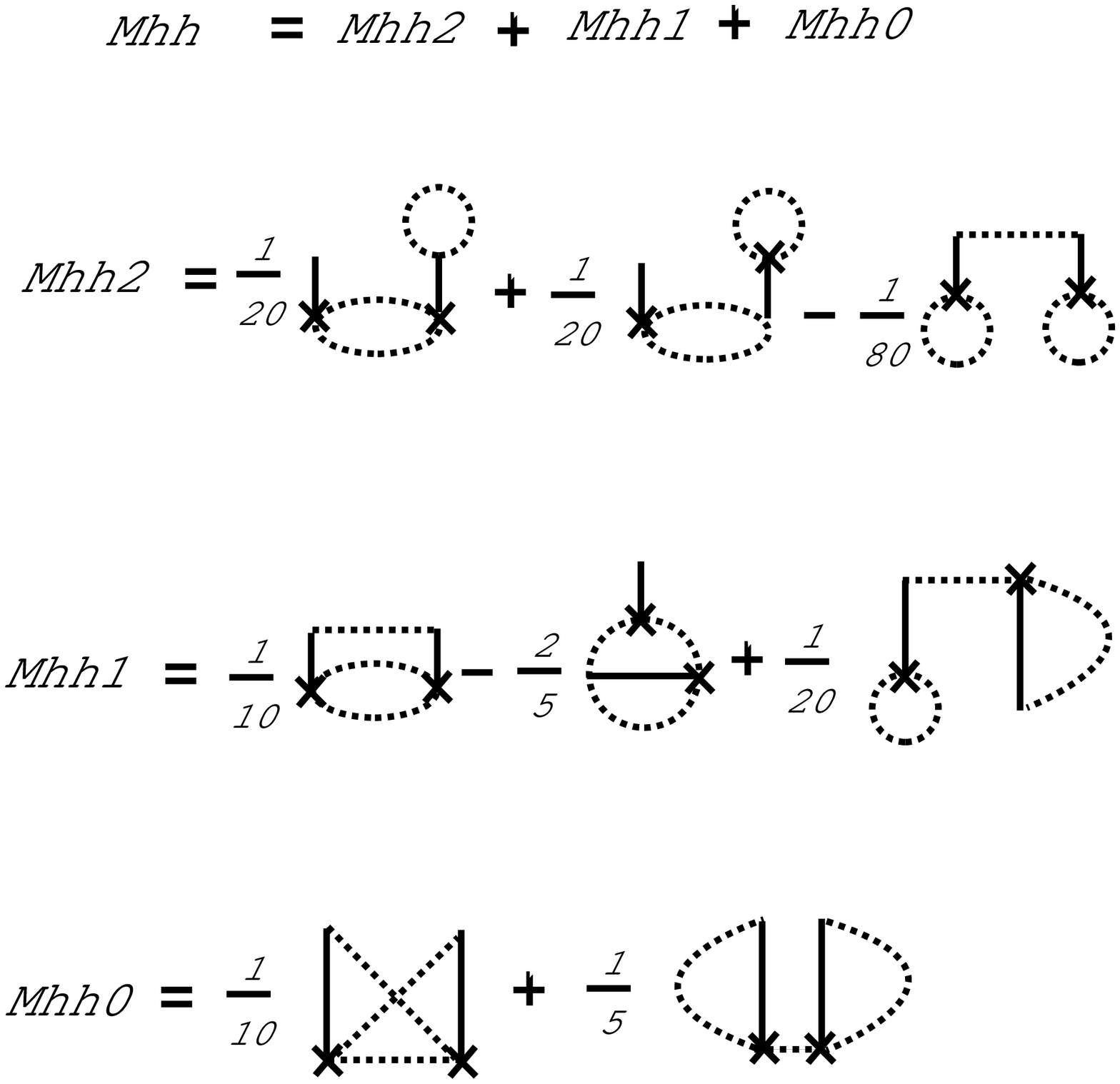}}
%\begin{center}\psbox[height=10cm]{Mhh}\end{center}
   \begin{center}
Fig.18\ 
$M$, Order of $h^2$.
%***Mhh.eps
   \end{center}
\end{figure}
%%%%%%%%%%%%%%%%%%%%%%%%%%%%%%%%%%%%%%%%%%%%%%%%%%%%%%%%%%%%%%%%%%%%%
%%%%%%%%%%%%%%%%%%%%%%%  Fig.19  %%%%%%%%%%%%%%%%%%%%%%%%%%%%%%%%%%%%
\begin{figure}
\centerline{\epsfysize=10cm\epsfbox{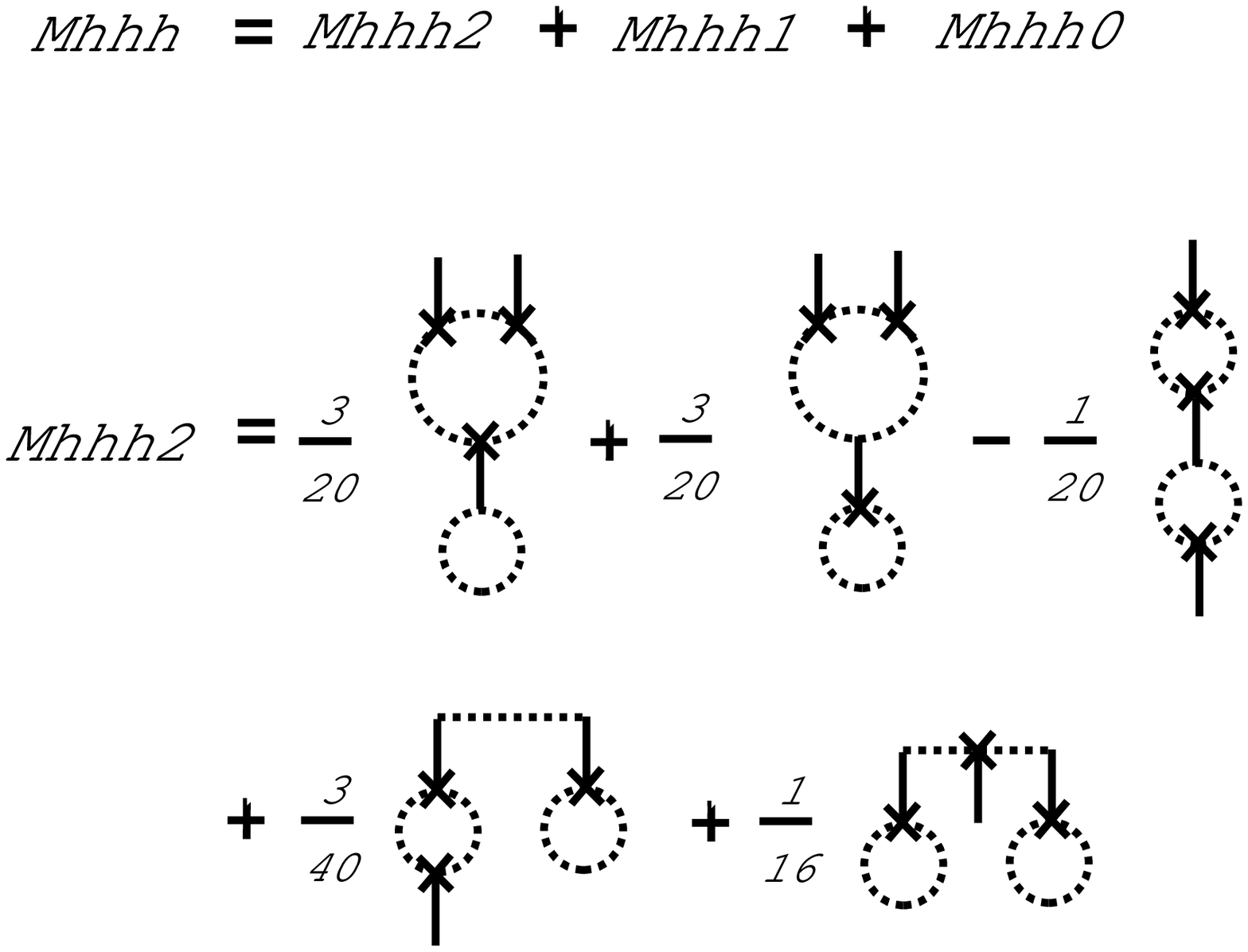}}
%\begin{center}\psbox[height=10cm]{Mhhh2}\end{center}
   \begin{center}
Fig.19\ 
$M$, Order of $h^3$ and loop no=2.
%***Mhhh2.eps
   \end{center}
\end{figure}
%%%%%%%%%%%%%%%%%%%%%%%%%%%%%%%%%%%%%%%%%%%%%%%%%%%%%%%%%%%%%%%%%%%%%
%%%%%%%%%%%%%%%%%%%%%%%  Fig.20  %%%%%%%%%%%%%%%%%%%%%%%%%%%%%%%%%%%%
\begin{figure}
\centerline{\epsfysize=10cm\epsfbox{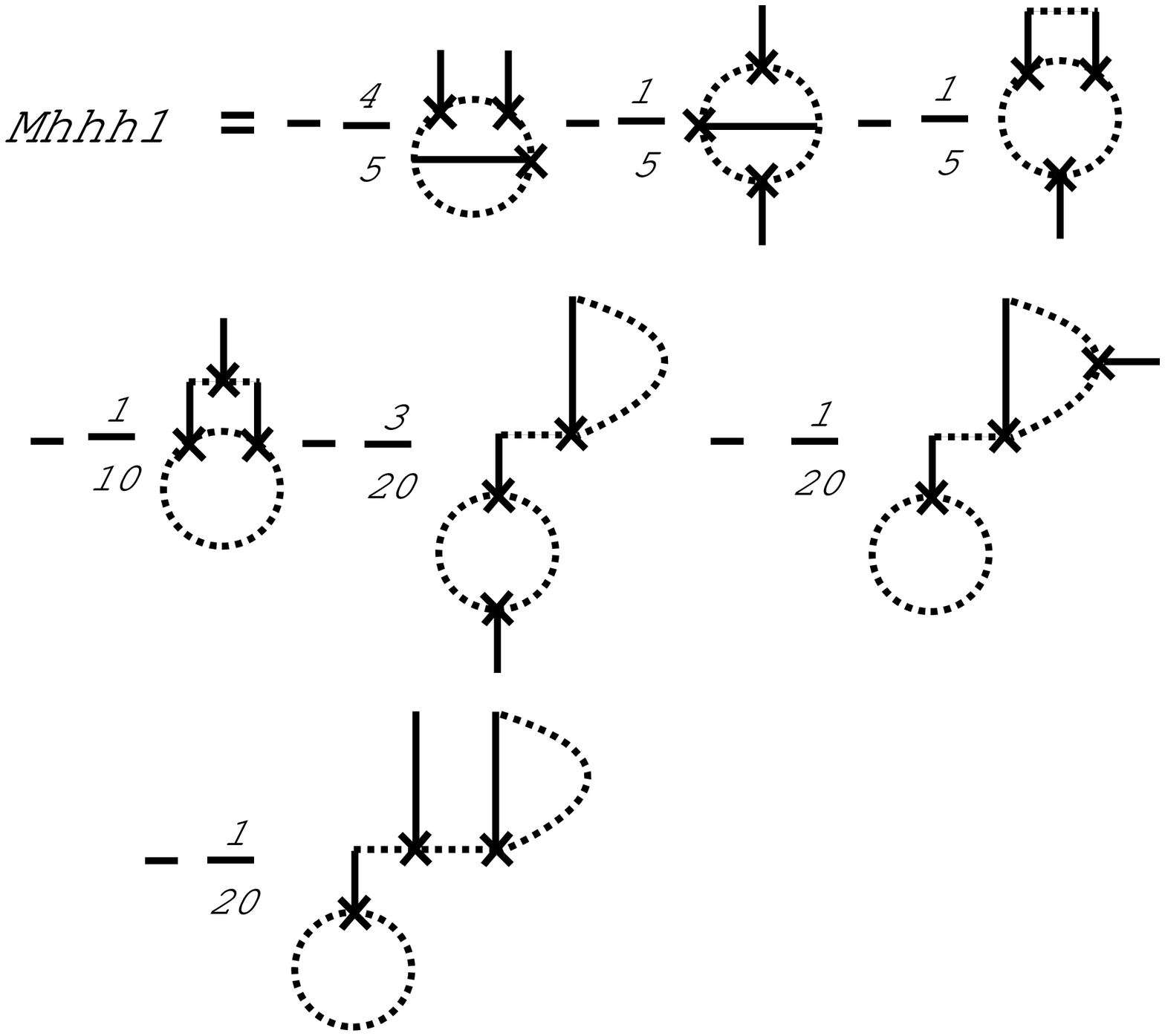}}
%\begin{center}\psbox[height=10cm]{Mhhh1}\end{center}
   \begin{center}
Fig.20\ 
$M$, Order of $h^3$ and loop no=1.
%***Mhhh1.eps
   \end{center}
\end{figure}
%%%%%%%%%%%%%%%%%%%%%%%%%%%%%%%%%%%%%%%%%%%%%%%%%%%%%%%%%%%%%%%%%%%%%
%%%%%%%%%%%%%%%%%%%%%%%  Fig.21  %%%%%%%%%%%%%%%%%%%%%%%%%%%%%%%%%%%%
\begin{figure}
\centerline{\epsfysize=4cm\epsfbox{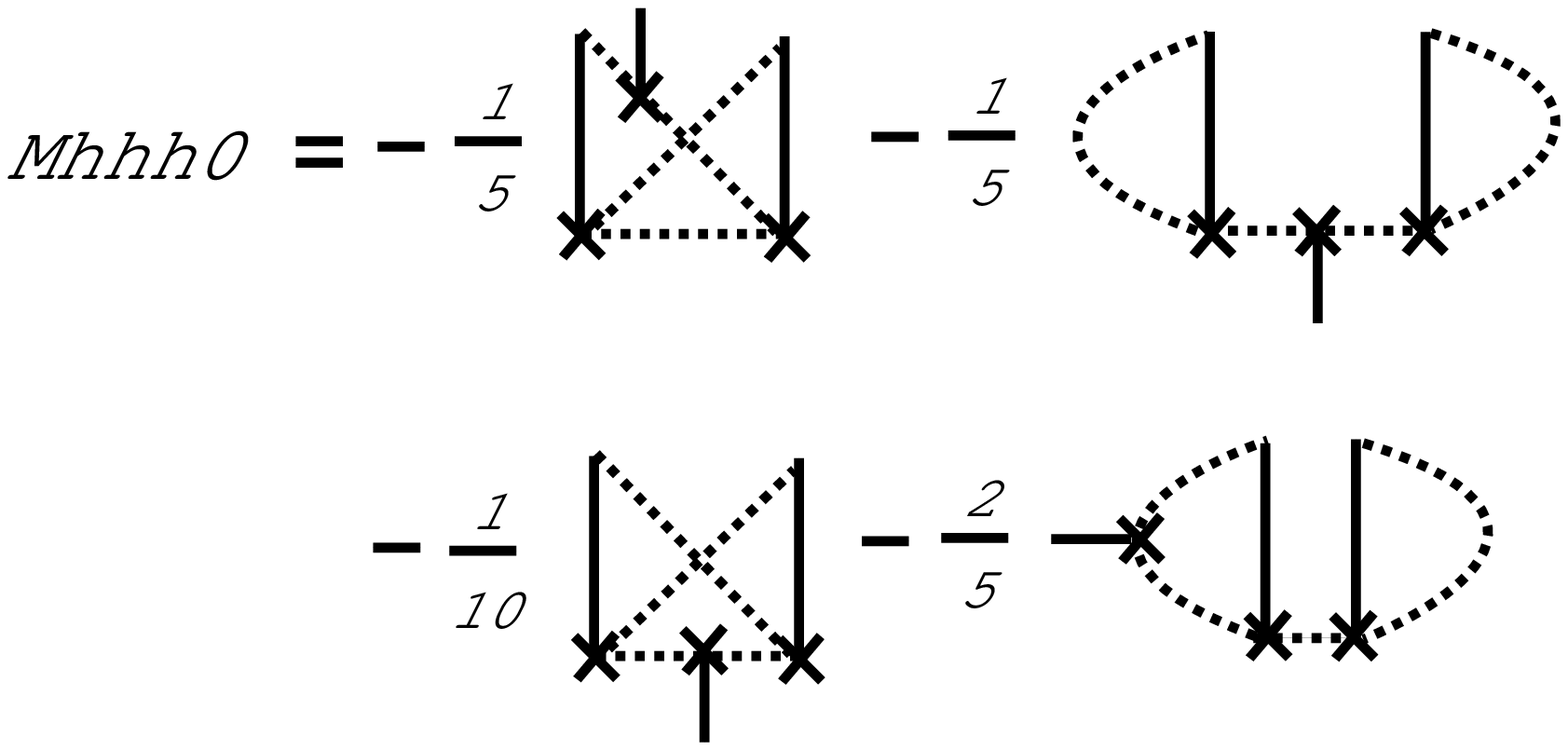}}
%\begin{center}\psbox[height=4cm]{Mhhh0}\end{center}
   \begin{center}
Fig.21\ 
$M$, Order of $h^3$ and loop no=0.
%***Mhhh0.eps
   \end{center}
\end{figure}
%%%%%%%%%%%%%%%%%%%%%%%%%%%%%%%%%%%%%%%%%%%%%%%%%%%%%%%%%%%%%%%%%%%%%

\newpage
%%%%%%%%%%%%%%%%%%%%%%%%%%  App. C.2  %%%%%%%%%%%%%%%%%%%%%%%%%%%%%%%
%%%%%%%%%%%%%%%%%%%%%%%%%%%%%%%%%%%%%%%%%%%%%%%%%%%%%%%%%%%%%%%%%%%
\begin{flushleft}
{\Large C.2\ Taylor Expansion of W,N and M}
\end{flushleft}

In Sec.4 of the text, we have Taylor-expanded $W_\mn, N_\m$ and
$M$, which are the background-functional appearing in the differential
operator $\Dvec$. 
%****(C2.1)%%%%%%%%%%%%%%%%%%%%  
\begin{eqnarray}
W_\mn(x+v)=W_\mn(x)+\pl_{\al_1}W_\mn\cdot v^{\al_1}
+\frac{1}{2\!}\pl_{\al_1}\pl_{\al_2}W_\mn\cdot v^{\al_1}v^{\al_2}+\cdots
                                             \com\nn\\
N_\m(x+v)=N_\m(x)+\pl_{\al_1}N_\m\cdot v^{\al_1}
+\frac{1}{2\!}\pl_{\al_1}\pl_{\al_2}N_\m\cdot v^{\al_1}v^{\al_2}+\cdots
                                             \com\nn\\
M(x+v)=M(x)+\pl_{\al_1}M\cdot v^{\al_1}
+\frac{1}{2\!}\pl_{\al_1}\pl_{\al_2}M\cdot v^{\al_1}v^{\al_2}+\cdots
                                             \pr
                                                    \label{C2.1}
\end{eqnarray}
%%%%%%%%%%%%%%%%%%%%%%%%%%%%%
We focus on those terms which have only
$\pl\pl h$-type ones. They are sufficient to determine the
Weyl anomaly, and are used in the text.
%%%%%%%%%%%%%%%%%%%%%%%%%%%%%%%%%%%%%%
\begin{flushleft}
(i)\ Expansion Terms appearing in $V_0$ of (\ref{exp.2})
\end{flushleft}

%****(C2.2)%%%%%%%%%%%%%%%%%%%%  
\begin{eqnarray}
V_0(x,v;w,r) &=\frac{1}{2!}\pl_{\al_1}\pl_{\al_2}W_\mn
\cdot v^{\al_1}v^{\al_2}
(-\frac{\del_\mn}{2r}+\frac{w_\m w_\n}{4r^2})
+\pl_{\al_1}N_\m\cdot v^{\al_1}(-\frac{w_\m}{2r})+M   \pr &
                                                    \label{C2.2}
\end{eqnarray}
%%%%%%%%%%%%%%%%%%%%%%%%%%%%%

%****(GraphExp.1), figC2a-c.eps   %%%%%%%%%%%%%%%%%%%%  
\begin{eqnarray}
\frac{1}{2!}\pl_{\al_1}\pl_{\al_2}W_\mn\cdot v^{\al_1}v^{\al_2}=
         \begin{array}{c}
\mbox{
%%%%%%%%%%%%%%  FIG  %%%%%%%%%%%%% 
\epsfysize=2cm\epsfbox{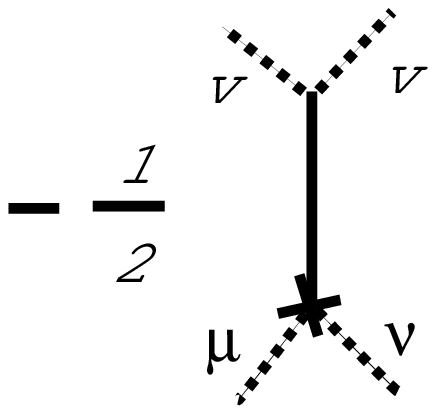}
%\psbox[height=2cm]{figC2a}
      }
         \end{array}
+\mbox{irrel. terms}\com                       \nn\\
\pl_{\al_1}N_\la\cdot v^{\al_1}=
         \begin{array}{c}
\mbox{
%%%%%%%%%%%%%  FIG  %%%%%%%%%%%%% 
\epsfysize=2cm\epsfbox{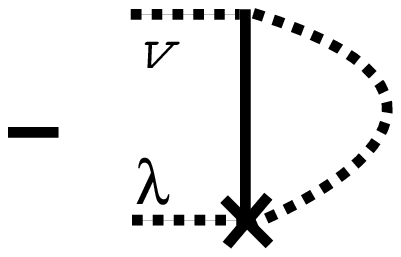}
%\psbox[height=2cm]{figC2b}
      }
         \end{array}
+\mbox{irrel. terms}\com                       \nn\\
M=
         \begin{array}{c}
\mbox{
%%%%%%%%%%%%%%%  FIG  %%%%%%%%%% 
\epsfysize=2cm\epsfbox{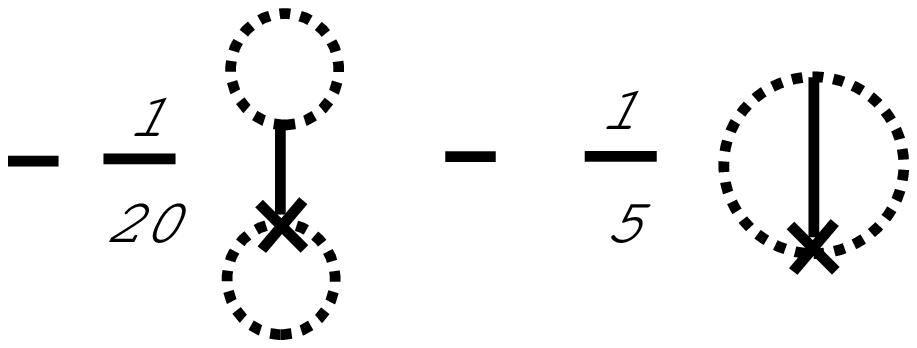}
%\psbox[height=2cm]{figC2c}
      }
         \end{array}
+\mbox{irrel. terms}\pr
                                                    \label{GraphExp.1}
\end{eqnarray}
%%%%%%%%%%%%%%%%%%%%%%%%%%%%%

%%%%%%%%%%%%%%%%%%%%%%%%%%%%%%%%%%%%%%
\begin{flushleft}
(ii)\ Expansion Terms appearing in $V_1$ of (\ref{exp.2})
\end{flushleft}

%****(C2.3)%%%%%%%%%%%%%%%%%%%%  
\begin{eqnarray}
V_1(x,v;w,r) &=\frac{1}{4!}\pl_{\al_1}\pl_{\al_2}\cdots\pl_{\al_4}W_\mn
\cdot v^{\al_1}v^{\al_2}\cdots v^{\al_4}
(-\frac{\del_\mn}{2r}+\frac{w_\m w_\n}{4r^2})              &\nn\\
 & +\frac{1}{3!}\pl_{\al_1}\pl_{\al_2}\pl_{\al_3}N_\m
\cdot v^{\al_1}v^{\al_2}v^{\al_3}(-\frac{w_\m}{2r})+
\frac{1}{2!}\pl_{\al_1}\pl_{\al_2}M
\cdot v^{\al_1}v^{\al_2}\pr                               &
                                                    \label{C2.3}
\end{eqnarray}
%%%%%%%%%%%%%%%%%%%%%%%%%%%%%

%****(GraphExp.2), figC2d-f.eps   %%%%%%%%%%%%%%%%%%%%  
\begin{eqnarray}
\frac{1}{4!}\pl_{\al_1}\pl_{\al_2}\cdots\pl_{\al_4}W_\mn
\cdot v^{\al_1}v^{\al_2}\cdots v^{\al_4}=
         \begin{array}{c}
\mbox{
%%%%%%%%%%%%%%  FIG  %%%%%%%%%% 
\epsfysize=3cm\epsfbox{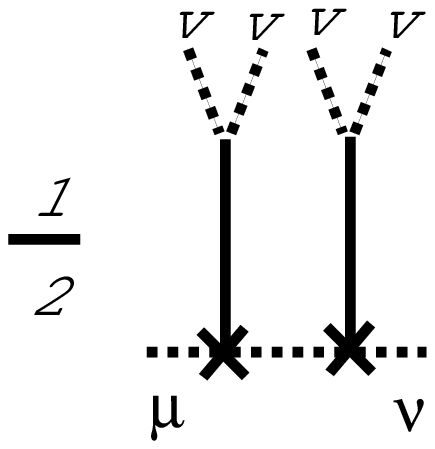}
%\psbox[height=3cm]{figC2d}
     }
         \end{array}
+\mbox{irrel. terms}\com                       \nn\\
\frac{1}{3!}\pl_{\al_1}\pl_{\al_2}\pl_{\al_3}N_\la
\cdot v^{\al_1}v^{\al_2}v^{\al_3}=
         \begin{array}{c}
\mbox{
%%%%%%%%%%%%%%%  FIG  %%%%%%%%%%% 
\epsfysize=3cm\epsfbox{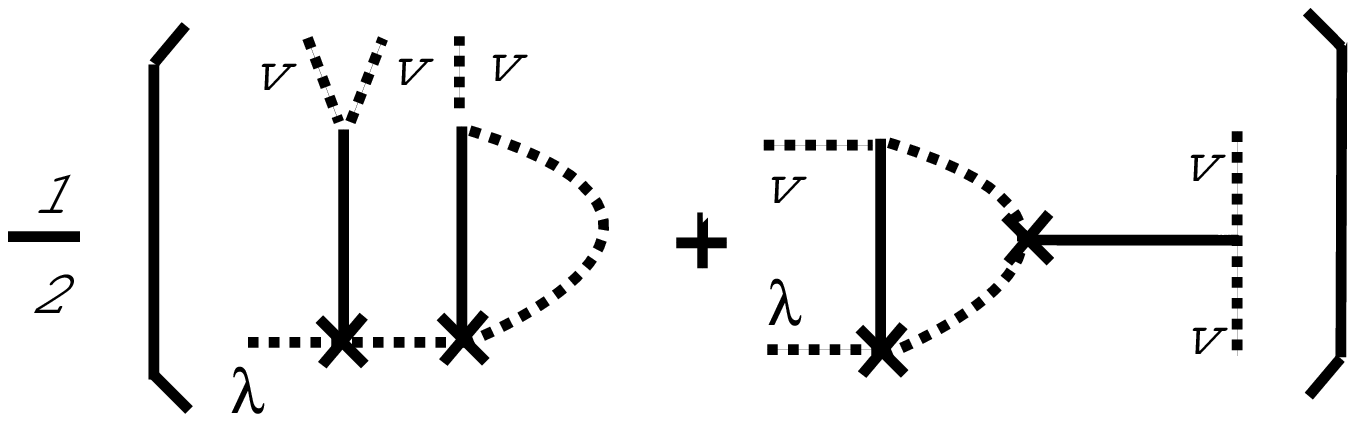}
%\psbox[height=3cm]{figC2e}
      }
         \end{array}                           \nn\\
+\mbox{irrel. terms}\com                       \nn\\
\frac{1}{2!}\pl_{\al_1}\pl_{\al_2}M
\cdot v^{\al_1}v^{\al_2}=                      \nn\\
         \begin{array}[t]{c}
\mbox{
%%%%%%%%%%%%%%%  FIG  %%%%%%%%%%% 
\epsfysize=9cm\epsfbox{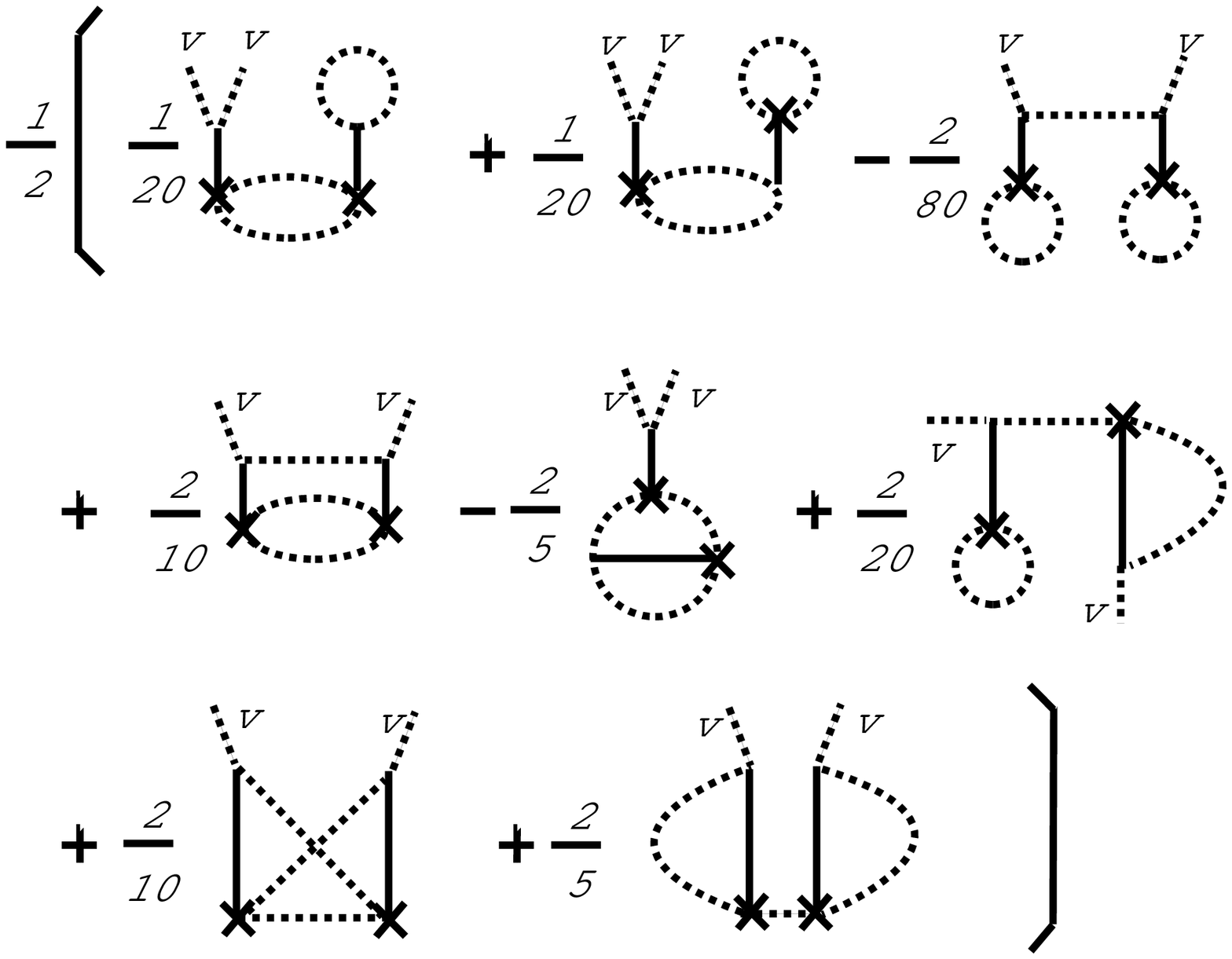}
%\psbox[height=9cm]{figC2f}
      }
         \end{array}                           \nn\\
+\mbox{irrel. terms}\pr
                                                    \label{GraphExp.2}
\end{eqnarray}
%%%%%%%%%%%%%%%%%%%%%%%%%%%%%

%%%%%%%%%%%%%%%%%%%%%%%%%%%%%%%%%%%%%%
\begin{flushleft}
(iii)\ Expansion Terms appearing in $V_2$ of (\ref{exp.2})
\end{flushleft}

%****(C2.4)%%%%%%%%%%%%%%%%%%%%  
\begin{eqnarray}
V_2(x,v;w,r) &=\frac{1}{6!}\pl_{\al_1}\pl_{\al_2}\cdots\pl_{\al_6}W_\mn
\cdot v^{\al_1}v^{\al_2}\cdots v^{\al_6}
(-\frac{\del_\mn}{2r}+\frac{w_\m w_\n}{4r^2})               &\nn\\
& +\frac{1}{5!}\pl_{\al_1}\pl_{\al_2}\cdots\pl_{\al_5}N_\m
\cdot v^{\al_1}v^{\al_2}\cdots v^{\al_5}
(-\frac{w_\m}{2r})                                     &\nn\\
& +\frac{1}{4!}\pl_{\al_1}\pl_{\al_2}\cdots\pl_{\al_4}M
\cdot v^{\al_1}v^{\al_2}\cdots v^{\al_4}\pr        &
                                                    \label{C2.4}
\end{eqnarray}
%%%%%%%%%%%%%%%%%%%%%%%%%%%%%

%****(GraphExp.3), figC2g-i.eps   %%%%%%%%%%%%%%%%%%%%  
\begin{eqnarray}
\frac{1}{6!}\pl_{\al_1}\pl_{\al_2}\cdots\pl_{\al_6}W_\mn
\cdot v^{\al_1}v^{\al_2}\cdots v^{\al_6}=
        \begin{array}{c}
\mbox{
%%%%%%%%%%%%%  FIG  %%%%%%%%%% 
\epsfysize=3cm\epsfbox{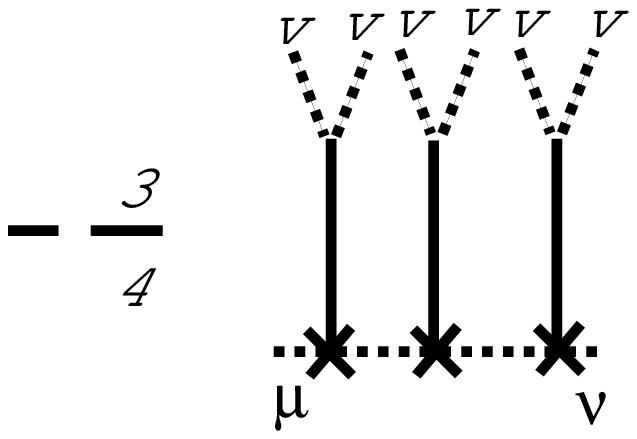}
%\psbox[height=3cm]{figC2g}
      }
        \end{array}
+\mbox{irrel. terms}\com                       \nn\\
\frac{1}{5!}\pl_{\al_1}\pl_{\al_2}\cdots\pl_{\al_5}N_\la
\cdot v^{\al_1}v^{\al_2}\cdots v^{\al_5}=      \nn\\
        \begin{array}{c}
\mbox{
%%%%%%%%%%%%  FIG %%%%%%%%%%%% 
\epsfysize=3cm\epsfbox{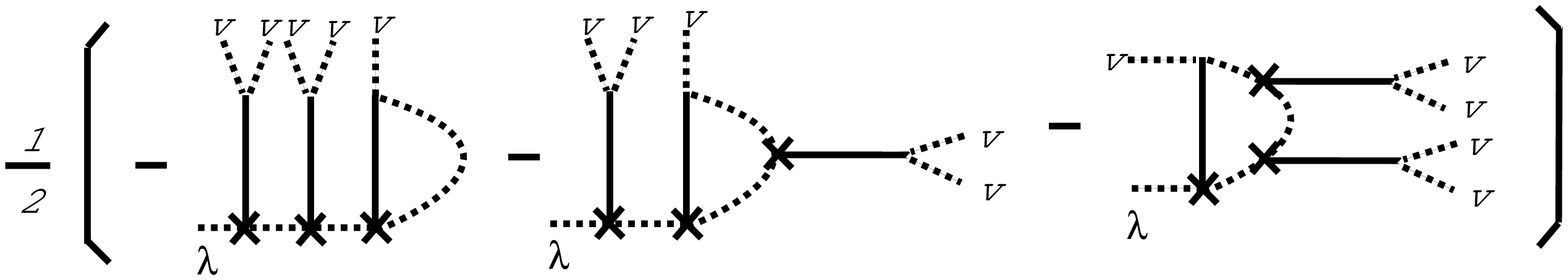}
%\psbox[height=3cm]{figC2h}
      }
        \end{array}                           \nn\\
+\mbox{irrel. terms}\com                       \nn\\
\frac{1}{4!}\pl_{\al_1}\pl_{\al_2}\cdots\pl_{\al_4}M
\cdot v^{\al_1}v^{\al_2}\cdots v^{\al_4}=      \nn\\
        \begin{array}[t]{c}
\mbox{
%%%%%%%%%%%%  FIG  %%%%%%%%%% 
\epsfysize=3cm\epsfbox{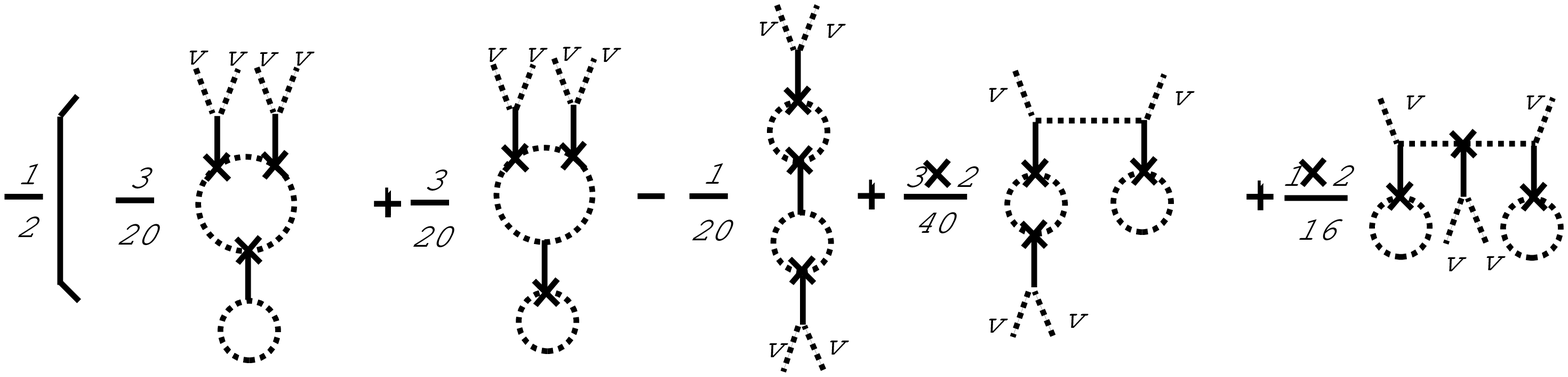}
%\psbox[height=3cm]{figC2i2L}
      }
        \end{array}                                                 \nn\\
        \begin{array}[t]{c}
\mbox{
%%%%%%%%%%%%%  FIG  %%%%%%%%%% 
\epsfysize=6cm\epsfbox{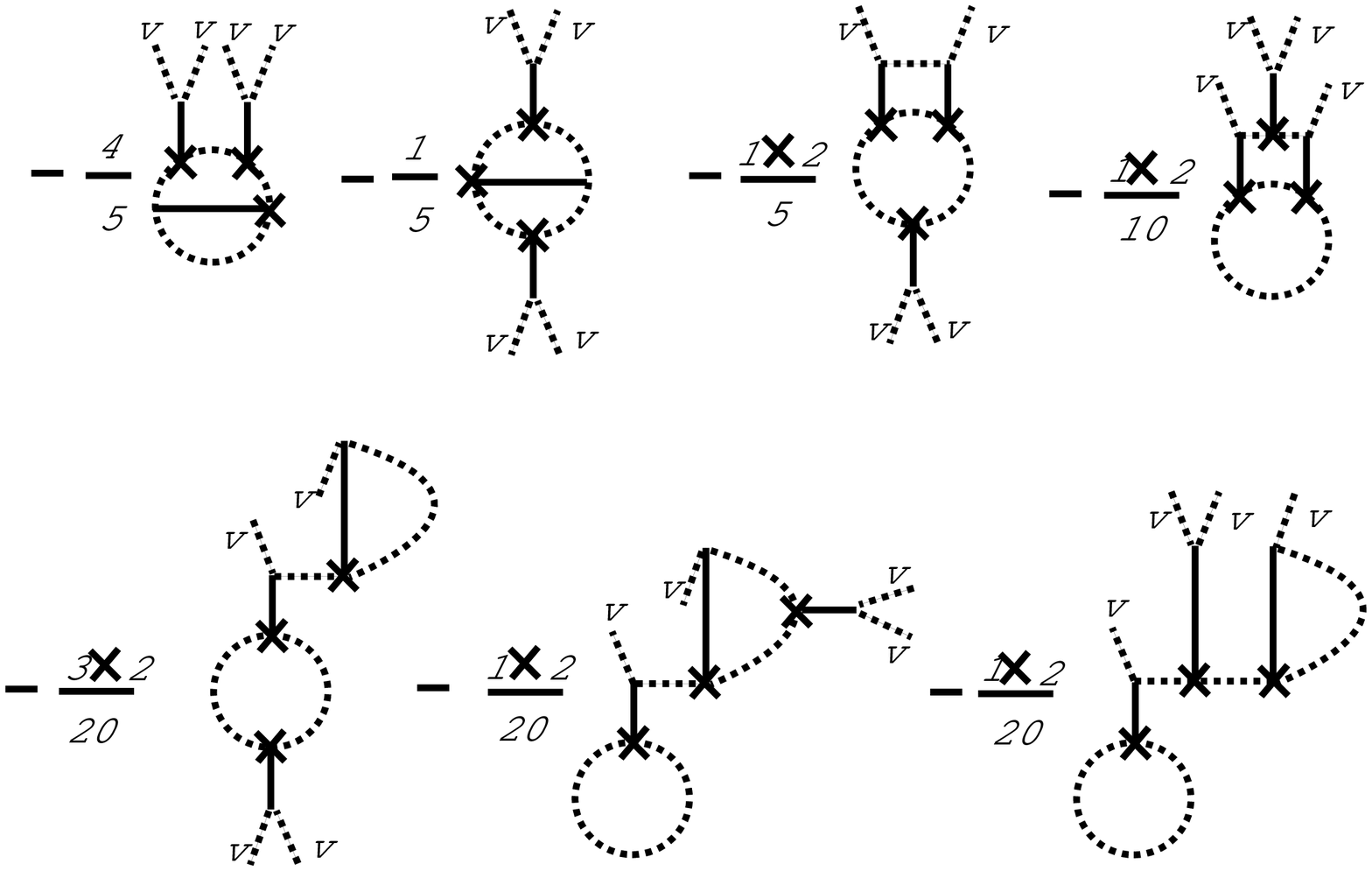}
%\psbox[height=6cm]{figC2i1L}
      } 
        \end{array}                           \nn\\
		\begin{array}[t]{c}                                                           
\mbox{
%%%%%%%%%%%%%%  FIG %%%%%%%%%% 
\epsfysize=3cm\epsfbox{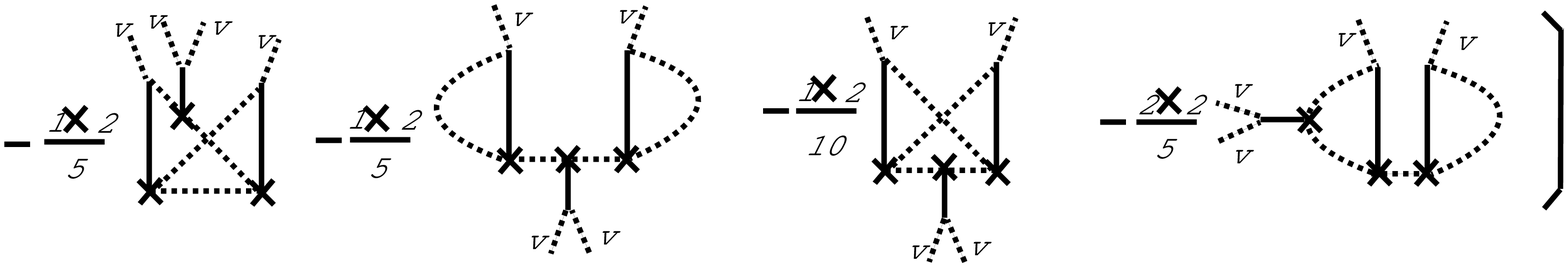}
%\psbox[height=3cm]{figC2i0L}
      }
        \end{array}                                     \nn\\                                                       
+\mbox{irrel. terms}\pr
                                                    \label{GraphExp.3}
\end{eqnarray}
%%%%%%%%%%%%%%%%%%%%%%%%%%%%%

\newpage
%%%%%%%%%%%%%%%%%%%%%%%%%%  App. D  %%%%%%%%%%%%%%%%%%%%%%%%%%%%%%%
%%%%%%%%%%%%%%%%%%%%%%%%%%%%%%%%%%%%%%%%%%%%%%%%%%%%%%%%%%%%%%%%%%%
\begin{flushleft}
{\Large\bf Appendix D.\ Weak-Expansion of Invariants with (Mass)$^6$-Dim }
\end{flushleft}

In 6 dim space, there are totally 17 independent general invariants
given in (\ref{inv.2}). In this appendix, we list the
coefficients of $(\pl\pl h)^3$-terms when the invariants are weak-field-
expanded ($g_\mn=\del_\mn+h_\mn$). 
They are obtained by a computer using a C-program\cite{SI98}. 
Among 17 ones, $O_i(i=1\sim 4)$ terms do not have $(\pl\pl h)^3$-terms.
As shown in the first column of the table, there are 90 independent
$(\pl\pl h)^3$-terms. Some terms (G3,G67,G1,G2) are given, in the ordinary
literal form, in (\ref{coef.0}) and graphically in Fig.9. Their complete
list is graphically given in \cite{II97}. For example the $A_1$ column says
%****(D.1)%%%%%%%%%%%%%%%%%%%%  
\begin{eqnarray}
A_1(R_{\mn\ls}R^{\si\la}_{~~~\tau\om}R^{\om\tau\n\m})
=(-1)\times \mbox{Graph3}(\pl_\si \pl_\tau h_\mn\cdot \pl_\n \pl_\la h_{\tau\om}\cdot
                                                      \pl_\om \pl_\m h_\ls )\nn\\
+(-3)\times \mbox{Graph13}(\pl_\m \pl_\n h_{\tau\si}\cdot \pl_\m \pl_\om h_{\ls}\cdot
                                                      \pl_\tau \pl_\om h_{\n\la} )\nn\\
+(1)\times \mbox{Graph14}(\pl_\m \pl_\n h_{\om\si}\cdot \pl_\n \pl_\la h_{\tau\om}\cdot
                                                      \pl_\la \pl_\m h_{\si\tau}  )\nn\\
+(3)\times \mbox{Graph15}(\pl_\m \pl_\n h_{\om\si}\cdot \pl_\tau \pl_\om h_{\n\la}\cdot
                                                      \pl_\la \pl_\m h_{\si\tau}  )
+O(h^4)
\pr       
                                                    \label{D.1}
\end{eqnarray}
%%%%%%%%%%%%%%%%%%%%%%%%%%%%%
The content of the table is 
fully used in Sec.5 and 6 of the text. Especially the choice of
the four graphs of Fig.9 relies on the table.

%%%%%%%%%%%%%%%%%%%%%%%%%%%%%%%%%%%%%%%%%%%%%%%%%%%%%%%%%%%%%%%%%%%%%%%%%%
%%%%%%%%%%%%%%%%%%%%%  Table G1-G13  %%%%%%%%%%%%%%%%%%%%%%%%%%%%%
%%%%%%%%%%%%%%%%%%%%%%%%%%%%%%%%%%%%%%%%%%%%%%%%%%%%%%%%%%%%%%%%%%%%%%%%%%
\begin{tabular}{|c||c|c|c|c|c|c|c|c||c|c|c|c||c|}
\hline
Graph  & $P_1$ & $P_2$ & $P_3$ & $P_4$ & $P_5$ & $P_6$ & $A_1$ & $B_1$ &
 $T_1$ & $T_2$ & $T_3$ & $T_4$ & $S$ \\
\hline
  $G1$ &  $0$  & $0$   & $0$   & $-\fourth$  & $0$   & $0$   & $0$   & $0$   &
 $0$   &  $0$  & $0$   & $0$   & $0$ \\
\hline
  $G2$ &  $0$  & $0$   & $0$   & $0$  & $\fourth$   & $0$   & $0$   & $0$   &
 $0$   & $0$   & $0$   & $0$   & $0$ \\
\hline
  $G3$ &  $0$  & $0$   & $0$   & $0$  & $0$   & $0$   & $-1$   & $-\fourth$   &
 $0$   & $0$   & $0$   & $0$   & $0$ \\
\hline
  $G4$ &  $0$  & $0$   & $0$ & $-\frac{3}{4}$  & $0$ & $0$   & $0$   & $0$   &
 $0$   &  $0$  & $0$   & $-2$   & $-4$ \\
\hline
  $G5$ &  $0$  & $0$   & $0$   & $0$  & $0$   & $\half$   & $0$   & $0$   &
 $0$   &  $0$  & $0$   & $-1$   & $-2$ \\
\hline
  $G6$ &  $0$  & $0$   & $0$   & $0$  & $0$   & $\half$   & $0$   & $0$   &
 $0$   &  $-1$  & $0$   & $0$   & $0$ \\
\hline
  $G7$ &  $0$  & $0$   & $0$   & $0$  & $0$   & $0$ & $0$ & $\frac{3}{2}$   &
 $0$   &  $0$  & $2$   & $0$   & $0$ \\
\hline
  $G8$ &  $0$  & $0$   & $0$   & $0$  & $\half$   & $0$   & $0$   & $0$   &
 $0$   & $-1$   & $0$   & $0$   & $0$ \\
\hline
  $G9$ &  $0$  & $0$   & $0$   & $0$  & $0$   & $\half$   & $0$   & $0$   &
 $0$   & $0$   & $0$   & $0$   & $0$ \\
\hline
 $G10$ &  $0$  & $0$   & $0$   & $0$  & $\fourth$   & $0$   & $0$   & $0$   &
 $0$   &  $-1$  & $0$   & $0$   & $-8$ \\
\hline
 $G11$ &  $0$  & $0$   & $0$   & $0$  & $0$   & $\half$   & $0$   & $0$   &
 $0$   &  $0$  & $0$   & $0$   & $-16$ \\
\hline
 $G12$ &  $0$  & $0$   & $0$   & $0$  & $0$   & $0$  & $0$ & $\frac{3}{2}$   &
 $0$   &  $0$  & $6$   & $0$   & $-8$ \\
\hline
 $G13$ &  $0$  & $0$   & $0$   & $0$  & $0$   & $0$ & $-3$ & $-\frac{3}{4}$   &
 $0$   &  $0$  & $-2$   & $0$   & $-4$ \\
\hline
\multicolumn{14}{c}{\q}                                                 \\
\multicolumn{14}{c}{Table 1\ \ Weak-Expansion of Invariants with (Mass)$^6$-Dim.:\  $(\pl\pl h)^3$-Part }\\
\multicolumn{14}{c}{G1-G13(\ul{l}[No of suffix-loops]=1)           }\\
\end{tabular}
%%%%%%%%%%%%%%%%%%%%%%%%%  END  of  Table ** %%%%%%%%%%%%%%%%%%%%%%%%%%%%%
%%%%%%%%%%%%%%%%%%%%%%%%%%%%%%%%%%%%%%%%%%%%%%%%%%%%%%%%%%%%%%%%%%%%%%%%%
\newpage
%%%%%%%%%%%%%%%%%%%%%%%%%%%%%%%%%%%%%%%%%%%%%%%%%%%%%%%%%%%%%%%%%%%%%%%%%%
%%%%%%%%%%%%%%%%%%%%%  Table G14-G42 %%%%%%%%%%%%%%%%%%%%%%%%%%%%%
%%%%%%%%%%%%%%%%%%%%%%%%%%%%%%%%%%%%%%%%%%%%%%%%%%%%%%%%%%%%%%%%%%%%%%%%%%
\begin{tabular}{|c||c|c|c|c|c|c|c|c||c|c|c|c||c|}
\hline
Graph  & $P_1$ & $P_2$ & $P_3$ & $P_4$ & $P_5$ & $P_6$ & $A_1$ & $B_1$ &
 $T_1$ & $T_2$ & $T_3$ & $T_4$ & $S$ \\
\hline
 $G14$ &  $0$  & $0$   & $0$   & $0$  & $0$   & $0$   & $1$   & $0$   &
 $0$   &  $0$  & $-1$   & $0$   & $12$ \\
\hline
 $G15$ &  $0$  & $0$   & $0$   & $0$  & $0$   & $0$   & $3$   & $0$   &
 $0$   &  $0$  & $-1$   & $0$   & $0$ \\
\hline
 $G16$ &  $0$  & $0$   & $0$   & $0$  & $-1$   & $0$   & $0$   & $0$   &
 $0$   &  $1$  & $0$   & $-2$   & $-4$ \\
\hline
 $G17$ &  $0$  & $0$   & $0$   & $0$  & $0$   & $-\half$   & $0$   & $0$   &
 $0$   &  $1$  & $0$   & $0$   & $-16$ \\
\hline
 $G18$ &  $0$  & $0$   & $0$   & $0$  & $0$   & $0$ & $0$ & $-\frac{3}{4}$   &
 $0$   &  $0$  & $-1$   & $0$   & $-4$ \\
\hline
 $G19$ &  $0$  & $0$   & $0$   & $0$  & $0$   & $-\half$   & $0$   & $0$   &
 $0$   &  $\half$  & $0$   & $\frac{3}{2}$   & $7$ \\
\hline
 $G20$ &  $0$  & $0$   & $0$   & $0$  & $0$   & $0$ & $0$ & $-\frac{3}{4}$   &
 $0$   &  $0$  & $-2$   & $0$   & $6$ \\
\hline
 $G21$ &  $0$  & $0$   & $0$   & $0$  & $0$   & $-\half$   & $0$   & $0$   &
 $0$   &  $0$  & $0$   & $0$   & $0$ \\
\hline
 $G22$ &  $0$  & $0$   & $0$   & $0$  & $0$   & $0$ & $0$ & $-\frac{3}{2}$   &
 $0$   &  $0$  & $-2$   & $0$   & $0$ \\
\hline
 $G23$ &  $0$  & $-\half$   & $0$   & $0$  & $0$   & $0$   & $0$   & $0$   &
 $-2$   &  $0$  & $0$   & $0$   & $0$ \\
\hline
 $G24$ &  $0$  & $0$   & $2$   & $0$  & $0$   & $0$   & $0$   & $0$   &
 $-1$   &  $0$  & $0$   & $0$   & $0$ \\
\hline
 $G25$ &  $0$  & $0$   & $0$   & $0$  & $0$   & $-\half$   & $0$   & $0$   &
 $0$   &  $0$  & $0$   & $0$   & $0$ \\
\hline
 $G26$ &  $0$  & $-\half$ & $0$ & $0$  & $0$   & $0$   & $0$   & $0$   &
 $0$   &  $0$  & $0$   & $0$   & $0$ \\
\hline
 $G27$ &  $0$  & $0$   & $0$   & $\frac{3}{8}$ & $0$ & $0$   & $0$   & $0$   &
 $0$   &  $0$  & $0$   & $1$   & $2$ \\
\hline
 $G28$ &  $0$  & $0$   & $0$   & $0$  & $0$   & $-\fourth$   & $0$   & $0$   &
 $0$   &  $0$  & $0$   & $\half$   & $1$ \\
\hline
 $G29$ &  $0$  & $0$   & $0$   & $0$  & $-\half$   & $0$   & $0$   & $0$   &
 $0$   &  $\frac{5}{4}$  & $0$   & $0$   & $8$ \\
\hline
 $G30$ &  $0$  & $0$   & $0$   & $0$      & $0$        & $-\half$   & $0$   & $0$   &
 $0$   &  $0$  & $0$   & $0$   & $8$ \\
\hline
 $G31$ &  $0$  & $0$   & $0$   & $0$  & $-\half$   & $0$   & $0$   & $0$   &
 $0$   &  $\fourth$  & $0$   & $0$   & $0$ \\
\hline
 $G32$ &  $0$  & $0$   & $0$   & $\frac{3}{4}$ & $0$ & $0$   & $0$   & $0$   &
 $0$   &  $0$  & $0$   & $1$   & $2$ \\
\hline
 $G33$ &  $0$  & $0$   & $0$   & $\frac{3}{8}$ & $0$ & $0$   & $0$   & $0$   &
 $0$   &  $0$  & $0$   & $1$   & $2$ \\
\hline
 $G34$ &  $0$  & $0$   & $0$   & $0$  & $0$   & $-\fourth$   & $0$   & $0$   &
 $0$   &  $\half$  & $0$   & $0$   & $0$ \\
\hline
 $G35$ &  $0$  & $0$   & $0$   & $\frac{3}{8}$ & $0$ & $0$   & $0$   & $0$   &
 $0$   &  $-\fourth$  & $0$   & $0$   & $-4$ \\
\hline
 $G36$ &  $0$  & $0$   & $0$   & $0$  & $0$   & $-\fourth$   & $0$   & $0$   &
 $0$   &  $\half$  & $\half$   & $0$   & $-2$ \\
\hline
 $G37$ &  $0$  & $0$   & $0$   & $0$  & $-\half$   & $0$   & $0$   & $0$   &
 $0$   &  $\half$  & $0$   & $0$   & $-8$ \\
\hline
 $G38$ &  $0$  & $0$   & $0$   & $0$  & $0$   & $-\half$   & $0$   & $0$   &
 $0$   &  $0$  & $1$   & $0$   & $-4$ \\
\hline
 $G39$ &  $0$  & $0$   & $0$   & $\frac{3}{8}$ & $0$ & $0$   & $0$   & $0$   &
 $0$   &  $-\fourth$  & $0$   & $1$   & $-2$ \\
\hline
 $G40$ &  $0$  & $0$   & $0$   & $0$  & $0$   & $-\fourth$   & $0$   & $0$   &
 $0$   &  $0$  & $\half$   & $\half$   & $-1$ \\
\hline
 $G41$ &  $0$  & $0$   & $0$   & $0$  & $-\half$   & $0$   & $0$   & $0$   &
 $0$   &  $-\half$  & $0$   & $0$   & $0$ \\
\hline
 $G42$ &  $0$  & $0$   & $0$   & $\frac{3}{4}$ & $0$ & $0$   & $0$   & $0$   &
 $0$   &  $-\half$  & $0$   & $0$   & $0$ \\
\hline
\multicolumn{14}{c}{\q}                                                 \\
\multicolumn{14}{c}{Table 1\ \ Weak-Expansion of Invariants with (Mass)$^6$-Dim.:\  $(\pl\pl h)^3$-Part }\\
\multicolumn{14}{c}{ G14-G42(\ul{l}=2)           }\\
\end{tabular}
%%%%%%%%%%%%%%%%%%%%%%%%%  END  of  Table ** %%%%%%%%%%%%%%%%%%%%%%%%%%%%%
%%%%%%%%%%%%%%%%%%%%%%%%%%%%%%%%%%%%%%%%%%%%%%%%%%%%%%%%%%%%%%%%%%%%%%%%%

\newpage
%%%%%%%%%%%%%%%%%%%%%%%%%%%%%%%%%%%%%%%%%%%%%%%%%%%%%%%%%%%%%%%%%%%%%%%%%%
%%%%%%%%%%%%%%%%%%%%%  Table G43-G69 %%%%%%%%%%%%%%%%%%%%%%%%%%%%%%%%%%%%%%%%
%%%%%%%%%%%%%%%%%%%%%%%%%%%%%%%%%%%%%%%%%%%%%%%%%%%%%%%%%%%%%%%%%%%%%%%%%%
\begin{tabular}{|c||c|c|c|c|c|c|c|c||c|c|c|c||c|}
\hline
Graph  & $P_1$ & $P_2$ & $P_3$ & $P_4$ & $P_5$ & $P_6$ & $A_1$ & $B_1$ &
 $T_1$ & $T_2$ & $T_3$ & $T_4$ & $S$ \\
\hline
 $G43$ &  $0$  & $0$   & $0$   & $0$  & $\fourth$   & $0$   & $0$   & $0$   &
 $0$   &  $\half$  & $0$   & $0$   & $-2$ \\
\hline
 $G44$ &  $0$  & $0$   & $0$   & $-\frac{3}{4}$ & $0$ & $0$   & $0$   & $0$   &
 $0$   &  $1$  & $0$   & $0$   & $-8$ \\
\hline
 $G45$ &  $0$  & $0$   & $0$   & $0$  & $0$   & $\fourth$   & $0$   & $0$   &
 $0$   &  $-\half$  & $-\half$   & $0$   & $14$ \\
\hline
 $G46$ &  $0$  & $0$   & $0$   & $0$  & $0$   & $\fourth$   & $0$   & $0$   &
 $0$   &  $-\fourth$  & $0$   & $-\frac{3}{4}$   & $-\frac{7}{2}$ \\
\hline
 $G47$ &  $0$  & $0$   & $0$   & $0$  & $0$   & $\fourth$   & $0$   & $0$   &
 $0$   &  $-\half$  & $0$   & $0$   & $8$ \\
\hline
 $G48$ &  $0$  & $0$   & $0$   & $0$  & $\half$   & $0$   & $0$   & $0$   &
 $0$   &  $-\fourth$  & $0$   & $1$   & $2$ \\
\hline
 $G49$ &  $0$  & $0$   & $0$   & $0$  & $0$   & $0$   & $0$ & $\frac{3}{4}$   &
 $0$   &  $0$  & $1$   & $0$   & $4$ \\
\hline
 $G50$ &  $0$  & $0$   & $-1$   & $0$  & $0$   & $0$   & $0$   & $0$   &
 $\frac{3}{2}$   &  $0$  & $0$   & $0$   & $0$ \\
\hline
 $G51$ &  $0$  & $0$   & $0$   & $-\frac{3}{4}$ & $0$ & $0$   & $0$   & $0$   &
 $0$   &  $\half$  & $0$   & $0$   & $4$ \\
\hline
 $G52$ &  $0$  & $0$   & $0$   & $0$  & $\half$   & $0$   & $0$   & $0$   &
 $0$   &  $\fourth$  & $0$   & $0$   & $4$ \\
\hline
 $G53$ &  $0$  & $\half$   & $0$   & $0$  & $0$   & $0$   & $0$   & $0$   &
 $2$   &  $0$  & $0$   & $0$   & $0$ \\
\hline
 $G54$ &  $0$  & $0$   & $-2$   & $0$  & $0$   & $0$   & $0$   & $0$   &
 $1$   &  $0$  & $0$   & $0$   & $0$ \\
\hline
 $G55$ &  $0$  & $0$   & $0$   & $-\frac{3}{4}$ & $0$ & $0$   & $0$   & $0$   &
 $0$   &  $\half$  & $0$   & $-1$   & $2$ \\
\hline
 $G56$ &  $0$  & $\half$   & $0$   & $0$  & $0$   & $0$   & $0$   & $0$   &
 $0$   &  $0$  & $0$   & $0$   & $0$ \\
\hline
 $G57$ &  $0$  & $0$   & $0$   & $0$  & $0$   & $\fourth$   & $0$   & $0$   &
 $0$   &  $-\fourth$  & $-\half$   & $-\frac{3}{4}$   & $\frac{9}{2}$ \\
\hline
 $G58$ &  $0$  & $0$   & $0$   & $0$  & $\half$   & $0$   & $0$   & $0$   &
 $0$   &  $0$  & $0$   & $1$   & $4$ \\
\hline
 $G59$ &  $0$  & $0$   & $0$   & $0$  & $0$   & $\half$   & $0$   & $0$   &
 $0$   &  $0$  & $-1$   & $0$   & $0$ \\
\hline
 $G60$ &  $0$  & $0$   & $0$   & $0$  & $\half$   & $0$   & $0$   & $0$   &
 $0$   &  $\half$  & $0$   & $1$   & $-2$ \\
\hline
 $G61$ &  $0$  & $1$   & $0$   & $0$  & $0$   & $0$   & $0$   & $0$   &
 $-2$   &  $0$  & $0$   & $0$   & $0$ \\
\hline
 $G62$ &  $0$  & $0$   & $0$   & $-\frac{3}{4}$ & $0$ & $0$   & $0$   & $0$   &
 $0$   &  $0$  & $0$   & $-\frac{3}{2}$   & $-3$ \\
\hline
 $G63$ &  $0$  & $0$   & $0$   & $0$  & $\fourth$   & $0$   & $0$   & $0$   &
 $0$   &  $-\fourth$  & $0$   & $0$   & $0$ \\
\hline
 $G64$ &  $0$  & $0$   & $0$   & $0$  & $0$   & $\half$   & $0$   & $0$   &
 $0$   &  $0$  & $0$   & $0$   & $0$ \\
\hline
 $G65$ &  $0$  & $0$   & $0$   & $0$  & $\half$   & $0$   & $0$   & $0$   &
 $0$   &  $-\half$  & $0$   & $1$   & $2$ \\
\hline
 $G66$ &  $0$  & $1$   & $0$   & $0$  & $0$   & $0$   & $0$   & $0$   &
 $2$   &  $0$  & $0$   & $0$   & $0$ \\
\hline
 $G67$ &  $0$  & $0$   & $0$   & $0$  & $0$   & $0$   & $0$   & $\fourth$   &
 $0$   &  $0$  & $0$   & $0$   & $0$ \\
\hline
 $G68$ &  $0$  & $0$   & $-1$   & $0$  & $0$   & $0$   & $0$   & $0$   &
 $0$   &  $0$  & $0$   & $0$   & $0$ \\
\hline
 $G69$ &  $-1$  & $0$   & $0$   & $0$  & $0$   & $0$   & $0$   & $0$   &
 $0$   &  $0$  & $0$   & $0$   & $0$ \\
\hline
\multicolumn{14}{c}{\q}                                                 \\
\multicolumn{14}{c}{Table 1\ \ Weak-Expansion of Invariants with (Mass)$^6$-Dim.:\  $(\pl\pl h)^3$-Part }\\
\multicolumn{14}{c}{G43-G69(\ul{l}=3)           }\\
\end{tabular}
%%%%%%%%%%%%%%%%%%%%%%%%%  END  of  Table G43-G69 %%%%%%%%%%%%%%%%%%%%%%%%%%%
%%%%%%%%%%%%%%%%%%%%%%%%%%%%%%%%%%%%%%%%%%%%%%%%%%%%%%%%%%%%%%%%%%%%%%%%%
\newpage
%%%%%%%%%%%%%%%%%%%%%%%%%%%%%%%%%%%%%%%%%%%%%%%%%%%%%%%%%%%%%%%%%%%%%%%%%%
%%%%%%%%%%%%%%%%%%%%%  Table G70-G90 %%%%%%%%%%%%%%%%%%%%%%%%%%%%%%%%%%%%%%%%
%%%%%%%%%%%%%%%%%%%%%%%%%%%%%%%%%%%%%%%%%%%%%%%%%%%%%%%%%%%%%%%%%%%%%%%%%%
\begin{tabular}{|c||c|c|c|c|c|c|c|c||c|c|c|c||c|}
\hline
Graph  & $P_1$ & $P_2$ & $P_3$ & $P_4$ & $P_5$ & $P_6$ & $A_1$ & $B_1$ &
 $T_1$ & $T_2$ & $T_3$ & $T_4$ & $S$ \\
\hline
 $G70$ &  $0$  & $0$   & $0$   & $\frac{1}{8}$ & $0$ & $0$   & $0$   & $0$   &
 $0$   &  $-\fourth$  & $0$   & $0$   & $4$   \\
\hline
 $G71$ &  $0$  & $0$   & $0$   & $0$  & $-\fourth$   & $0$   & $0$   & $0$   &
 $0$   &  $-\frac{1}{8}$  & $0$   & $-\half$   & $-2$  \\
\hline
 $G72$ &  $0$  & $0$   & $1$   & $0$  & $0$   & $0$   & $0$   & $0$   &
 $-\frac{3}{2}$   &  $0$  & $0$   & $0$   & $0$   \\
\hline
 $G73$ &  $0$  & $0$   & $0$   & $\frac{3}{8}$ & $0$ & $0$   & $0$   & $0$   &
 $0$   &  $-\half$  & $0$   & $0$   & $4$   \\
\hline
 $G74$ &  $0$  & $-1$   & $0$   & $0$  & $0$   & $0$   & $0$   & $0$   &
 $2$   &  $0$  & $0$   & $0$   & $0$   \\
\hline
 $G75$ &  $0$  & $0$   & $0$   & $0$  & $-\fourth$   & $0$   & $0$   & $0$   &
 $0$   &  $-\half$  & $0$   & $-\half$   & $3$   \\
\hline
 $G76$ &  $0$  & $-\fourth$   & $0$   & $0$  & $0$   & $0$   & $0$   & $0$   &
 $1$   &  $0$  & $0$   & $0$   & $0$   \\
\hline
 $G77$ &  $0$  & $0$   & $0$   & $\frac{1}{8}$  & $0$   & $0$   & $0$   & $0$   &
 $0$   &  $0$  & $0$   & $\fourth$   & $\half$   \\
\hline
 $G78$ &  $0$  & $0$   & $0$   & $0$  & $-\fourth$   & $0$   & $0$   & $0$   &
 $0$   &  $\frac{1}{8}$  & $0$   & $-\half$   & $-1$   \\
\hline
 $G79$ &  $0$  & $-\fourth$   & $0$   & $0$  & $0$   & $0$   & $0$   & $0$   &
 $-\half$   &  $0$  & $0$   & $0$   & $0$   \\
\hline
 $G80$ &  $0$  & $0$   & $0$   & $\frac{3}{8}$  & $0$   & $0$   & $0$   & $0$   &
 $0$   &  $-\fourth$  & $0$   & $\fourth$   & $-\half$   \\
\hline
 $G81$ &  $0$  & $-1$   & $0$   & $0$  & $0$   & $0$   & $0$   & $0$   &
 $-2$   &  $0$  & $0$   & $0$   & $0$   \\
\hline
 $G82$ &  $0$  & $0$   & $0$   & $0$  & $-\fourth$   & $0$   & $0$   & $0$   &
 $0$   &  $-\fourth$  & $0$   & $-\half$   & $1$   \\
\hline
 $G83$ &  $0$  & $-\half$   & $0$   & $0$  & $0$   & $0$   & $0$   & $0$   &
 $1$   &  $0$  & $0$   & $0$   & $0$   \\
\hline
 $G84$ &  $0$  & $0$   & $1$   & $0$  & $0$   & $0$   & $0$   & $0$   &
 $0$   &  $0$  & $0$   & $0$   & $0$   \\
\hline
 $G85$ &  $3$  & $0$   & $0$   & $0$  & $0$   & $0$   & $0$   & $0$   &
 $0$   &  $0$  & $0$   & $0$   & $0$   \\
\hline
\hline
 $G86$ &  $0$  & $\fourth$   & $0$   & $0$  & $0$   & $0$   & $0$   & $0$   &
 $-1$   &  $0$  & $0$   & $0$   & $0$   \\
\hline
 $G87$ &  $0$  & $\fourth$   & $0$   & $0$  & $0$   & $0$   & $0$   & $0$   &
 $\half$   &  $0$  & $0$   & $0$   & $0$   \\
\hline
 $G88$ &  $-3$  & $0$   & $0$   & $0$  & $0$   & $0$   & $0$   & $0$   &
 $0$   &  $0$  & $0$   & $0$   & $0$   \\
\hline
 $G89$ &  $0$  & $\half$   & $0$   & $0$  & $0$   & $0$   & $0$   & $0$   &
 $-1$   &  $0$  & $0$   & $0$   & $0$   \\
\hline
\hline
 $G90$ &  $1$  & $0$   & $0$   & $0$  & $0$   & $0$   & $0$   & $0$   &
 $0$   &  $0$  & $0$   & $0$   & $0$   \\
\hline
\multicolumn{14}{c}{\q}                                                 \\
\multicolumn{14}{c}{Table 1\ \ Weak-Expansion of Invariants with (Mass)$^6$-Dim.:\  $(\pl\pl h)^3$-Part }\\
\multicolumn{14}{c}{G70-G85(\ul{l}=4),G86-G89(\ul{l}=5),G90(\ul{l}=6)     }\\
\end{tabular}
%%%%%%%%%%%%%%%%%%%%%%%%%  END  of  Table G70-G90 %%%%%%%%%%%%%%%%%%%%%%%%%%%%%
%%%%%%%%%%%%%%%%%%%%%%%%%%%%%%%%%%%%%%%%%%%%%%%%%%%%%%%%%%%%%%%%%%%%%%%%%

\vs 1
%%%%%%%%%%%%%%%%%%%%%%%%%%%%%%%%%%%%%%%%%%%%%%%%%%%%%%%%%%%%%%%%%%
%%%%%%%%%%%%%%%%%%%%%%%% reference %%%%%%%%%%%%%%%%%%%%%%%%%%%%%%%
%%%%%%%%%%%%%%%%%%%%%%%%%%%%%%%%%%%%%%%%%%%%%%%%%%%%%%%%%%%%%%%%%%

%%%%%%%%%%%%%%%%%%%%%%%%%%%%%

\newpage

\begin{flushleft}
{\Large\bf Figure Captions }
\end{flushleft}

\begin{itemize}
\item
Fig.1\ Graph of $G_0(x-y;t)$
\item
Fig.2\ Graph of $G_1(x,y;t)$
\item
Fig.3\ Graph of $G_2(x,y;t)$
\item
Fig.4\ Graph of $G_3(x,y;t)$
\item
Fig.5\ Graph of $G_0(0;t)$
\item
Fig.6\ Graph of $G_1(x,x;t)$
\item
Fig.7\ Graph of $G_2(x,x;t)$
\item
Fig.8\ Graph of $G_3(x,x;t)$
\item
Fig.9\ 
Four Graphs of 3,67,1 and 2. See the (\ref{graph.1})
 for the definition of the graph.
\item
Fig.10\ 
Two important $(\pl\pl h)^4$-graphs in 8 dim. 
\item
Fig.11\ 
Two important $(\pl\pl h)^5$-graphs in 10 dim. 
\item
Fig.12\ 
Propagator Rule 1, Eq.(\ref{a.5}).
\item
Fig.13\ 
Propagator Rule 2, Eq.(\ref{a.6x}).
\item
Fig.14\ 
Propagator Rule 3, Eq.(\ref{a.6y}).
\item
Fig.15\ 
Propagator Rule 4, Eq.(\ref{a.6z}).
\item
Fig.16\ 
$W^\mn$ and $N_\la$.
\item
Fig.17\ 
$M$, Order of $h$.
\item
Fig.18\ 
$M$, Order of $h^2$.
\item
Fig.19\ 
$M$, Order of $h^3$ and loop no=2.
\item
Fig.20\ 
$M$, Order of $h^3$ and loop no=1.
\item
Fig.21\ 
$M$, Order of $h^3$ and loop no=0.
\end{itemize}


\begin{thebibliography}{99}
\bibitem{foot1}   %*foot1*
{
Other examples of the coordinate space approach are
the operator product expansion method and the lattice field theory.
The former was exploited in the 2 dim conformal field theory.
Traditionally the coordinate space has been used for the analysis of
conformally invariant or covariant Green functions. For example,
see a recent progress of \cite{PS98}.
}
\bibitem{foot1bb} %*foot1bb*
{
The heat-kernel is formulated in the coordinate space, 
using the quatum-mechanical non-linear sigma model,
and the Feynman rules are given
in \cite{fB91,BN93,BPSN95,BPSN96}.
}
\bibitem{foot1b}  %*foot1b*
{
It must be compared with the dimensional regularization.
}
\bibitem{foot2}    %*foot2*
{
The case of special $W^{ij}_\mn$~(Riemann geometry) was done by
Gilkey\cite{Gil}, where he analyzed using the normal coordinates in order to
keep the general covariance manifestly.
}
\bibitem{foot4}    %*foot4*
{
These properties come only from the physical dimensions of
$W_\mn,N_\m$ and $M$, not the speciality of the present model. The 
relevant-part selection in this section is valid for general theories.
         }
\bibitem{foot5}    %*foot5*  
{
The divergent parts $t^{-1},t^{-2},\cdots$ are considered to
be canceled by the divergent counter-terms. They generally
correspond
to power-divergences (not log-divergence ) and the counter-terms are 
those operators which have massive couplings: ex. the cosmological term,
the Einstein term.
}
\bibitem{foot5b}   %*foot5b*  
{The
$\wbar^8$-term gives disconnected graphs, and $\wbar^6$-terms give
graphs which conflict with the second one of the common features
of Graph3 and 67.
}
\bibitem{foot7}   %*foot7*
{
The total independent $(\pl\pl h)^3$-terms are given by 90 terms
shown in the first column of the table in App.D. 
}
\end{thebibliography}
\end{document}